\documentclass[structabstract]{aa}                                           

\usepackage{natbib}
\bibpunct{(}{)}{;}{a}{}{,} 

\usepackage{graphicx}
\usepackage{framed}

\usepackage{txfonts}
\begin{document}
%
\def\etal{et al.~\/}
\def\eg{{\it e.g.~\/}}
\def\etc{{\it etc.~\/}}
\def\ie{{\it i.e.~\/}}
\def\cf{{\it c.f.~\/}}  

\def\mic{{$\mu$m}}

\def\fout{{$S_{out}$~\/}}
\def\fin{{$S_{in}$~\/}}
\def\deltaf{{$\Delta S$~\/}}
\def\dfin{{$\Delta S/S_{in}$~\/}}
\def\dfout{{$|\Delta S|/S_{out}$~\/}}
\def\absdfin{{$\left\vert \Delta S/S_{in} \right\vert$~\/}}
\def\absdf{{$\left\vert \Delta S \right\vert$~\/}}

\def\f24{{S$_{24\mu}$m}}
\def\f100{{S$_{100\mu}$m}}
\def\f250{{S$_{250\mu}$m}}
\def\f350{{S$_{350\mu}$m}}
\def\f500{{S$_{500\mu}$m}}

\def\My{{M$_{\odot}$yr$^{-1}$}}
\def\LIR{{$L_{\rm IR}$}}
\def\Msun{{M$_{\odot}$\,}}
\def\Lsun{{L$_{\odot}$\,}}
\def\FIR{{$L_{\rm FIR}$}}
\def\L24{{$L_{24\mu}$m}}
\def\L8{{$L_{8\mu}$m}}
\def\zz{{\it z~\/}} 

\def\hst{{\it HST~\/}}
\def\herschel{{\it Herschel}}
\def\spitzer{{\it Spitzer}}
\def\goodsh{GOODS-{\it Herschel~\/}}
\def\spire{{SPIRE~\/}}
\def\pacs{{PACS~\/}}
\def\goods{{GOODS~\/}}


\def\cirba{65\%}   
\def\cirbb{94\%}   
\def\cirbc{74\%}   
\def\cirbd{37\%}   
\def\cirbe{8\%}     

\title{Starbursts \textit{in} and \textit{out} of the star-formation main sequence} 
\author{D.~Elbaz \inst{1}
\and R.~Leiton \inst{2,3,1}
\and N.~Nagar \inst{2}
\and K.~Okumura\inst{1}
\and M.~Franco \inst{1}
\and C.~Schreiber \inst{4,1}
\and M.~Pannella \inst{5,1}
\and T.~Wang\inst{1,6}
\and M.~Dickinson\inst{7}
\and T.~D\'iaz-Santos\inst{8}
\and L.~Ciesla\inst{1}
\and E.~Daddi\inst{1}
\and F.~Bournaud\inst{1}
\and G.~Magdis\inst{9,10}
\and L.~Zhou\inst{1,11}
\and W.~Rujopakarn\inst{12,13,14}
}
\institute{Laboratoire AIM-Paris-Saclay, CEA/DRF/Irfu - CNRS - Universit\'e Paris Diderot, CEA-Saclay, pt courrier 131, F-91191 Gif-sur-Yvette, France\\ %
\email{delbaz@cea.fr}  
\and Department of Astronomy, Universidad de Concepci\'on, Casilla 160-C Concepci\'on, Chile
\and Instituto de F\'{i}sica y Astronom\'{i}a, Universidad de Valpara\'{i}so, Avda. Gran Breta\~na 1111, Valparaiso, Chile
\and Leiden Observatory, Leiden University, NL-2300 RA Leiden, The Netherlands %
\and Fakult\"at f\"ur Physik der Ludwig-Maximilians-Universit\"at, D-81679 M\"unchen, Germany 
\and Institute of Astronomy, The University of Tokyo, 2-21-1 Osawa, Mitaka, Tokyo, 181-0015 Japan
\and National Optical Astronomy Observatory, 950 North Cherry Avenue, Tucson, AZ 85719, USA
\and N\'ucleo de Astronom\'ia de la Facultad de Ingenier\'ia, Universidad Diego Portales, Av. Ej\'ercito Libertador 441, Santiago, Chile
\and Dark Cosmology Centre, Niels Bohr Institute, University of Copenhagen, Juliane Mariesvej 30, 2100, Copenhagen, Denmark
\and Institute for Astronomy, Astrophysics, Space Applications and Remote Sensing, National Observatory of Athens, GR-15236 Athens, Greece
\and School of Astronomy and Space Science, Nanjing University, Nanjing 210093, China
\and Department of Physics, Faculty of Science, Chulalongkorn University, 254 Phayathai Road, Pathumwan, Bangkok 10330, Thailand
\and National Astronomical Research Institute of Thailand (Public Organization), Donkaew, Maerim, Chiangmai 50180, Thailand
\and Kavli Institute for the Physics and Mathematics of the Universe (WPI),The University of Tokyo Institutes for Advanced Study, The University of Tokyo, Kashiwa, Chiba 277-8583, Japan
}
   \date{Received ; accepted }

 
  \abstract
   {}
 {We use high-resolution continuum images obtained with the Atacama Large Millimeter Array (ALMA) to probe the surface density of star-formation in $z$$\sim$2 galaxies and study the different physical properties between galaxies \textit{within} and \textit{above} the star-formation main sequence of galaxies.}
{We use ALMA images at 870\,$\mu$m with 0.2 arcsec resolution in order to resolve star-formation in a sample of eight star-forming galaxies at $z$$\sim$2 selected among the most massive \textit{Herschel} galaxies in the GOODS-\textit{South} field. This sample is supplemented with eleven galaxies from the public data of the 1.3 mm survey of the Hubble Ultra-Deep Field, HUDF. We derive dust and gas masses for the galaxies, compute their depletion times and gas fractions and study the relative distributions of rest-frame UV and far-infrared light.}
{ALMA reveals systematically dense concentrations of dusty star-formation close to the center of the stellar component of the galaxies. We identify two different starburst regimes: \textit{(i)} the classical population of starbursts located above the SFR-M$_\star$ main sequence, with enhanced gas fractions and short depletion times and \textit{(ii)} a sub-population of galaxies located \textit{within} the scatter of the main sequence that experience compact star formation with depletion timescales typical of starbursts of $\sim$150 Myr. In both starburst populations, the far infrared and UV are distributed in distinct regions and dust-corrected star formation rates estimated using UV-optical-NIR data alone underestimate the total star formation rate. Starbursts \textit{hidden} in the main sequence show instead the lowest gas fractions of our sample and could represent the last stage of star-formation prior to passivization. Being \textit{Herschel}-selected, these main sequence galaxies are located in the high-mass end of the main sequence, hence we do not know whether these "starbursts hidden in the main sequence" also exist below 10$^{11}$ M$_{\odot}$. Active galactic nuclei are found to be ubiquitous in these compact starbursts, suggesting that the triggering mechanism also feeds the central black hole or that the active nucleus triggers star formation.
}
{}

  \authorrunning{D. Elbaz et al.}
\titlerunning{Starbursts \textit{in} and \textit{out} of the main sequence}
   \maketitle
%

\section{Introduction}
\label{SEC:intro}
During the six billion years that have passed between a redshift of $z$$\sim$2.5 and 0.5, galaxies formed 75\,\% of their present stellar mass (see Fig.11 of \citealt{madau14}) following a star-formation mode in which most of the UV starlight was absorbed by interstellar dust and re-radiated in the mid to far infrared (mid-IR and far-IR respectively, see e.g., \citealt{magnelli09,magnelli13}, \citealt{lefloch05}, \citealt{burgarella13}, \citealt{madau14} and references therein). 

The galaxies that contributed most to the cosmic star formation rate (SFR) density therefore radiated most of their light in the infrared domain and at the peak epoch of cosmic star-formation, the so-called "cosmic noon" around $z$$\sim$2, these galaxies belonged to the class of luminous and ultraluminous infrared galaxies; LIRGs and ULIRGs, (U)LIRGs hereafter, with total infrared luminosities of $L_{\rm IR}$=10$^{11}$ -- 10$^{12}$ L$_{\odot}$ and $L_{\rm IR}$$>$10$^{12}$ L$_{\odot}$ respectively. It is therefore of prime importance to understand the star formation mode of $z$$\sim$2 dusty star-forming galaxies to trace back the origin of present-day stars and galaxies.

Contrary to their local siblings, the distant (U)LIRGs do not systematically exhibit the signature of merger-driven starbursts with compact star formation and depletion times of the order of $\sim$150 Myr. Instead, they appear to be in majority forming stars through a secular mode of star-formation (see e.g., \citealt{elbaz10, elbaz11}, \citealt{daddi10a}, \citealt{rujopakarn11}, \citealt{wuyts11b}) with depletion times, $\tau_d$=$M_{\rm gas}$/SFR$\sim$600 Myr \citep{tacconi18}. Here $\tau_d$ is the time it would take for the galaxy to exhaust its molecular gas reservoir assuming a constant SFR. It is the inverse of the star formation efficiency, SFE. This evolution from a local population of rare violent merger-driven local (U)LIRGs to a common population of secularly evolving star-forming galaxies at $z$$\sim$2 is for the most part a natural result of the fast rise of the gas fraction of (U)LIRGs with increasing redshift (see e.g., \citealt{daddi10b}, \citealt{magdis12b}, \citealt{tacconi10,tacconi18}). 

This change in the nature of (U)LIRGs as a function of cosmic time can also be seen in the framework of the global evolution of the correlation between the SFR and stellar mass followed by star-forming galaxies, the so-called "star-formation main sequence" (MS, hereafter). This tight correlation between the SFR and stellar mass (M$_{\star}$) is followed by the majority of star-forming galaxies from $z$$\sim$0 up to at least $z$$\sim$3.5 (\citealt{elbaz07}, \citealt{noeske07}, \citealt{daddi07}, \citealt{schreiber15,schreiber17}, \citealt{pannella09,pannella15}, \citealt{karim11}, \citealt{wuyts11a}, \citealt{rodighiero14}, \citealt{whitaker12,whitaker14}, \citealt{renzini15}). While the existence of a correlation between SFR and M$_{\star}$ is natural, the fact that 68\,\% of the star-forming galaxies of a given stellar mass formed their stars with the same SFR within a factor 2 (0.3~dex--\textit{rms}) during 85\,\% of cosmic history (\citealt{schreiber15}) does appear as a surprise and a challenge for models. 

This implies, in particular, that galaxies more massive than M$_{\star}$=10$^{10}$ M$_{\odot}$ were LIRGs and galaxies with M$_{\star}$$\geq$1.4$\times$10$^{11}$ M$_{\odot}$ were ULIRGs at $z$$\sim$2. (U)LIRGs therefore represented a common phase among distant massive galaxies and studying their nature is equivalent to studying the origin of massive galaxies. And indeed the studies of the dust and gas content of $z$$\sim$2 star-forming galaxies did reveal much longer typical depletion times for MS galaxies at all masses, including (U)LIRGs at the high mass end of the MS which also present depletion times of about 600 Myr (\citealt{magdis12b}, \citealt{tacconi18}, \citealt{genzel15}, \citealt{bethermin15}).

The existence of a SFR -- M$_{\star}$ MS is commonly used to disentangle a secular--universal star-formation mode of galaxies \textit{within} the MS from a stochastic star-formation mode of galaxies \textit{out} of the MS, in which \textit{starbursts} systematically lie above the MS (see e.g., \citealt{rodighiero11}, \citealt{elbaz11}, \citealt{schreiber15} and references therein). The fact that the proportion of starbursts -- defined as galaxies experiencing star formation 3 or 4 times above the median of the MS SFR -- remains limited to a few percent at all redshifts and masses (\citealt{rodighiero11}, \citealt{schreiber15}) is puzzling when one considers that the observed (e.g., \citealt{kartaltepe07}) and modeled (e.g., \citealt{hopkins10}) merger rate rapidly rises with increasing redshift.

What physical processes sustain the secular star-formation of the MS? What role did mergers play around cosmic noon? Are starbursts limited to the small population of galaxies with an extremely large specific SFR (sSFR=SFR/M$_{\star}$, e.g., \citealt{rodighiero11}) or can there be starbursts "hidden" within the MS? Should one interpret the MS of star-forming galaxies as evidence that galaxies within it unequivocally form their stars following a common universal mode that is, in particular, unaffected by mergers?

We address these questions in this paper by taking advantage of the exquisite angular resolution of the Atacama Large Millimeter Array (ALMA). We use ALMA to probe the distribution of dusty star formation in $z$$\sim$2 (U)LIRGs and compare it with that derived from rest-frame UV light. We compare the spatial locations of UV-transparent and dusty star formation and discuss the presence or absence of spatial correlations between both with other galaxy properties, such as their depletion time and star-formation compactness. 
Here we identify a population of galaxies that lie \textit{within} the MS but that exhibit enhanced star formation typical of starbursts -- in terms of star-formation efficiency, SFE=SFR/M$_{\rm gas}$, or equivalently depletion time, $\tau_{\rm dep}$=1/SFE. We consider several possible scenarios that may provide an explanation for the existence of these \textit{starbursts hidden in the MS}, study a possible link with the presence of an active galactic nucleus (AGN) and discuss implications on the formation of compact early-type galaxies as observed at $z$$\sim$2 (e.g., \citealt{vanderwel14}).

Throughout this paper we use a \cite{salpeter55} initial mass function (IMF), and adopt a $\Lambda$CDM cosmology with $\Omega_{\rm M}$ = 0.3, $\Omega_\Lambda$ = 0.7 and
$H_0$ = 70 km s$^{-1}$ Mpc$^{-1}$ . As a matter of notation, we refer to the rest-frame GALEX far-ultraviolet (FUV) bandpass and to the total integrated IR light in the range 8--1000\,$\mu$m when using the subscripts "UV" and "IR", respectively.

\section{Data}
\label{SEC:data}
An ensemble of 8 galaxies with \textit{Herschel} photometry defines the core sample of this study for which deep 870\,$\mu$m ALMA (band 7) continuum images were obtained (40--50 min on source, Cycle 1, P.I. R.Leiton). These galaxies are complemented with 11 galaxies observed at 1.3 mm from public ALMA data in the Hubble Ultra-Deep Field, HUDF ($\sigma_{1.3}$$\sim$ 35\,$\mu$Jy; \citealt{dunlop17}, \citealt{rujopakarn16}). The resulting sample of 19 galaxies at $z$$\sim$2 is described below.

\subsection{Sample selection}
\label{SEC:sample}
The main sample of galaxies used for this paper comes from the ALMA project 2012.1.00983.S (P.I.R.Leiton, Cycle 1) which was observed from August 29 to September 1st, 2014. It consists of eight $z$$\sim$2 ULIRGs (ultra-luminous infrared galaxies, L$_{\rm IR}$$\geq$10$^{12}$ L$_{\odot}$) that were selected from a sample of \textit{Herschel} galaxies detected in the GOODS-\textit{South} field from the GOODS-{\it Herschel} open time key program \citep{elbaz11}. 

These galaxies were selected in a way to avoid being heavily biased towards the minor population of starburst galaxies well above the MS, but at the same time to reach a high enough signal-to-noise ratio in the high-resolution ALMA images at 870\,$\mu$m (i.e., 290\,$\mu$m rest-frame). The image quality and resolution were set with the goal of being able to determine the compactness and clumpiness of star-formation in these galaxies. This resulted in the requirements listed below that limited the sample to only 8 galaxies with a median stellar mass of M$_\star$=1.9$\times$10$^{11}$ M$_{\odot}$.

Starting from the GOODS-\textit{Herschel} galaxy catalog (described in \citealt{elbaz11}), we selected the ALMA targets under the following conditions:

\textit{(i)} a redshift -- either spectroscopic or photometric -- of 1.5$<$$z$$<$2.6 to ensure that the MIPS-24\,$\mu$m band encompasses the 8\,$\mu$m wavelength to allow the determination of a rest-frame 8\,$\mu$m luminosity, necessary to compute the $IR8$ color index. This color index, $IR8$=$L_{\rm IR}/L_{8\mu m}$, was found to exhibit a tight correlation with the surface density of mid and far-infrared luminosity by \cite{elbaz11}. Here $L_{8\mu m}$ is the $\nu L_\nu$ broadband luminosity integrated in the \textit{Spitzer}--IRAC band 4 centered at 8\,$\mu$m and $L_{\rm IR}$ is the total infrared luminosity integrated from 8 to 1000\,$\mu$m. $IR8$ provides an independent tracer of dusty star formation compactness. This redshift encompasses the key epoch of interest here, the cosmic noon of the cosmic SFR density, and is large enough to bring the central wavelength, 870\,$\mu$m, of ALMA band 7 (345 GHz) close to the peak of the far-infrared emission. 

\textit{(ii)} a sampling of the far infrared spectral energy distribution (SED) with measurements in at least four far-IR bands (100, 160, 250, 350\,$\mu$m). This requirement is mainly constrained by the condition to have a 3--$\sigma$ detection in the 250 and 350\,$\mu$m bands together with the condition that the \textit{Herschel}--SPIRE measurements are not heavily affected by contamination from close neighbors. The latter condition is determined through the use of a "clean index" (defined in \citealt{elbaz10,elbaz11}). The clean index is used to reject sources with highly uncertain flux densities due to confusion by only selecting sources with at most one neighbor closer than 0.8$\times$FWHM(250$\mu$m)$=$18$\arcsec$ and brighter than half the 24\,$\mu$m flux density of the central object. This was done using the list of sources detected at 24\,$\mu$m above 20\,$\mu$Jy. Simulations using realistic infrared SED and galaxies spatial distributions together with the \textit{Herschel} noise showed that this criterion ensures a photometric accuracy better than 30\,\% in at least 68\,\% of the cases for SPIRE detections \citep{leiton15}.

\textit{(iii)} we rejected sources with unphysical SEDs, i.e., for which one or more of the flux densities from 24 to 350\,$\mu$m presented a non physical jump. This smoothness condition on the SED was required to reject sources with blending effects, affecting mainly the longest \textit{Herschel} wavelengths even after imposing the clean index criterion.

For the sake of simplicity, we labeled the eight sources GS1 to GS8. We also provide their CANDELS ID, $ID_{CLS}$, from \cite{guo13} in Table~\ref{TAB:coord}. We note that all of the \textit{Herschel} sources studied here were found to be associated with a single ALMA source, none was resolved into two or more ALMA sources.

\begin{table}[htp]
\caption{ALMA sources.}
\begin{center}
\begin{tabular}{lrcclcrrrrrrrc}
\hline
ID & ID$_{CLS}$& Ra$_{CLS}$ , Dec$_{CLS}$ & offset \\
   &           &    3h32m...,$-$27$^{\circ}$...      & $arcsec$ \\
(1)&(2)& (3)  &  (4) \\ \hline
\hline
  GS1 &  3280 & 14.55s , 52$\arcmin$56.54$\arcsec$ & 0.176,$-$0.188 \\
  GS2 &  5339 & 54.69s , 51$\arcmin$40.70$\arcsec$ & 0.052,$-$0.232 \\
  GS3 &  2619 & 39.25s , 53$\arcmin$25.71$\arcsec$ & 0.083,$-$0.173 \\
  GS4 &  7184 & 17.21s , 50$\arcmin$37.07$\arcsec$ & 0.161,$-$0.280 \\
  GS5 &  9834 & 35.72s , 49$\arcmin$16.04$\arcsec$ & 0.089,$-$0.248 \\
  GS6 & 14876 & 28.51s , 46$\arcmin$58.14$\arcsec$ & 0.091,$-$0.269 \\
  GS7 &  8409 & 37.74s , 50$\arcmin$00.41$\arcsec$ & 0.067,$-$0.225 \\
  GS8 & 5893b & 46.84s , 51$\arcmin$21.12$\arcsec$ & 0.081,$-$0.184 \\
\hline
 UDF1 & 15669 & 44.03s , 46$\arcmin$35.70$\arcsec$ & 0.066,$-$0.277 \\
 UDF2 & 15639 & 43.53s , 46$\arcmin$39.00$\arcsec$ & 0.057,$-$0.277 \\
 UDF3 & 15876 & 38.54s , 46$\arcmin$34.06$\arcsec$ & 0.083,$-$0.243 \\
 UDF4 & 15844 & 41.03s , 46$\arcmin$31.45$\arcsec$ & 0.063,$-$0.250 \\
 UDF5 & 13508 & 36.97s , 47$\arcmin$27.21$\arcsec$ & 0.101,$-$0.239 \\
 UDF6 & 15010 & 34.43s , 46$\arcmin$59.57$\arcsec$ & 0.100,$-$0.254 \\
 UDF7 & 15381 & 43.32s , 46$\arcmin$46.80$\arcsec$ & 0.050,$-$0.250 \\
 UDF8 & 16934 & 39.74s , 46$\arcmin$11.25$\arcsec$ & 0.043,$-$0.289 \\
UDF11 & 12624 & 40.05s , 47$\arcmin$55.46$\arcsec$ & 0.090,$-$0.242 \\
UDF13 & 15432 & 35.08s , 46$\arcmin$47.58$\arcsec$ & 0.098,$-$0.260 \\
UDF16 & 14638 & 42.38s , 47$\arcmin$07.61$\arcsec$ & 0.069,$-$0.242 \\
\hline
\end{tabular}
\end{center}
\begin{small}
\textit{\textbf{Notes:}} 
The upper part of the table lists the 8 galaxies (GS1 to GS8) observed with ALMA at 870\,$\mu$m at a 0.2 arcsec resolution in our ALMA program. The lower part lists the 11 galaxies (UDF\#) from the 1.3mm ALMA survey of the HUDF by \cite{dunlop17} at a resolution of 0.35 arcsec.
Col.(1) Simplified ID. For the UDF galaxies, we use the same IDs as in \cite{dunlop17}.
Cols.(2) and (3) CANDELS ID and coordinates from \cite{guo13}. GS8, initially associated with the galaxy with the CANDELS ID 5893, was found to be associated with a background galaxy that we will call 5893b (see Section~\ref{SEC:dark}).
Col.(4) offset to be applied to the \textit{HST} CANDELS coordinates to match the ALMA astrometry. 
\end{small}
\label{TAB:coord}
\end{table}

\subsection{Supplementary sample from the Hubble Ultra Deep Field}
\label{SEC:hudf}
We supplemented our sample with a reference sample of galaxies detected with ALMA at 1.3 mm with a resolution of $\sim$0.35 arcsec within the 4.5 arcmin$^2$ survey of the Hubble Ultra Deep Field (HUDF) down to $\sigma_{1.3}$$\sim$ 35\,$\mu$Jy (see \citealt{dunlop17}, \citealt{rujopakarn16}). We use here the 11 galaxies listed in Table 2 of \cite{rujopakarn16} (see Section~\ref{SEC:alma}). The galaxies are labeled UDF\# in Table~\ref{TAB:coord} as in \cite{dunlop17}.

\subsection{ALMA observations and data reduction}
\label{SEC:alma}
\subsubsection{ALMA observations}
\label{SEC:observations}
Each one of the eight targeted galaxies was observed with a single pointing with a total of 36 antennas in band 7 (345 GHz, 870$\mu$m) at an angular resolution of 0.2\arcsec (ALMA synthesized beam of 0.2\arcsec$\times$0.16\arcsec). The integration time on each science target ranges from 37 to 50 minutes, resulting in typical signal-to-noise ratios at 870\,$\mu$m of S/N$\sim$35 and up to 75 for the brightest one.
The integration time was defined in order to reach a minimum S/N=10 on 20\,\% of the predicted 870\,$\mu$m ALMA flux density (extrapolated from \textit{Herschel}) or equivalently 50-$\sigma$ on the total flux in order to be able to measure an effective radius even for the most compact galaxies and to individually detect the major clumps of star formation when they exist and produce at least 20\,\% of the total ALMA flux density. For the typical predicted flux density of F$_{870}$$\sim$2.5 mJy of the sample, this led to a total observing time of at least 35 minutes/object. Accounting for the predicted flux densities of the galaxies, we used slightly different integration times of 36.5, 38.8 and 49.5 minutes (excluding overheads) for [GS4, GS5, GS6, GS8], [GS1, GS7] and [GS2, GS3] respectively. The observed standard deviation of the noise spans $rms$=40-70\,$\mu$Jy. Accounting for the obtained S/N, the accuracy on the size measurements given by CASA corresponds to FWHM/$\sqrt{(S/N)}$$\sim$0.034 arcsec that represents a theoretically expected precision, if we assume that the sources have a Gaussian profile, corresponding to $\sim$280 pc at $z$$\sim$2.

\subsubsection{Data reduction, flux and size measurements}
\label{SEC:size}
The data reduction was carried out with CASA, and the final images were corrected for the primary beam, although all our targets are located at the center of the ALMA pointings. Flux densities and sizes were both measured in the $uv$ plane, using the \textit{uvmodelfit} code in CASA, and in the image plane using the \textit{GALFIT} code \citep{peng02}. Since \textit{uvmodelfit} only allows 2D Gaussian profile fitting, we computed Gaussian and S\'ersic profiles with \textit{GALFIT} to compare both results.
The Gaussian semi-major axis ($R_{1/2}$=0.5$\times$FWHM of the major axis) derived in the $uv$ and image planes -- with $uvmodelfit$ and \textit{GALFIT} respectively -- agree within 15\,\%, with only a 5\,\% systematic difference (larger sizes in the \textit{GALFIT} measurements in the image plane). However, the uncertainties estimated by \textit{GALFIT} in the image plane are 45\,\% smaller (median,  with values that can reach more than a factor 2). We consider that the uncertainties measured in the \textit{uv} plane are most probably more realistic, and anyway more conservative, hence we have decided to use the \textit{uvmodelfit} measurements for our analysis (see Table~\ref{TAB:alma}). 

We compared the Gaussian semi-major axis from \textit{uvmodelfit}, $R_{1/2}$, with the effective radius, $R_e$, obtained with a S\'ersic profile fit either leaving the S\'ersic index free, $n^{\rm Sersic}_{ALMA}$, or imposing $n^{\rm Sersic}_{ALMA}$=1 (exponential disk profile). We find that $R_{1/2}$ and $R_e$ agree within 20\,\% in both cases with no systematic difference when imposing $n^{\rm Sersic}_{ALMA}$=1 and 4\,\% smaller sizes when the S\'ersic index is left free. 
We obtain a S/N ratio greater than 3 for the S\'ersic indices (Col.(7) in Table~\ref{TAB:alma}) of all the GS sources except GS2 and GS3 (ID CANDELS 5339 and 2619). 

Hence even though we did perform S\'ersic profile fittings and determined $n^{\rm Sersic}_{ALMA}$, the light distribution of our galaxies does not seem to show very strong departure from a 2D Gaussian. As a result both $R_{1/2}$ and $R_e$ provide an equally good proxy for the half-light radius, encompassing 50\,\% of the IR luminosity. We did measure some moderate asymmetries quantified by the minor (\textit{b}) over major (\textit{a}) axis ratio, \textit{b/a} (Col.(5) in Table~\ref{TAB:alma}) that we used to derive circularized half-light radii, $R_{ALMA}^{circ}$ (listed in kpc in the Col.(6) of Table~\ref{TAB:alma}) following Eq.~\ref{EQ:conv}. 
\begin{equation}
R_{ALMA}^{circ}[{\rm kpc}]=R_{1/2} \times \sqrt{ \frac{b}{a} } \times \rm{Conv(\arcsec\,to\,kpc)}
\label{EQ:conv}
\end{equation}
Conv(\arcsec\,to kpc) is the number of proper kpc at the redshift of the source and is equal to 8.46, 8.37 and 8.07 kpc/\arcsec at $z$=1.5, 2 and 2.5 respectively.
We then used $R_{ALMA}^{circ}$ -- that encompasses 50\,\% of the IR luminosity -- to compute the IR luminosity surface densities of our galaxies as in Eq.~\ref{EQ:Sir}.
\begin{equation}
\Sigma_{\rm IR}[{\rm L}_{\odot}~{\rm kpc}^{-2}] = \frac{L_{\rm IR} / 2}{\pi (R_{ALMA}^{circ})^2}
\label{EQ:Sir}
\end{equation}

\begin{table*}[htp]
\caption{Galaxy properties derived from the ALMA and \textit{HST}--WFC3 $H$ band (1.6\,$\mu$m) data.}
\begin{center}
\begin{tabular}{llccccrrrrr}
\hline
ID   & $z$ &  F$_{ALMA}$ & R$_{1/2maj}$& $b/a$ & R$_{ALMA}^{circ}$ & $n^{\rm Sersic}_{ALMA}$ & R$_{H}^{circ}$ & $n^{\rm Sersic}_{H}$  & $\Sigma_{\rm IR}^{(a)}$ &  $IR8$   \\
     &     &  ($\mu$Jy)  & (arcsec) & & (kpc) &  &(kpc) &  & ($\times$10$^{11}$L$_{\odot}$kpc$^{-2}$) & \\
(1)&(2)& (3)  &  (4) & (5) & (6) & (7)  & (8) & (9)  & (10) & (11)  \\ 
\hline
  GS1 & 2.191        & 1190 $\pm$ 120 &  0.145 $\pm$  0.016 & 0.72 &  0.87$\pm$0.09 & 0.63$\pm$0.18 & 0.73 & 3.75 &  3.73 $\pm$ 1.00 &  9.3 $\pm$ 0.8 \\
  GS2 & 2.326        & 1100 $\pm$  70 &  0.163 $\pm$  0.031 & 0.87 &  1.16$\pm$0.22 & 1.07$\pm$0.99 & 1.90 & 0.72 &  2.63 $\pm$ 1.21 &  7.2 $\pm$ 0.9 \\
  GS3 & 2.241        & 1630 $\pm$  70 &  0.150 $\pm$  0.016 & 0.42 &  0.52$\pm$0.05 & 4.69$\pm$1.65 & 2.81 & 2.03 & 14.96 $\pm$ 3.90 &  8.1 $\pm$ 0.6 \\
  GS4 & 1.956$^{sp}$ & 2100 $\pm$  70 &  0.140 $\pm$  0.010 & 0.82 &  0.97$\pm$0.07 & 1.88$\pm$0.59 & 3.93 & 1.77 &  3.54 $\pm$ 0.81 & 13.4 $\pm$ 1.8 \\
  GS5 & 2.576$^{sp}$ & 4420 $\pm$  70 &  0.139 $\pm$  0.006 & 0.92 &  1.03$\pm$0.04 & 1.27$\pm$0.22 & 2.57 & 1.08 & 10.12 $\pm$ 1.38 & -- \\
  GS6 & 2.309$^{sp}$ & 5210 $\pm$  70 &  0.120 $\pm$  0.004 & 1.00 &  0.98$\pm$0.03 & 1.15$\pm$0.21 & 2.10 & 0.25 & 10.86 $\pm$ 1.27 & 27.3 $\pm$ 2.4 \\
  GS7 & 1.619$^{sp}$ & 2320 $\pm$  70 &  0.194 $\pm$  0.012 & 0.84 &  1.38$\pm$0.09 & 1.78$\pm$0.38 & 3.66 & 1.04 &  2.33 $\pm$ 0.41 & 19.8 $\pm$ 1.7 \\
    GS8 & 3.240        & 6420 $\pm$ 140 &  0.142 $\pm$  0.003 & 0.62 &  0.67$\pm$0.02 & 0.67$\pm$0.04 & 1.63 & 3.53 & 20.86 $\pm$ 2.25 &             -- \\
\hline
 UDF1 & 3.000        & 924 $\pm$ 76 &  0.195 $\pm$  0.020 & 0.85 &  1.24$\pm$0.18 & -- & 0.54 & 7.16 &  5.95 $\pm$ 2.00 &             -- \\
 UDF2 & 2.794$^{sp}$ & 996 $\pm$ 87 &  0.265 $\pm$  0.030 & 0.85 &  1.77$\pm$0.18 & -- & 3.24 & 0.86 &  1.32 $\pm$ 0.33 &             -- \\
 UDF3 & 2.543$^{sp}$ & 863 $\pm$ 84 &  0.375 $\pm$  0.045 & 0.36 &  1.59$\pm$0.27 & -- & 1.55 & 0.81 &  3.35 $\pm$ 1.28 &  9.9 $\pm$ 0.9 \\
 UDF4 & 2.430        & 303 $\pm$ 46 &  0.270 $\pm$  0.060 & 0.52 &  1.42$\pm$0.35 & -- & 2.79 & 0.20 &  0.84 $\pm$ 0.47 &  7.9 $\pm$ 1.3 \\
 UDF5 & 1.759$^{sp}$ & 311 $\pm$ 49 &  0.480 $\pm$  0.125 & 0.20 &  1.59$\pm$0.62 & -- & 2.24 & 0.71 &  0.48 $\pm$ 0.40 &  7.2 $\pm$ 0.7 \\
 UDF6 & 1.411$^{sp}$ & 239 $\pm$ 49 &  0.530 $\pm$  0.205 & 0.19 &  1.77$\pm$1.06 & -- & 3.71 & 0.48 &  0.46 $\pm$ 0.58 & -- \\
 UDF7 & 2.590        & 231 $\pm$ 48 & \textit{ 0.120 $\pm$  0.060} &  --  & \textit{ 2.65$\pm$1.33} & -- & 4.24 & 0.77 & \textit{ 0.25 $\pm$ 0.27} & -- \\
 UDF8 & 1.546$^{sp}$ & 208 $\pm$ 46 &  0.675 $\pm$  0.225 & 0.53 &  3.81$\pm$1.24 & -- & 5.57 & 3.04 &  0.11 $\pm$ 0.08 &  4.7 $\pm$ 0.4 \\
UDF11 & 1.998$^{sp}$ & 186 $\pm$ 46 &  0.715 $\pm$  0.285 & 0.48 &  3.72$\pm$1.50 & -- & 4.40 & 1.41 &  0.26 $\pm$ 0.22 &  6.4 $\pm$ 0.5 \\
UDF13 & 2.497$^{sp}$ & 174 $\pm$ 45 &  0.430 $\pm$  0.170 & 0.55 &  2.30$\pm$0.97 & -- & 1.14 & 1.86 &  0.22 $\pm$ 0.20 &  6.9 $\pm$ 1.3 \\
UDF16 & 1.319$^{sp}$ & 155 $\pm$ 44 & \textit{ 0.115 $\pm$  0.058} &  --  & \textit{ 2.74$\pm$1.37} & -- & 3.15 & 2.16 & \textit{ 0.07 $\pm$ 0.07} & -- \\
\hline
\end{tabular}
\end{center}
\begin{small}
\textit{\textbf{Notes:}} 
Col.(1) Simplified ID. 
Col.(2) photometric redshift, except for the galaxies marked with $^{(sp)}$ for which a spectroscopic redshift is available.
Col.(3) F$_{\rm ALMA}$ is the continuum flux density at 870\,$\mu$m for the GS1 to GS8 sources and at 1.3mm for the UDF1 to UDF16 sources.
Cols.(4) and (5) Semi-major axis, R$_{1/2maj}$ in arcsec, and axis ratio, $b/a$, of the ALMA sources measured from \textit{uvmodelfit} in CASA for the GS sources and from \cite{rujopakarn16} for the UDF galaxies. The consistency of the GS and UDF was checked in the direct images using \textit{GALFIT}. The sizes of UDF7 and UDF16 (in italics) are measured at the 2-$\sigma$ level. 
Col.(6) circularized effective ALMA radius, $R_{\rm ALMA}^{circ}$, in kpc, as defined in Eq.~\ref{EQ:conv}. 
Col.(7) S\'ersic index, $n_{\rm ALMA}^{\rm Sersic}$, derived from the S\'ersic fit to the ALMA 870\,$\mu$m image for the GS sources using \textit{GALFIT} on the direct images. The S/N of the UDF sources is not high enough to allow the fit of a S\'ersic index.
Cols.(8) and (9) are the circularized effective S\'ersic radius, $R_{\rm H}^{\rm circ}$ in kpc, and index, $n_{\rm H}^{\rm circ}$, derived from the S\'ersic fit to WFC3 $H$ band images by \cite{vanderwel12}.
Col.(10) IR surface density in L$_{\odot}$kpc$^{-2}$, $\Sigma_{\rm IR}$=(L$_{\rm IR}$/2)/[$\pi$(R$_{\rm ALMA}^{circ}$)$^2$], where L$_{\rm IR}$ is given in Table~\ref{TAB:sample} and $R_{\rm ALMA}^{circ}$ in Col.(6).
Col.(11) $IR8$=L$_{\rm IR}$/L$_{8\mu \rm m}$ color index. The 8\,$\mu$m rest-frame luminosities were derived from the observed \textit{Spitzer}-MIPS 24\,$\mu$m photometry as in \cite{elbaz11}. L$_{8\mu \rm m}$, hence also $IR8$, can only be determined from the observed 24\,$\mu$m luminosity for galaxies with 1.5$\leq$$z$$\leq$2.5.
\end{small}
\label{TAB:alma}
\end{table*}

The sizes of the HUDF galaxies were measured as well using a 2D elliptical Gaussian fitting by \cite{rujopakarn16} who used the PyBDSM\footnote{http://www.astron.nl/citt/pybdsm} code. We analyzed the public ALMA image of the HUDF and found that the quality of the images did not permit to constrain both a S\'ersic effective radius and index, hence we do not provide S\'ersic indices for the HUDF galaxies. We fitted with \textit{GALFIT} 2D Gaussian elliptical profiles on the 11 HUDF sources listed in \cite{rujopakarn16} and found a good agreement between our measured Gaussian FWHM values and those quoted in \cite{rujopakarn16} with a median ratio of exactly 1 and an \textit{rms} of 16\,\% for the sources with S/N$>$5. Below this threshold, the measured sizes agree within the error bars which start to be quite large. We quote in Table~\ref{TAB:alma} the sizes listed in \cite{rujopakarn16}. 

The flux densities of the GS galaxies were computed using our 2D elliptical Gaussian fitting in the \textit{uv} plane. For the HUDF galaxies, they correspond to those listed in \cite{rujopakarn16} consistent with our own measurements.
 \begin{figure}
 \centering
      \includegraphics[width=8.7cm]{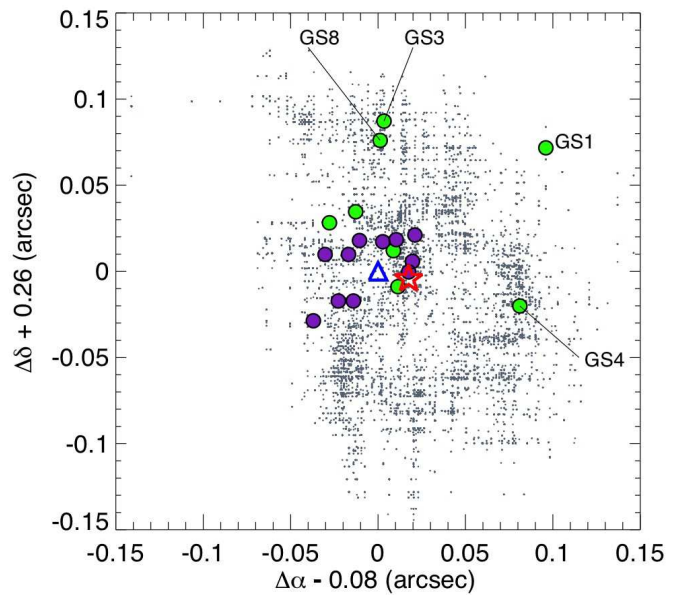}     
      \caption{Astrometric offsets to be applied to the positions of the CANDELS HST catalog in GOODS-\textit{South} to match the positions of the sources detected in Pan-STARRS1 (small grey dots). We centered the diagram on the systematic astrometric correction of [0.08\arcsec, $-$0.26\arcsec] introduced by \cite{dunlop17} and \cite{rujopakarn16} for the HUDF, marked by an open blue triangle. The open red star marks the median of the systematic astrometric correction over the whole 10$\arcmin$$\times$15$\arcmin$ GOODS-\textit{South} field [0.095\arcsec, $-$0.264\arcsec]. A detailed description of these astrometric offsets will be provided in Dickinson et al. (in prep.). The large green and purple dots mark the 8 GS sources and 11 UDF sources detected with ALMA at 870\,$\mu$m and 1.3mm respectively.}
     \label{FIG:astrom}
  \end{figure}
\subsubsection{ALMA vs \textit{HST} astrometry}
\label{SEC:astrometry}
The ALMA and \textit{HST} coordinates present a small systematic offset in the GOODS-\textit{South} field.
This offset does not exist between ALMA and other observatories such as 2MASS, JVLA, GAIA or Pan-STARRS but it affects the astrometry of the \textit{HST} sources. A comparison of the positions of \textit{HST} sources in the HUDF with 2MASS \citep{dunlop17} and JVLA \citep{rujopakarn16} showed that the \textit{HST} positions needed to be corrected by $-$0.26 arcsec in Declination and $+$0.08 arcsec in Right Ascension. This implies that the \textit{HST} coordinates (in decimal degrees) must be systematically corrected by [$+$2.51,$-$7.22] $\times$10$^{-5}$ degree (including the cos($\delta$) factor). 

This offset is too small to change the \textit{HST} counterparts of the ALMA detections. 
However, it has an impact on the detailed comparison of the location and shape of the ALMA millimeter emission with that of the \textit{HST} optical light that will be discussed in the following sections. Hence we decided to extend further our analysis of this astrometric issue by searching for possible local offsets added to the global one mentioned above. A detailed description of the resulting analysis will be presented in Dickinson et al. (in prep.). We just briefly summarize here the main lines of this process and its implications on our analysis.

The main reasons for this astrometric issue can be traced back to the astrometric references that were used to build the \textit{HST} mosaics of the GOODS-\textit{South} field. At the time, the astrometric reference used for GOODS-\textit{South} was an ESO 2.2m Wide Field Imager (WFI) image, itself a product of a combination of different observing programs (the ESO Imaging Survey, EIS and COMBO-17 among others). The GOODS \textit{HST} team subsequently re-calibrated the WFI astrometry to match the \textit{HST} Guide Star Catalog (GSC2).

More modern astrometric data are now available in this field such as Pan-STARRS1 \citep{chambers16}. We used the PanStarr DR1 catalogue provided by \cite{flewelling2016} to search for possible offsets in the different regions of the whole 10$\arcmin$$\times$15$\arcmin$ GOODS-\textit{South} field. 

We found residual distortions that we believe to be due to some combination of distortions in the WFI mosaic images and in the GSC2 positions, and zonal errors registering the \textit{HST} ACS images to the WFI astrometry. These residual local distortions are plotted in Fig.~\ref{FIG:astrom} after having corrected the \textit{HST} positions for the global offset mentioned above and marked with the open blue triangle. The distortion pattern was determined using a 2.4 arcmin diameter circular median determination of the offset in order to avoid artificial fluctuations due to the position uncertainty on the individual objects. This pattern was then applied to the 34,930 \textit{HST} WFC3-$H$ sources of the CANDELS catalog in GOODS-\textit{South} (\citealt{guo13}; shown as grey dots in Fig.~\ref{FIG:astrom}).

The 11 galaxies detected by ALMA in the HUDF are all well centered on this position with residual offsets of the order of 0.02\arcsec (large filled purple dots). These extra corrections are truly negligible, since they correspond to 160 pc at the redshifts of the sources. However, the 8 GS galaxies are spread over a wider area in GOODS-\textit{South} including parts where the residual offsets can be as large as $\sim$0.07\arcsec, i.e., 0.6 kpc. This is the case of GS1, GS3, GS4 and GS8. 

We found that these local offsets did not not affect the associations with optical counterparts and that they were smaller than the difference between the positions of the rest-frame UV and FIR light distributions that we discuss in the following sections. Except in the case of the galaxy GS4, where the peak of the ALMA emission presented an offset with respect to the \textit{HST}-WFC3 $H$-band centroid before applying the local correction for the \textit{HST} astrometry and fell right on the $H$-band center after correction. 

\subsection{Dust, gas and stellar masses}
\label{SEC:sed}
 \begin{figure}[h!]
 \centering
      \includegraphics[width=4cm]{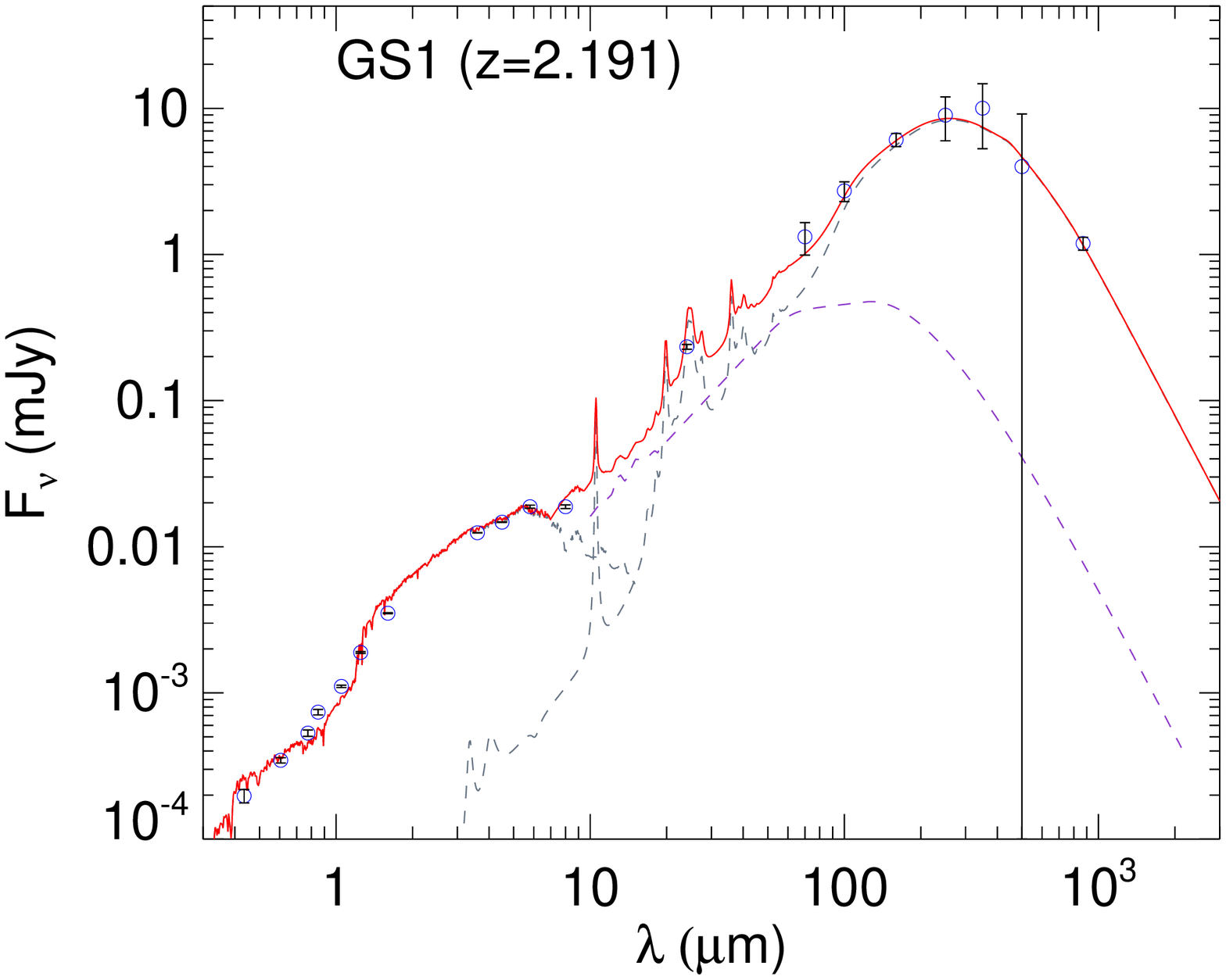}     
      \includegraphics[width=4cm]{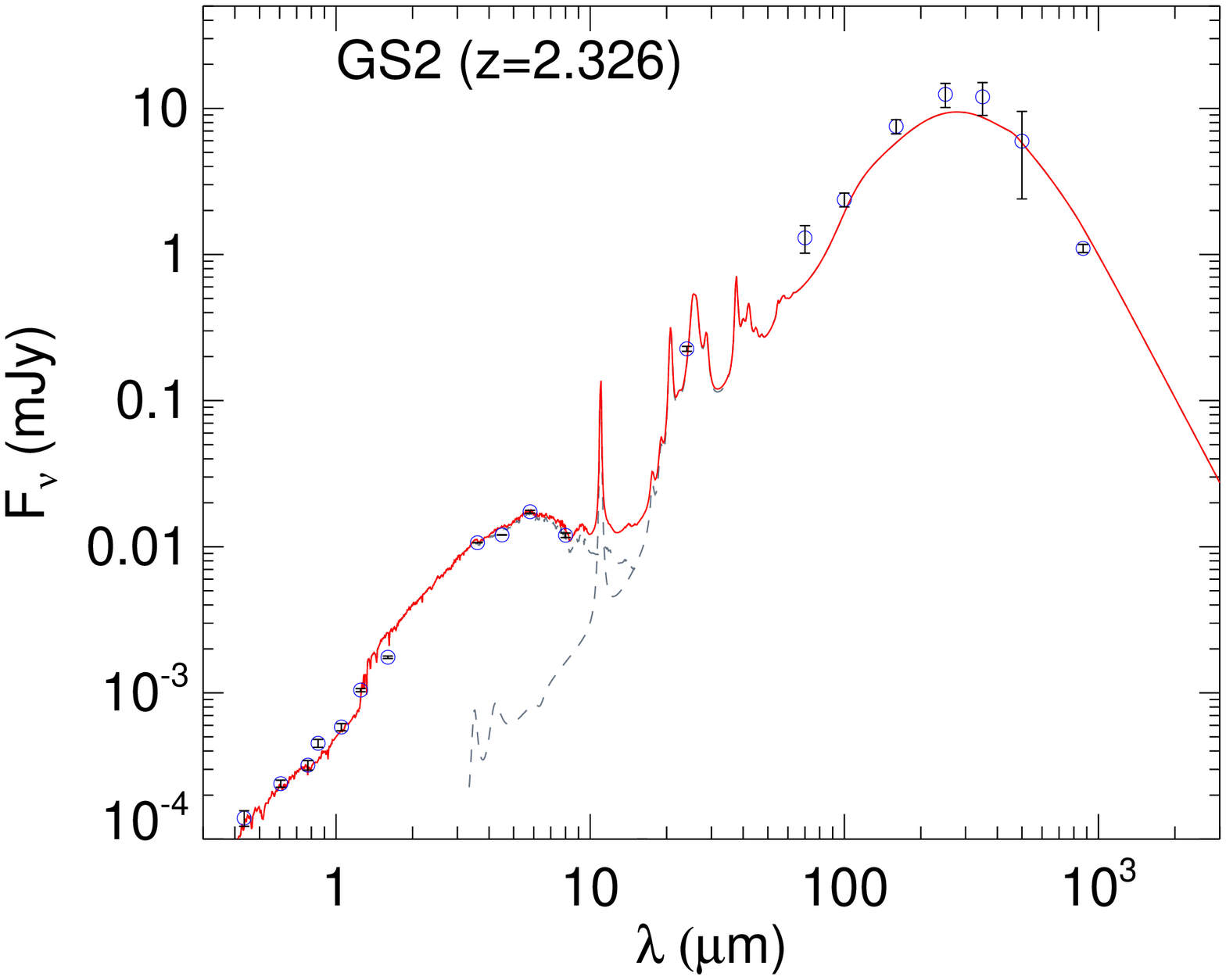}     
      \includegraphics[width=4cm]{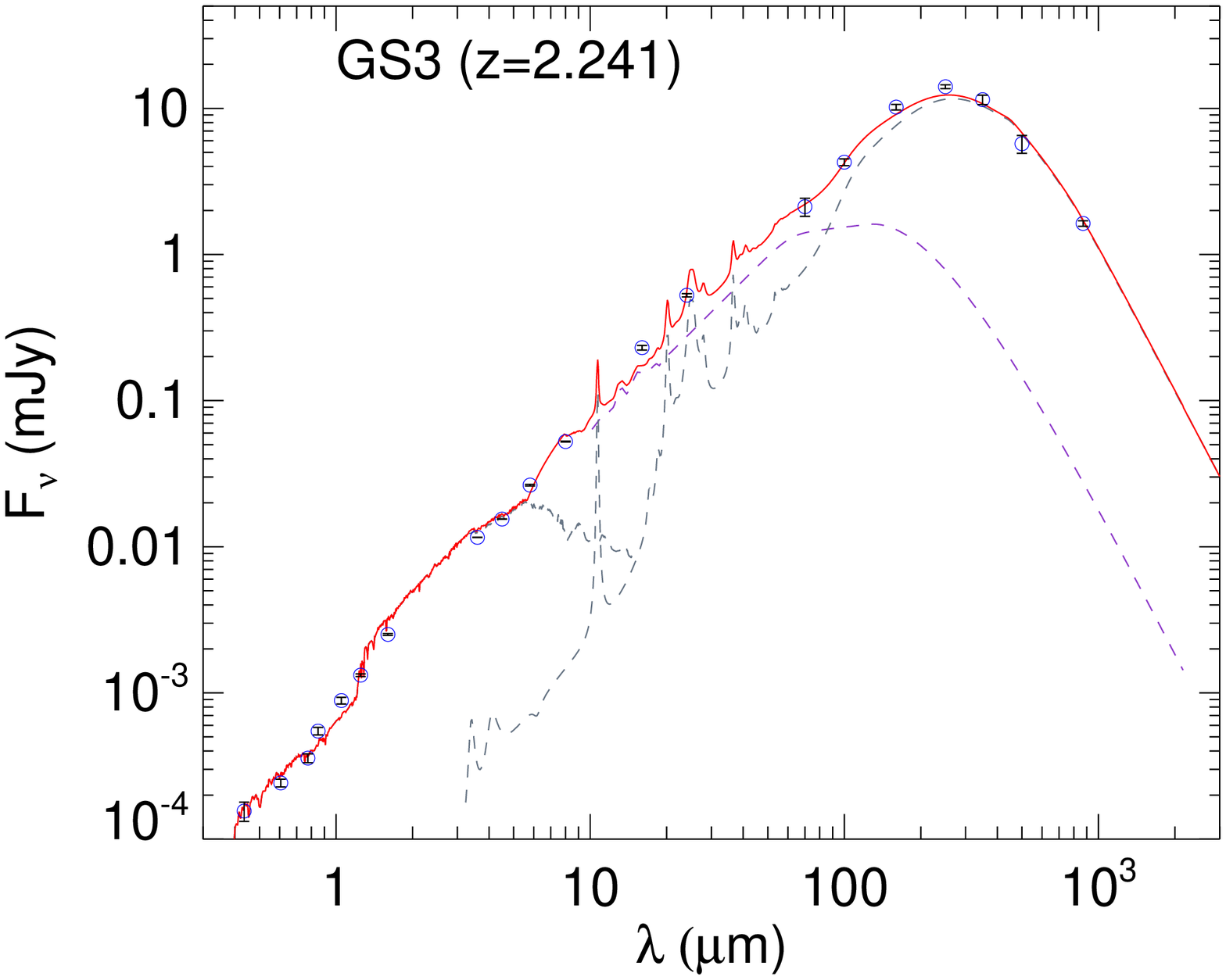}     
      \includegraphics[width=4cm]{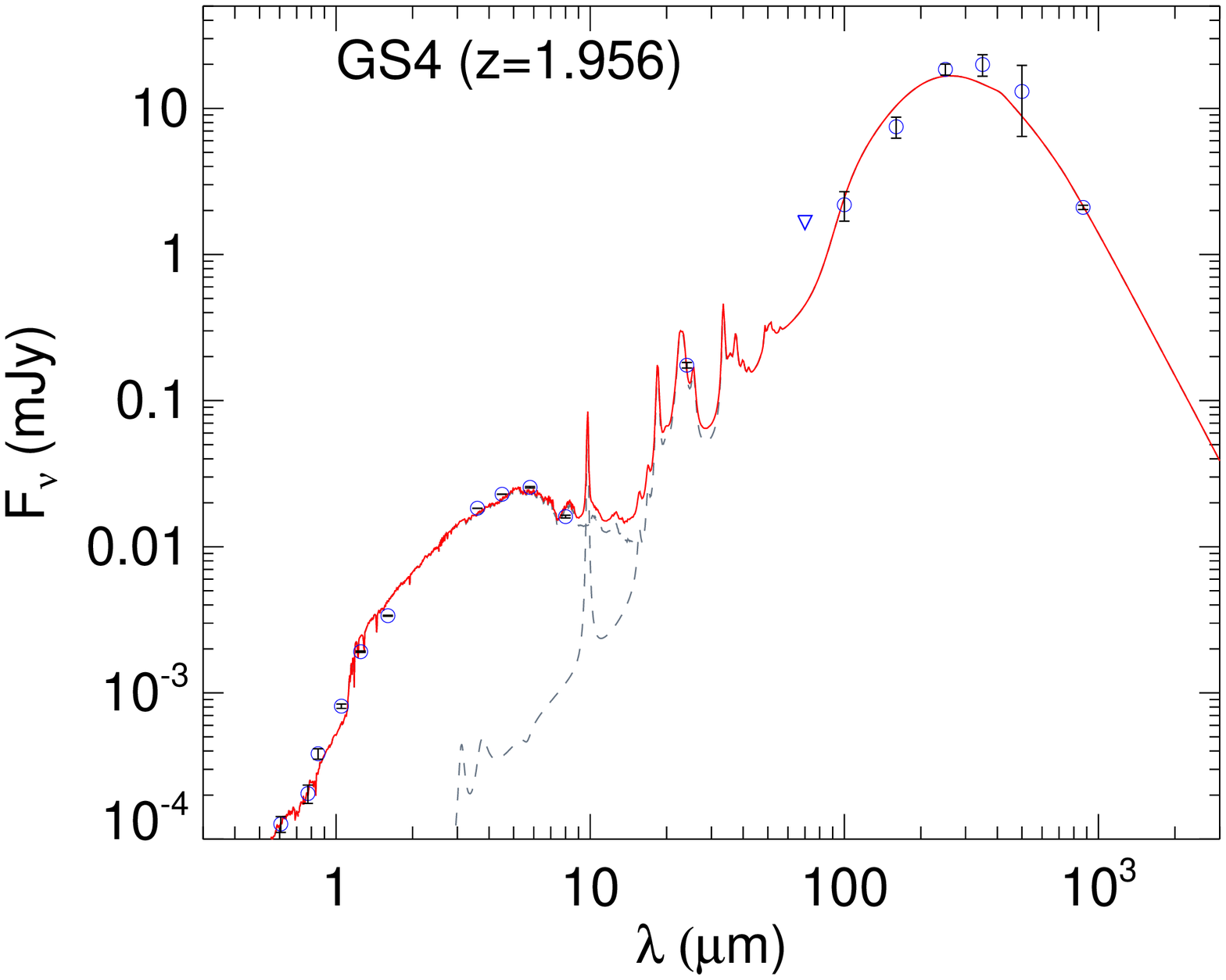}     
      \includegraphics[width=4cm]{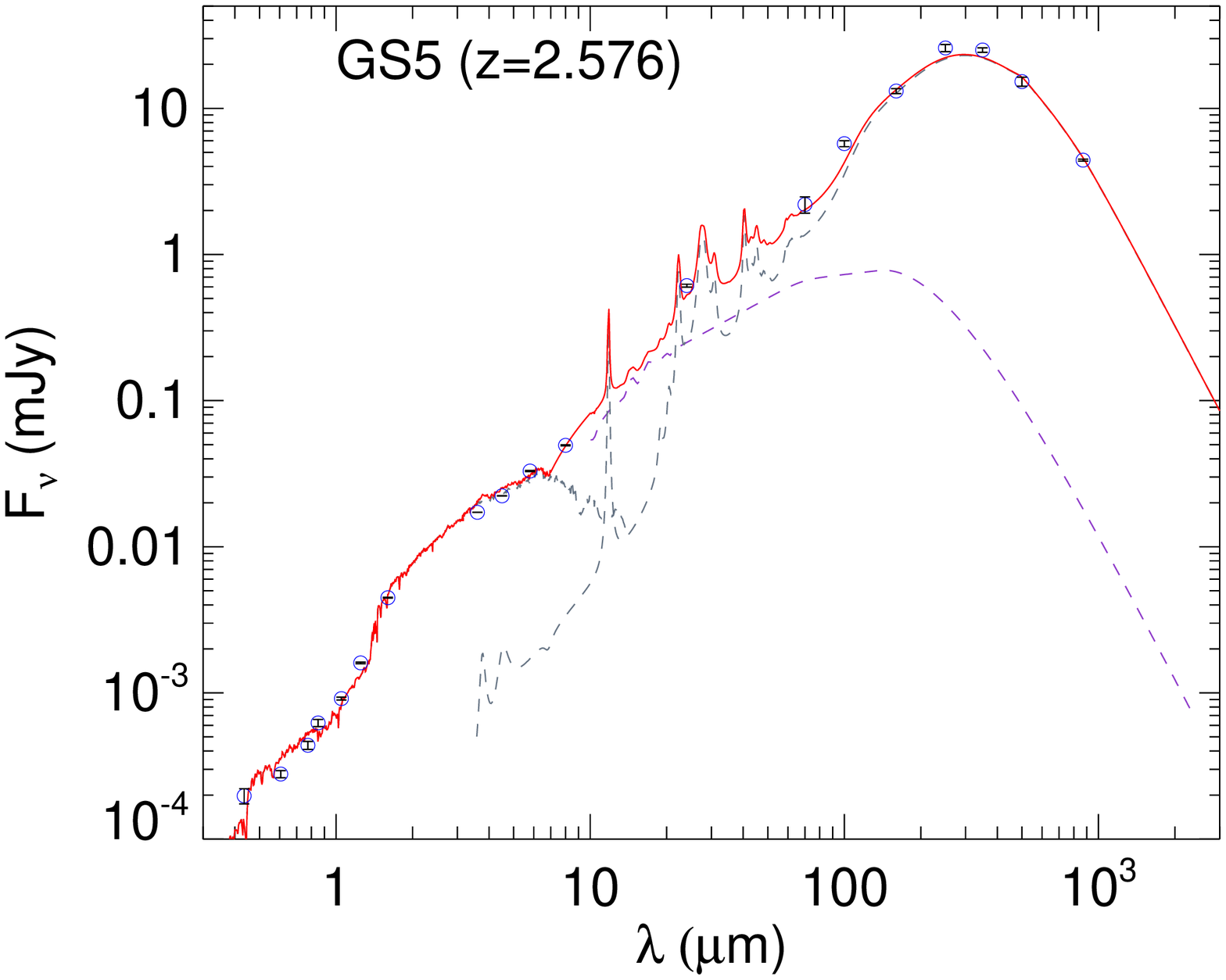}     
      \includegraphics[width=4cm]{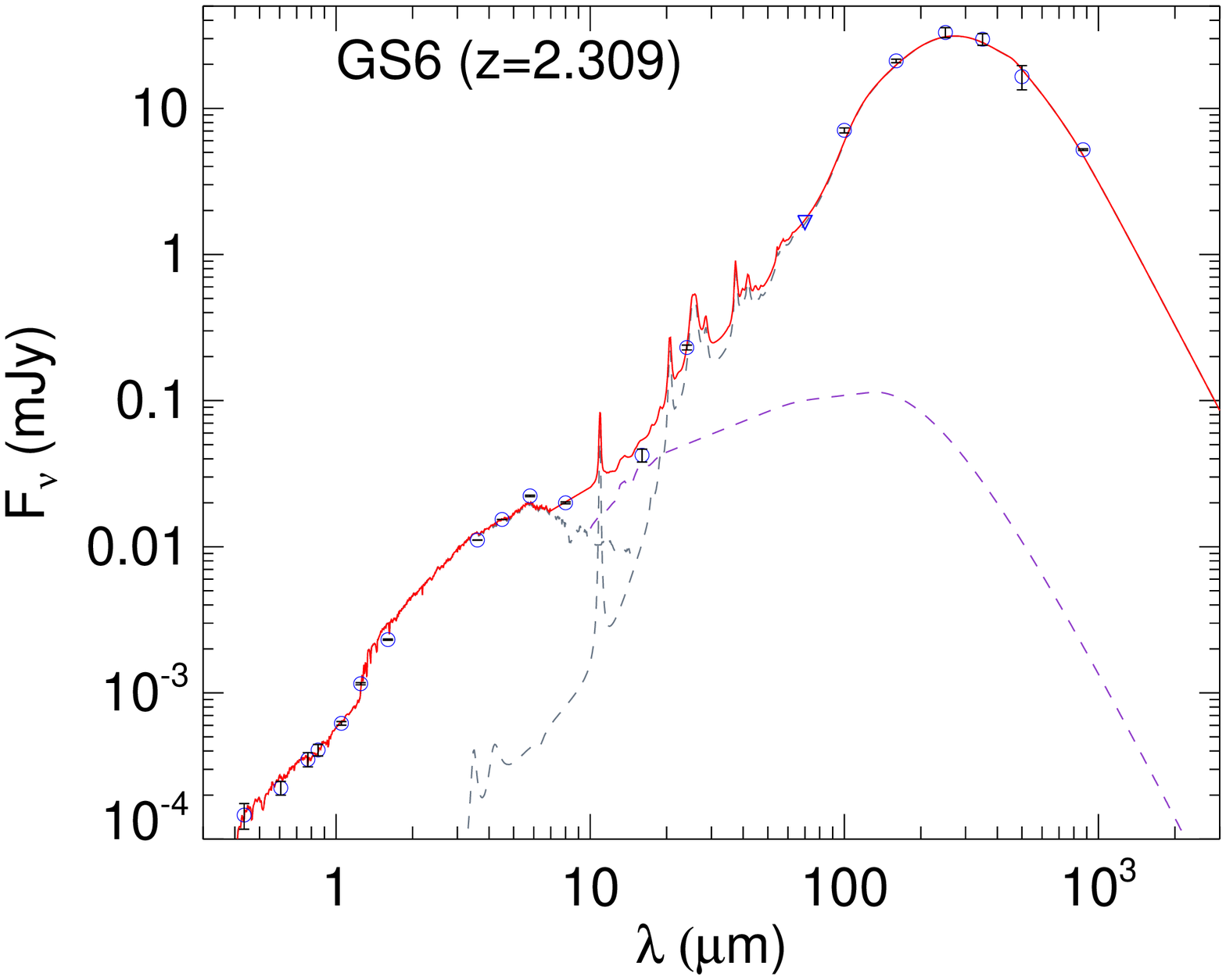}     
      \includegraphics[width=4cm]{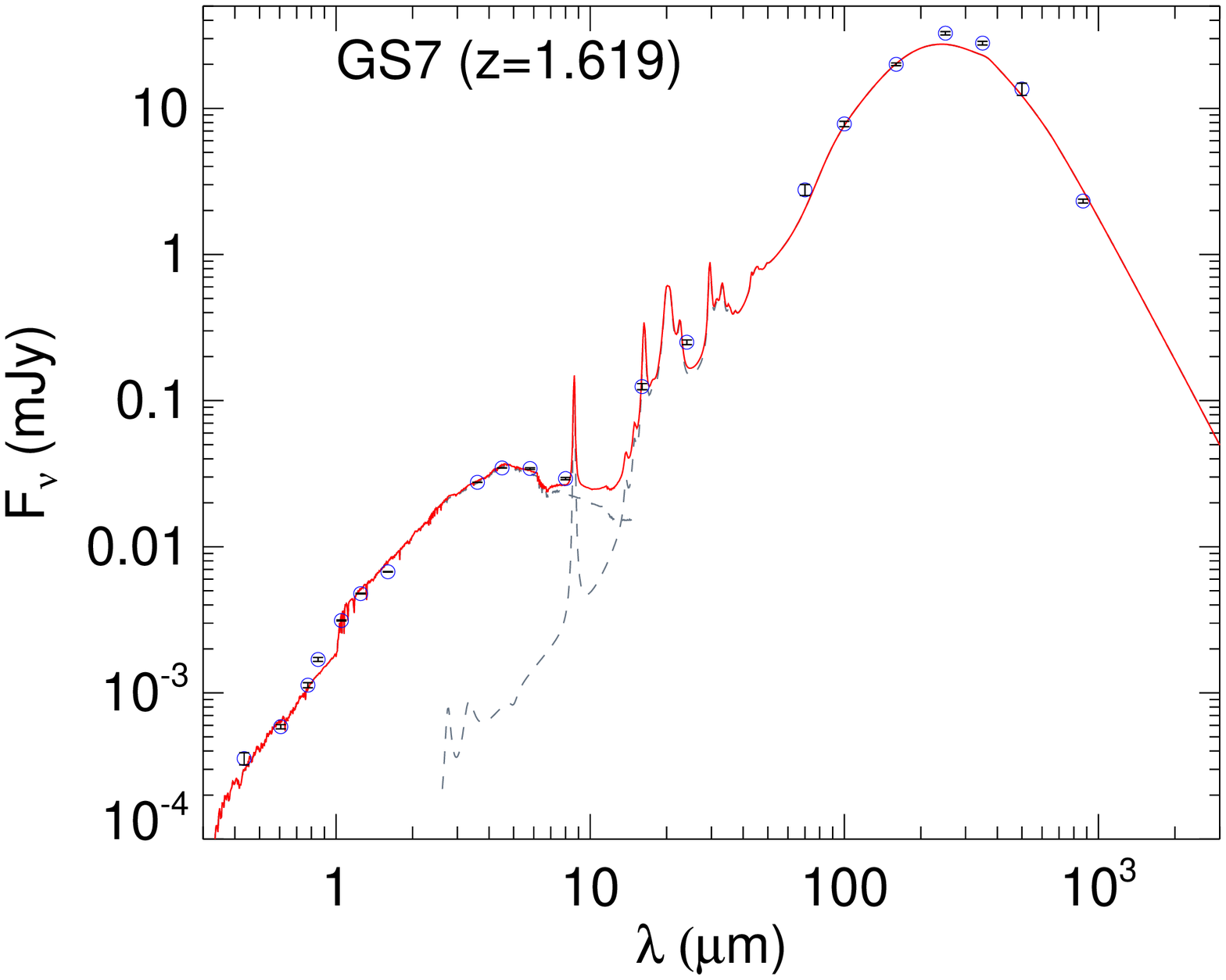}     
      \includegraphics[width=4cm]{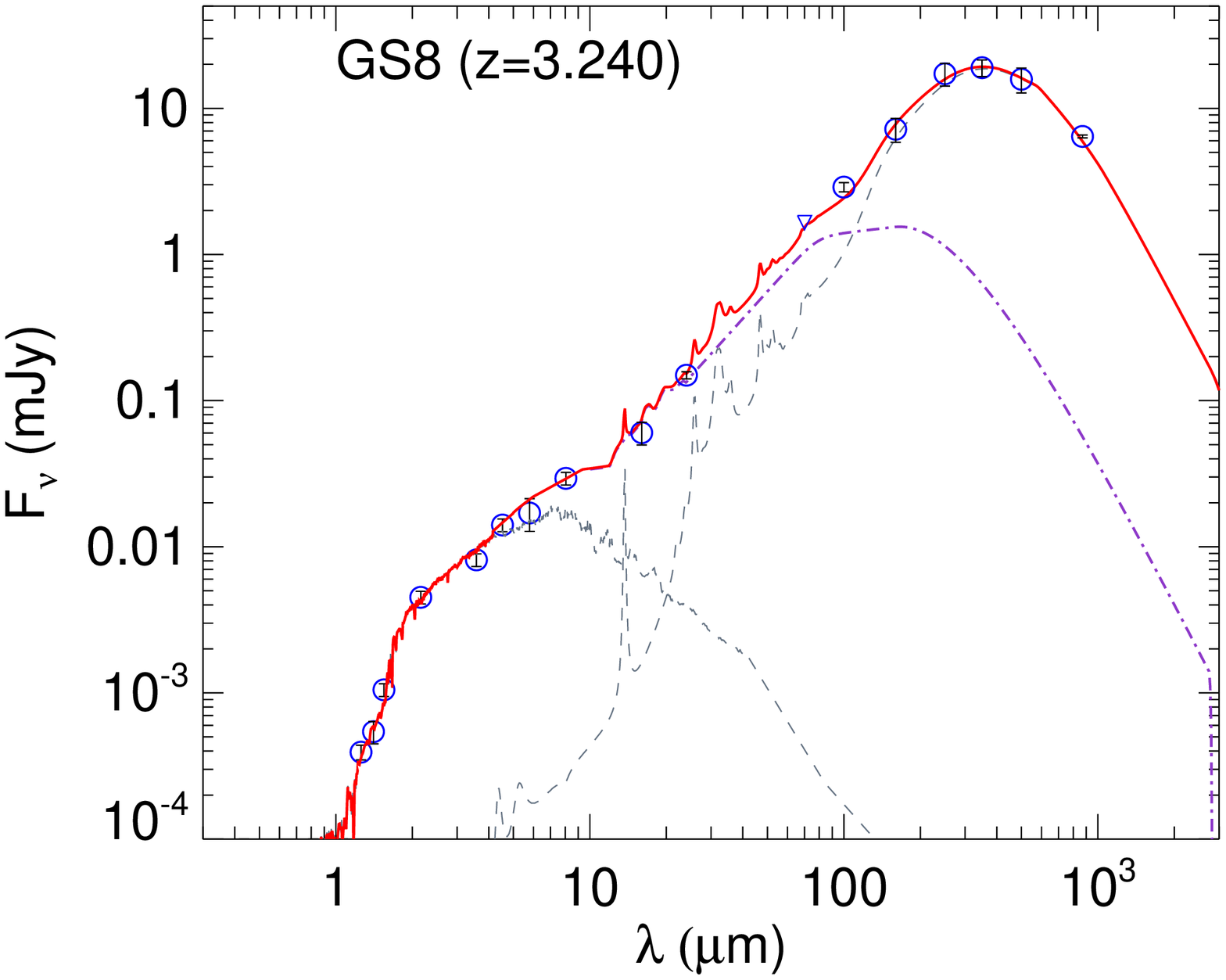}     
      \caption{Spectral energy distributions (SEDs) of the 8 GS galaxies. The solid red line shows the combination of the model fit of the \textit{(i)} optical-NIR side of the SEDs done with the FAST code, \textit{(ii)} IR energy distribution from the best-fitting \cite{draine07} model (grey dashed line), and when necessary \textit{(iii)} the warm dust continuum heated by an AGN using the \cite{mullaney11} code \textit{decompIR} (purple dashed line). The specific case of GS8 for which the optical counterpart is nearly undetected is discussed in Section~\ref{SEC:dark}.}
     \label{FIG:SEDs_GS}
  \end{figure}

 \begin{figure}[h!]
 \centering
 \includegraphics[width=4cm]{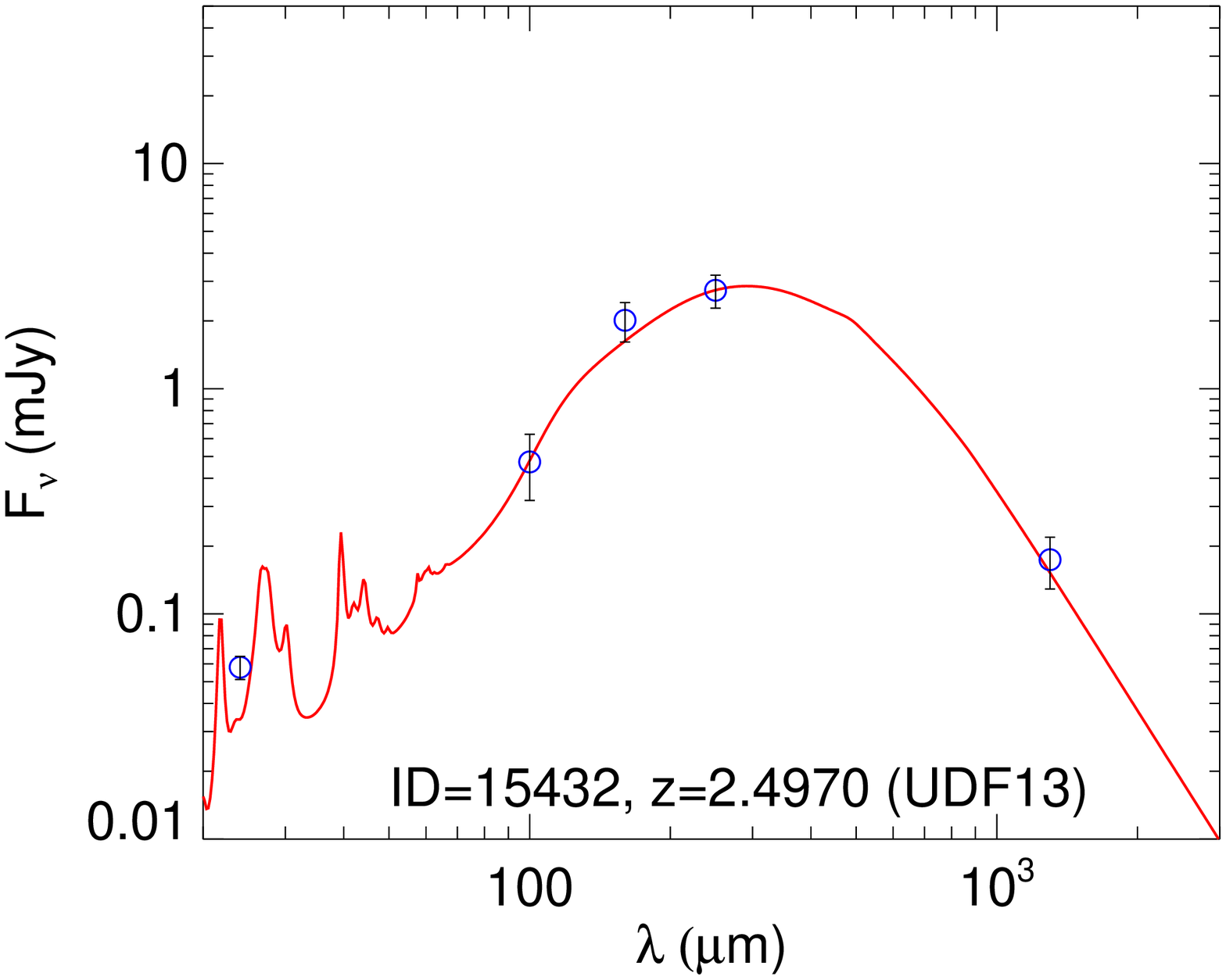}
\includegraphics[width=4cm]{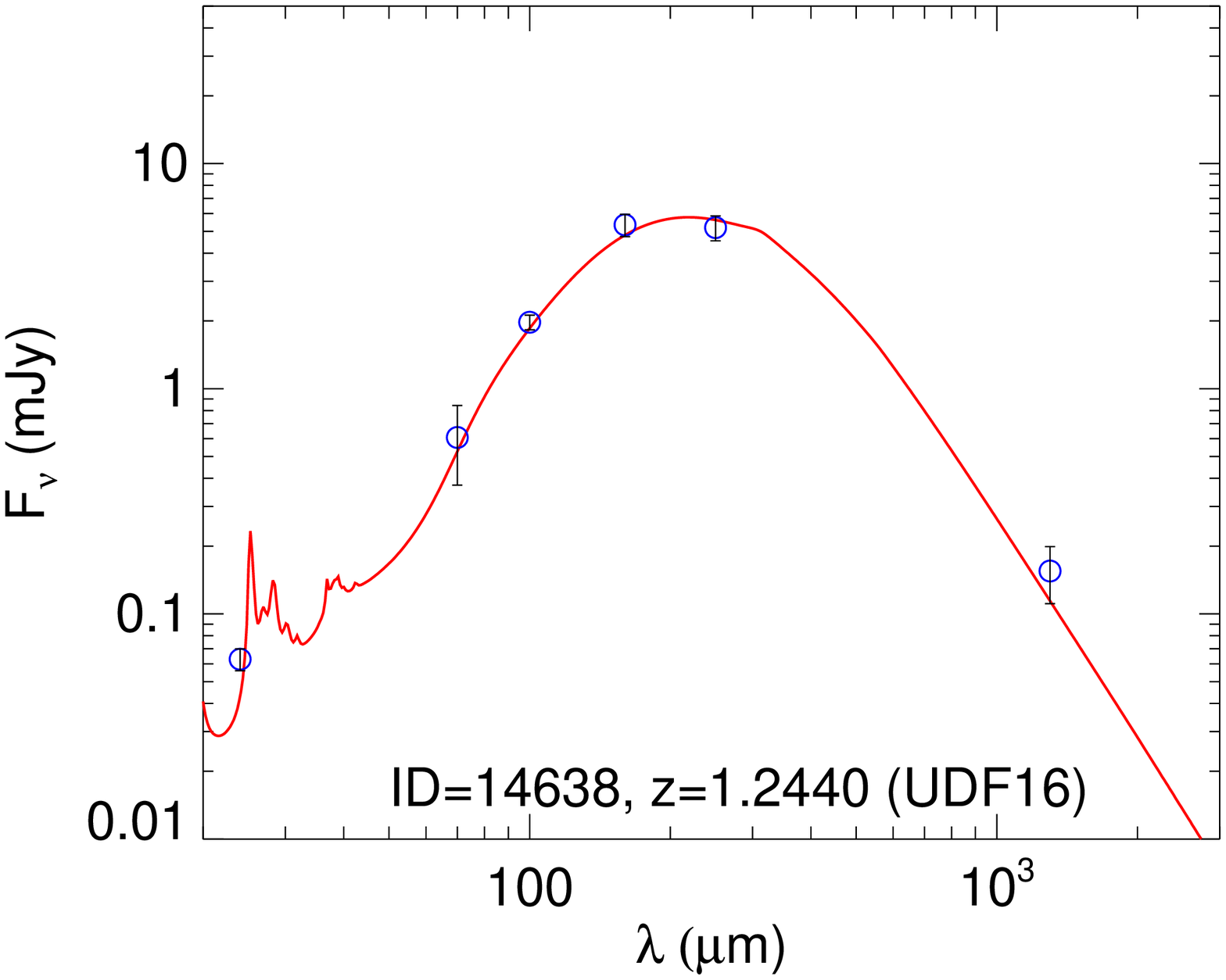}
\includegraphics[width=4cm]{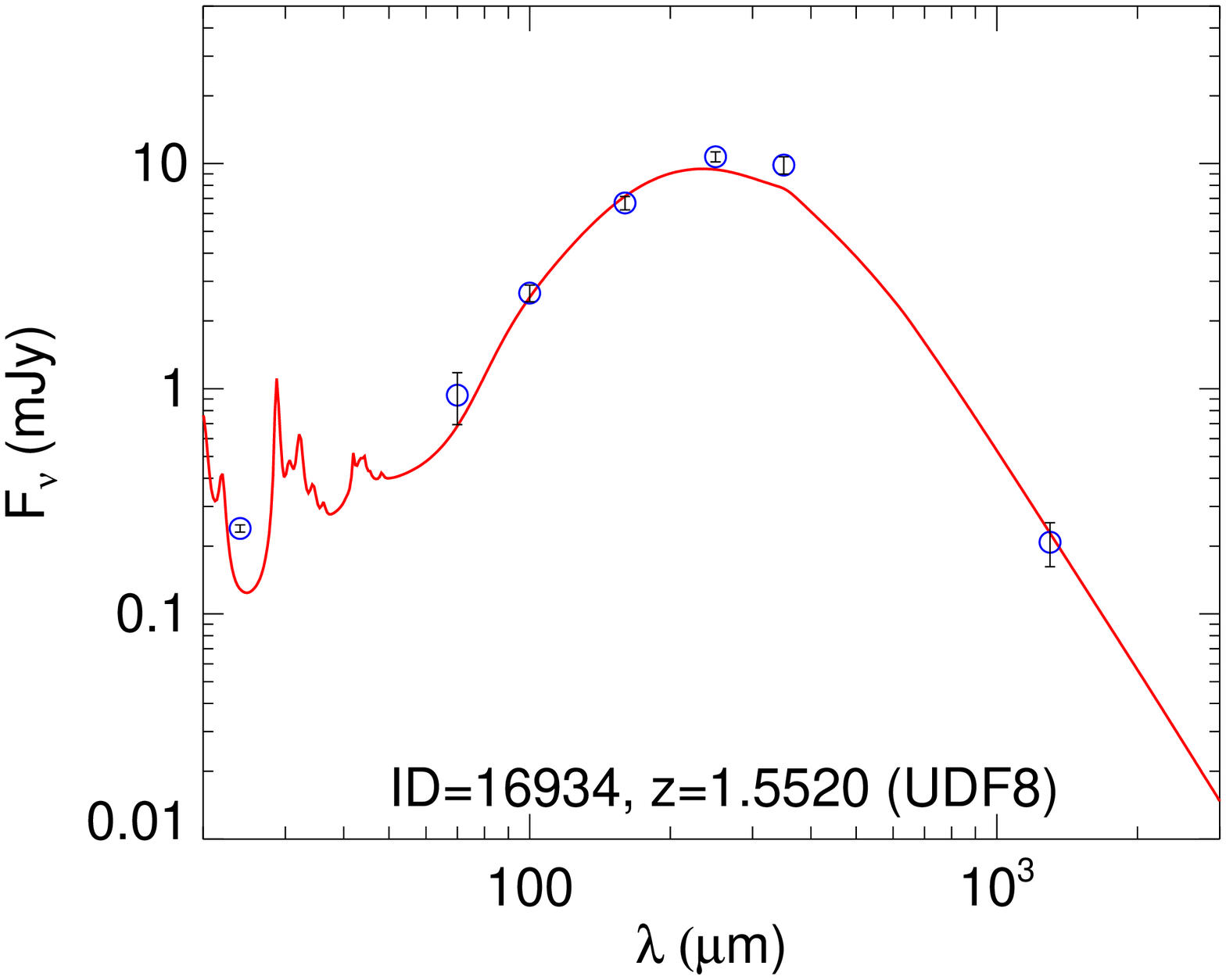}
\includegraphics[width=4cm]{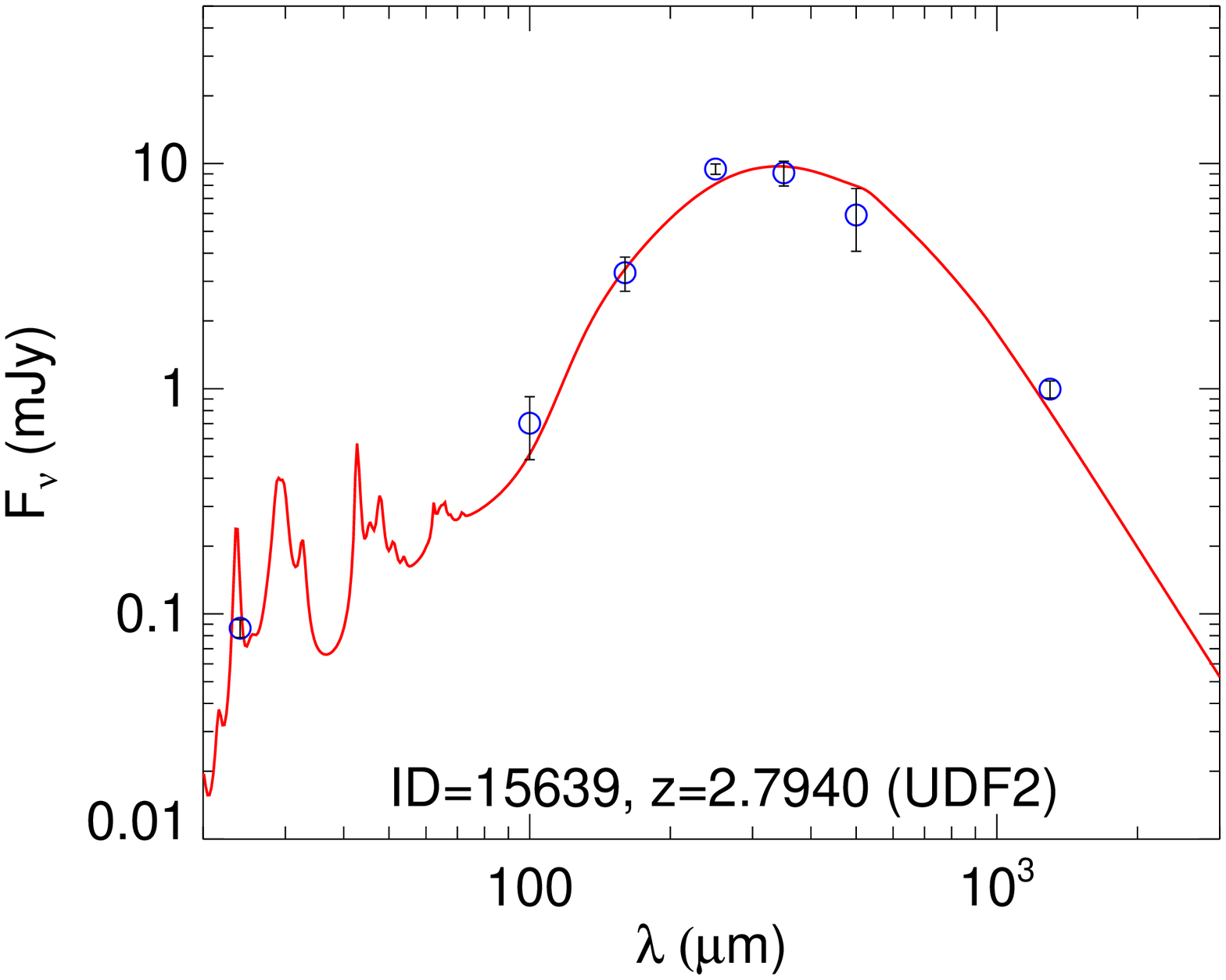}
\includegraphics[width=4cm]{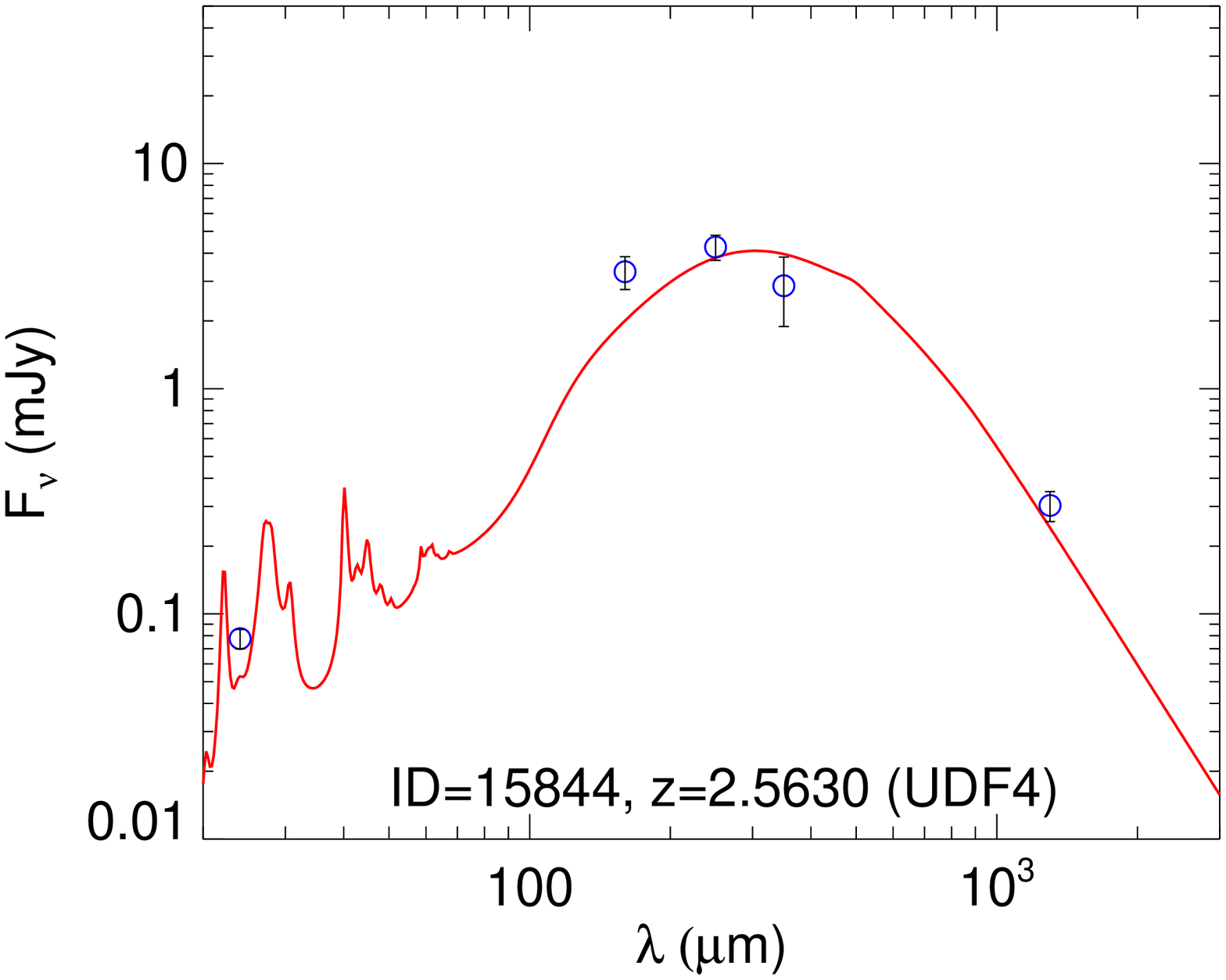}
\includegraphics[width=4cm]{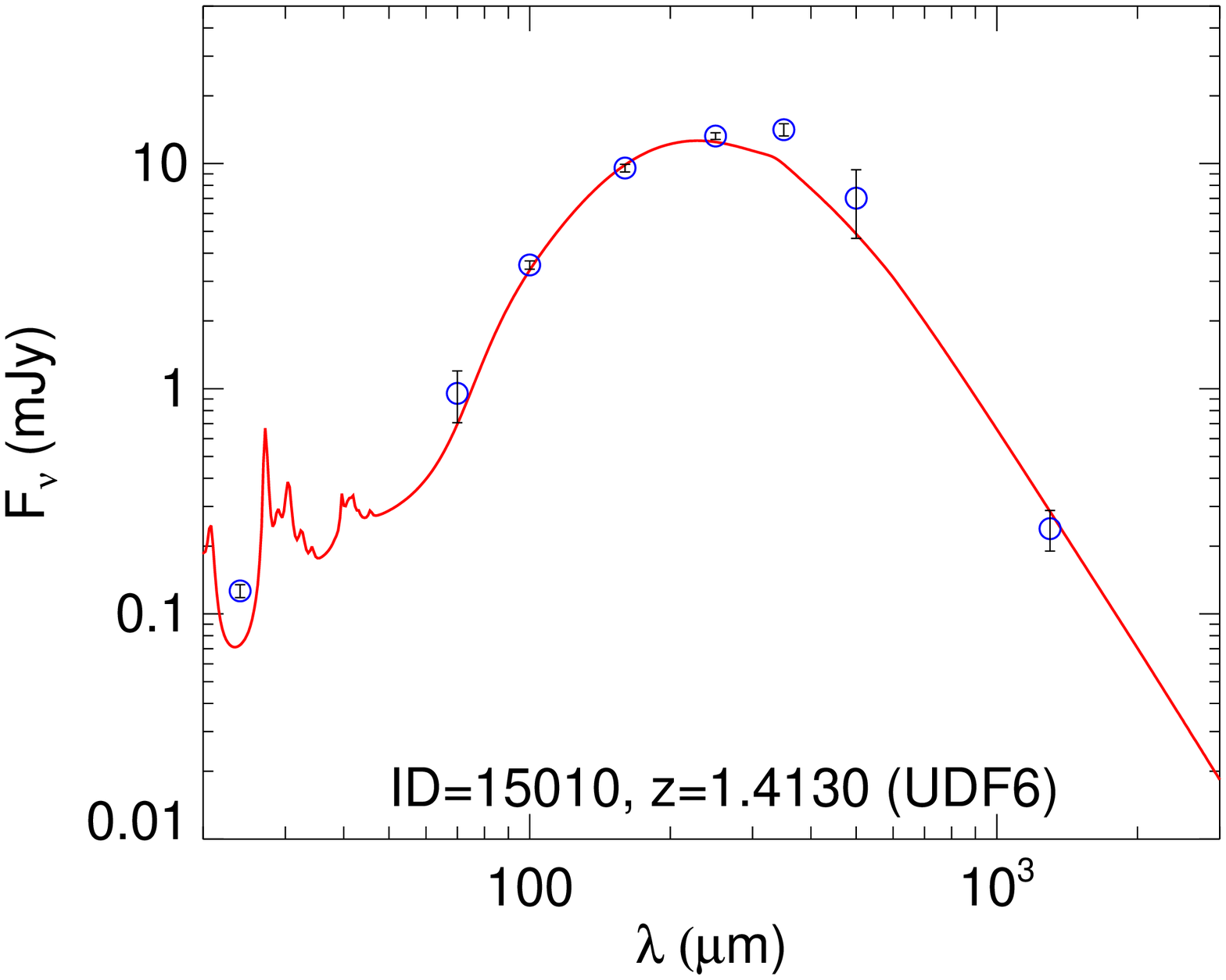}
\includegraphics[width=4cm]{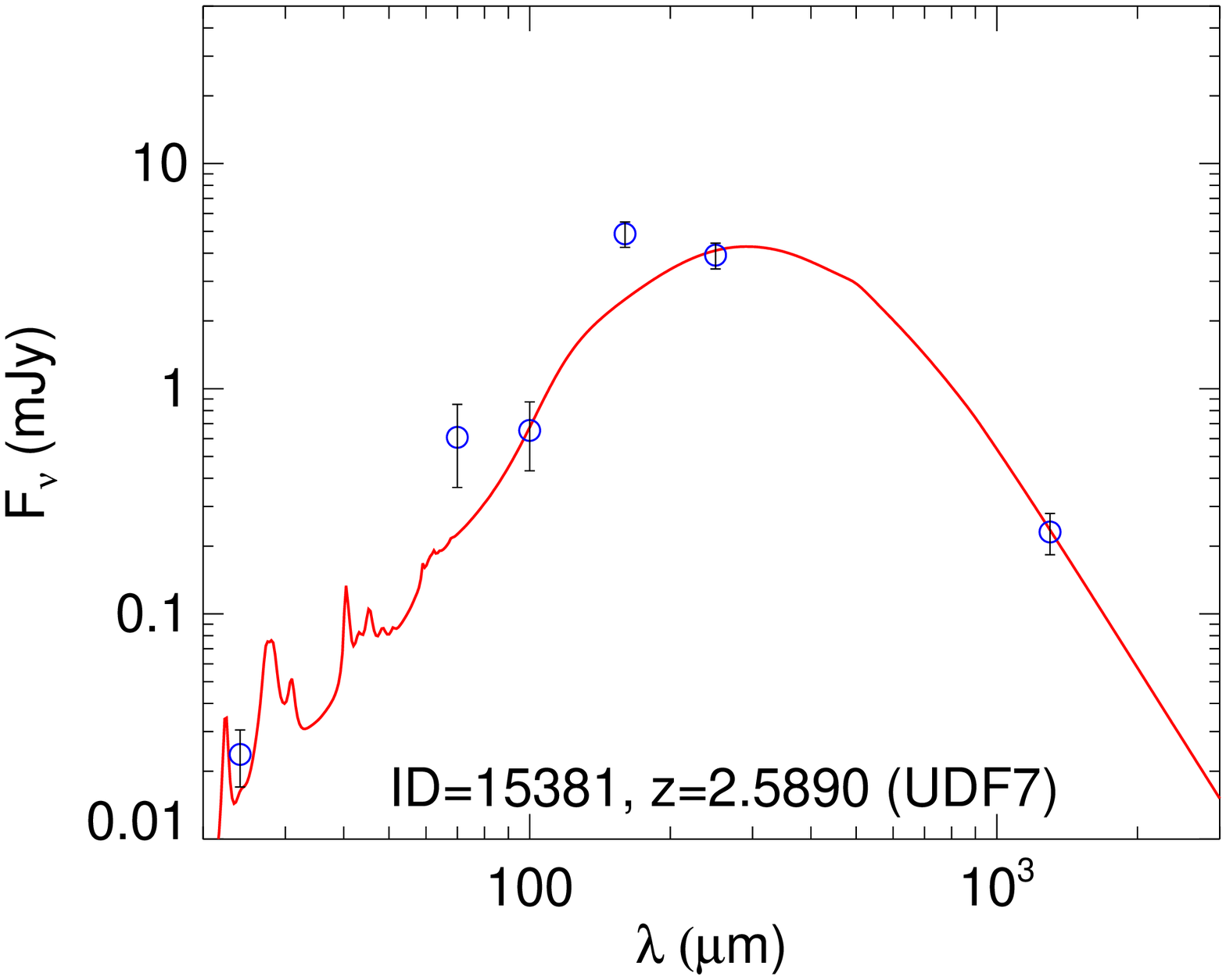}
\includegraphics[width=4cm]{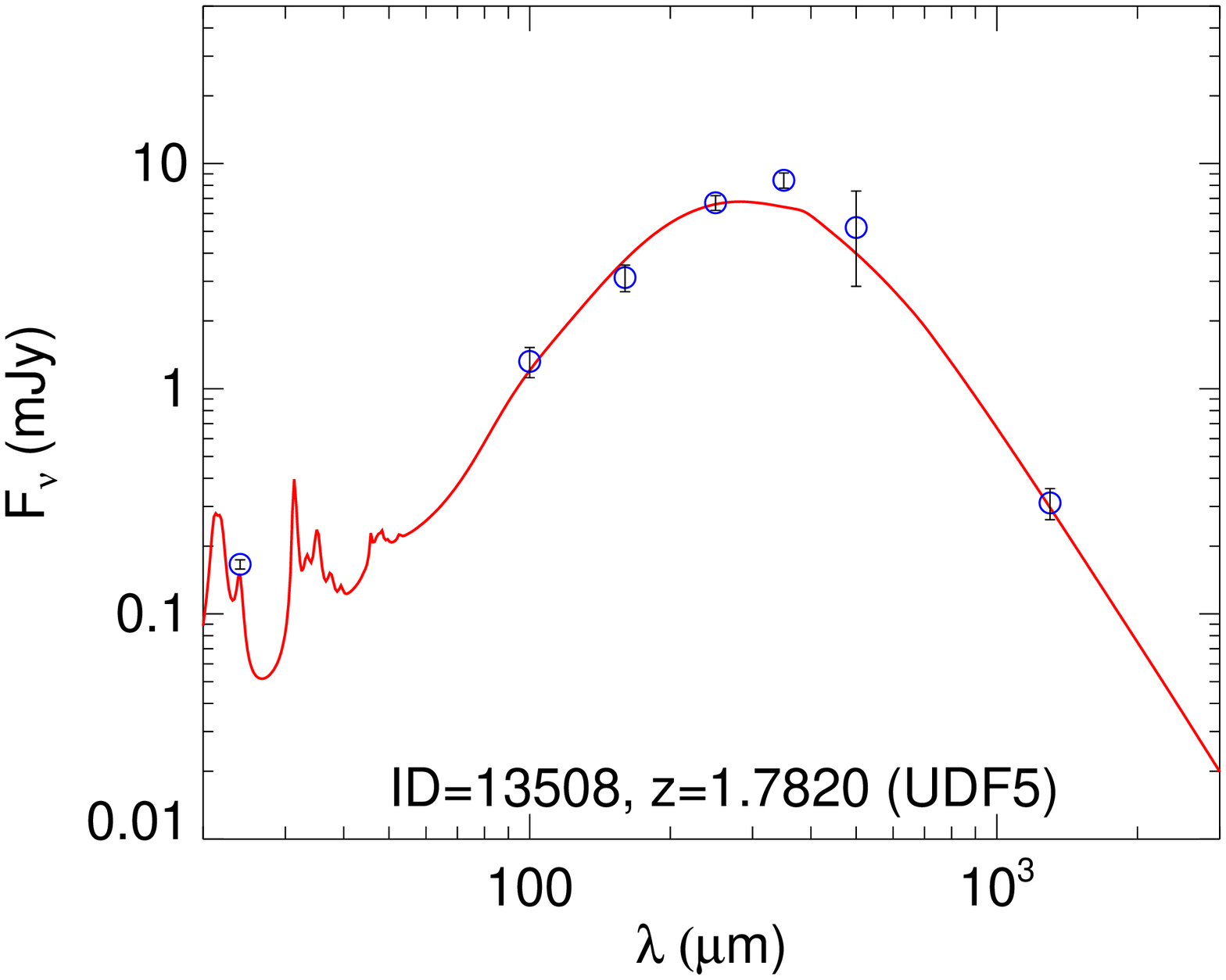}
\includegraphics[width=4cm]{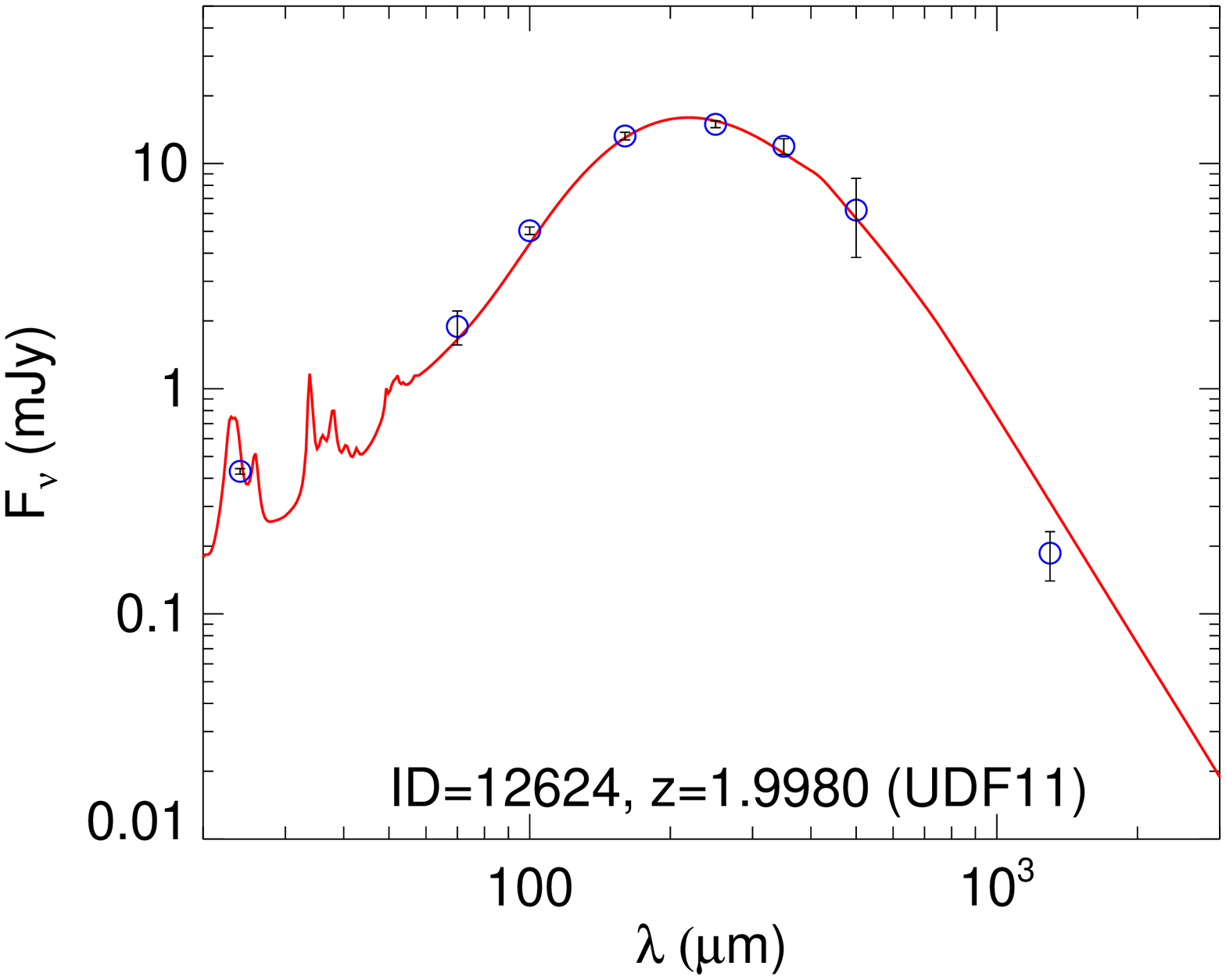}
\includegraphics[width=4cm]{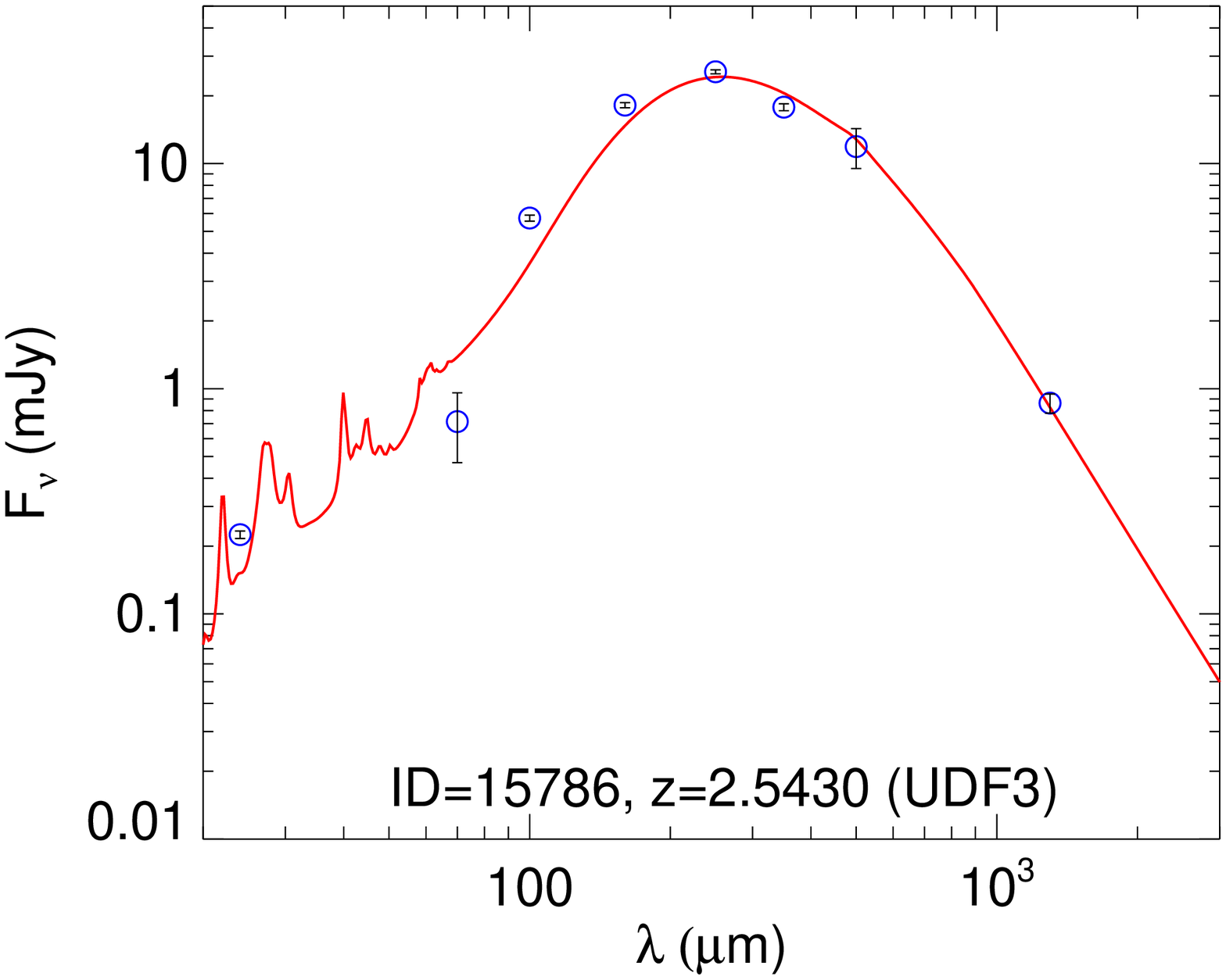}
\includegraphics[width=4cm]{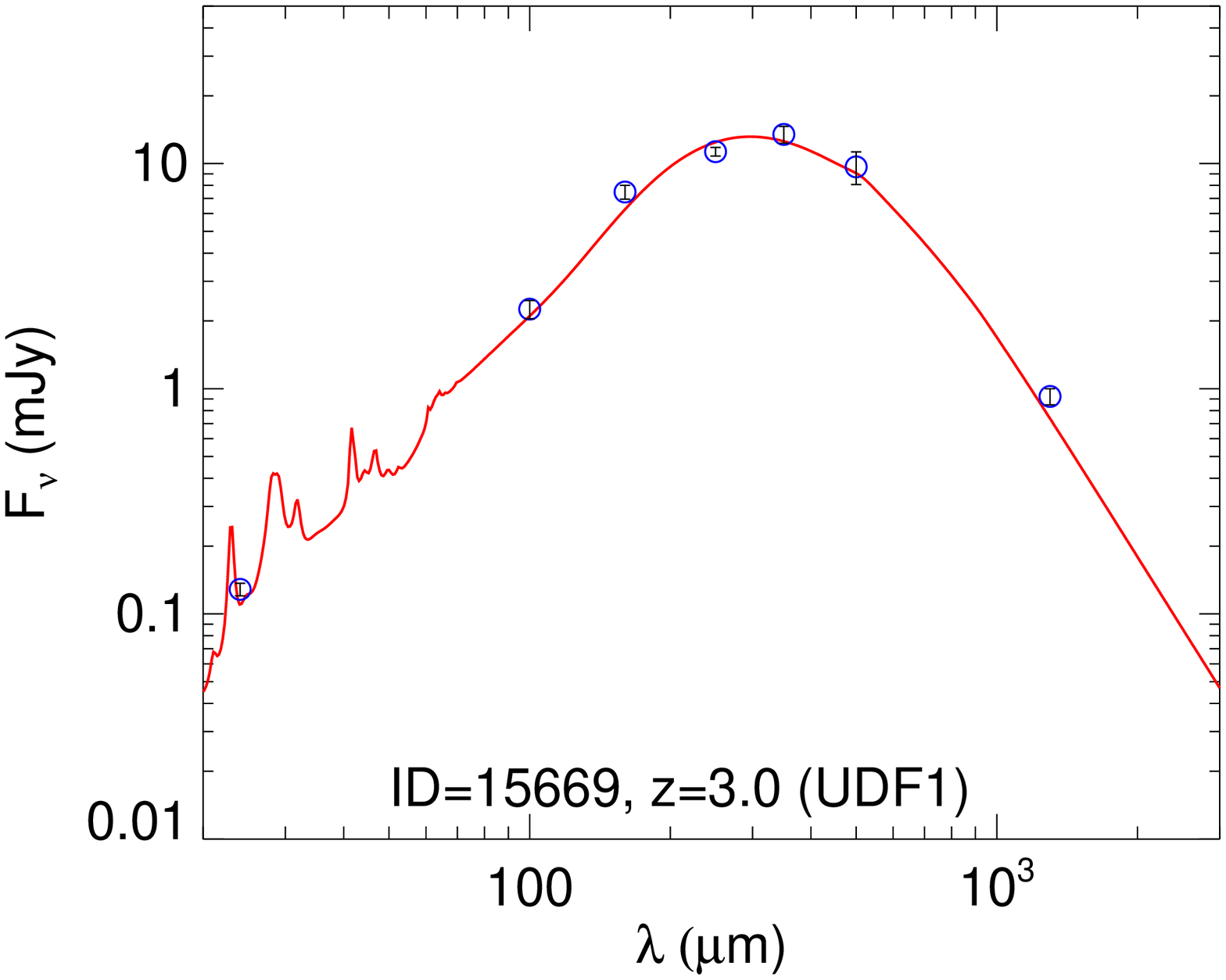}
      \caption{Mid-IR to sub-millimeter SEDs of the 11 UDF galaxies fitted by the best-fitting IR energy distribution from \cite{draine07}.}
     \label{FIG:SEDs_UDF}
  \end{figure}
The ALMA sources were cross-matched with the catalog of GOODS-\textit{Herschel} sources described in \cite{elbaz11}. All of the sources discussed in the present paper are detected with both \textit{Herschel} photometers PACS and SPIRE with a S/N$>$3. The 8 GS sources are detected in the two PACS bands and at 250 and 350\,$\mu$m with SPIRE (including 3 at 500\,$\mu$m). The 11 UDF sources are detected in the PACS-160\,$\mu$m and SPIRE-250\,$\mu$m bands, 9 are detected at 100\,$\mu$m, 7 at 350\,$\mu$m and 2 at 500\,$\mu$m. The 500\,$\mu$m is obviously mainly limited by the large beam size at this wavelength with \textit{Herschel} that imposes a hard confusion limit.

The full SEDs including the optical, near-IR, mid-IR, far-IR and sub-millimeter flux densities of the 8 GS galaxies are presented in Fig.~\ref{FIG:SEDs_GS} together 
with spectral model fits to the data. The fit of the stellar side of the galaxies was used to determine their photometric redshifts with the EAzY\footnote{Publicly available at http://www.github.com/gbrammer/eazy-photoz} code \citep{brammer08} and stellar masses with the FAST\footnote{Publicly available at http://astro.berkeley.edu/~mariska/FAST.html} code that is compatible with EAzY (see the Appendix of \citealt{kriek09}). For the galaxies with spectroscopic redshifts (GS4, GS5, GS6 and GS7), we computed the stellar masses at these spectroscopic values. The case of GS8 is peculiar and will be discussed in detail in Section~\ref{SEC:dark}. 

For the UDF galaxies, we used the same redshifts as \cite{dunlop17} and \cite{rujopakarn16} for consistency. We present their dust SEDs in Fig.~\ref{FIG:SEDs_UDF}. We computed the stellar masses of the UDF galaxies at those redshifts. 

Following \cite{pannella15}, stellar masses were computed using a delayed exponentially declining star formation history  with the \cite{bruzual03} stellar population synthesis model to fit the observed photometry up to the IRAC 4.5\,$\mu$m band. We assumed a solar metallicity, a \cite{salpeter55} IMF and a \cite{calzetti00} attenuation law with $A_V$ ranging from 0 to 4. 

The dust side of the SED of the galaxies (from the \textit{Spitzer} IRS-16\,$\mu$m and MIPS-24\,$\mu$m to the ALMA flux densities) was modeled using the IR emission spectra for dust heated by stellar light from \cite{draine07} by running the code CIGALE\footnote{Publicly available at http://cigale.lam.fr} \citep{noll09}. Following \cite{draine07b}, we fixed the slope of the distribution of intensities of the interstellar radiation field (ISRF, $U$), $\alpha$, to $\alpha$=2 and adopted an upper limit of $U_{\rm max}$=10$^6$ $U_{\odot}$ for the ISRF in units of the solar ISRF. 

The best-fitting SED was used to determine for each galaxy its total dust emission, $L_{\rm IR}$, and a dust mass, M$_{\rm dust}$, listed in Table~\ref{TAB:sample}. To derive a gas mass, we determined the total gas-to-dust ratio ($\delta_{\rm GDR}$=M$_{\rm gas}$/M$_{\rm dust}$) using Eq.~\ref{EQ:leroy} from \cite{leroy11} (given in the text of their Section 5.2) that links $\delta_{\rm GDR}$ with metallicity for local galaxies, hence assuming that this relation holds at all redshifts. 
\begin{equation}
\begin{array}{lcl}
\log_{10}\left(\delta_{\rm GDR}\right)&=& \log_{10}\left(\frac{M_{HI} + M_{H_2}}{M_{\rm dust}}\right) \\
\\
&= & (9.4\pm1.1) - (0.85\pm0.13) \left[ 12 + \log_{10} \left( O/H \right) \right]
\end{array}
\label{EQ:leroy}
\end{equation}

 \begin{figure}
 \centering
      \includegraphics[width=8.8cm]{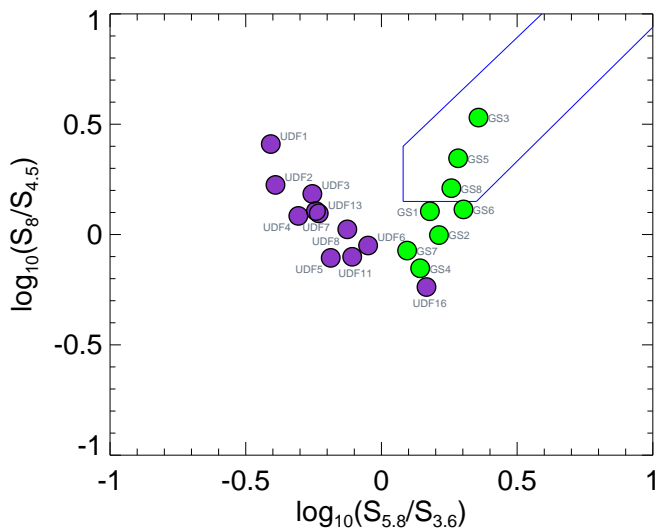}     
      \caption{Mid-IR color-color diagram to search for potential power-law AGNs. Galaxies within the solid blue line are considered as candidate power-law AGNs by \cite{donley12}.}
     \label{FIG:AGNs}
  \end{figure}

Metallicities for our $z$$\sim$2 sample of ALMA galaxies were inferred using the mass - metallicity relation described in Eq.~\ref{EQ:genzel} taken from \citealt{genzel12} (see their Section 2.2) for galaxies at $z$=1.5 -- 3, based on a combination of datasets including the data of \cite{erb06}. 
\begin{equation}
\begin{array}{ll}
12+\log_{10} (O/H) = \\
-4.51+2.18~\log_{10}\left( M_{\star}/1.7 \right) - 0.0896\left[\log_{10}\left( M_{\star}/1.7 \right)\right]^2 
\end{array}
\label{EQ:genzel}
\end{equation}
We replaced M$_{\star}$ in Eq.~\ref{EQ:genzel}  by M$_{\star}$/1.7 since \cite{genzel12} used a Chabrier IMF whereas we are here using a Salpeter IMF (M$_{\star}^{\rm Chabrier}$=M$_{\star}^{\rm Salpeter}$/1.7). We note a potential caveat in this analysis that is related to the metallicity of dusty starbursts. If the metallicity of starbursts above the main sequence was found to be systematically higher than the one of galaxies of similar masses within the main sequence (as suggested by \citealt{silverman15}, \citealt{puglisi17}), we would underestimate their gas content and overestimate their depletion time.

Three galaxies are classified as power-law active galactic nuclei (AGN) following the color-color diagram definition of \cite{donley12} (blue solid line in Fig.~\ref{FIG:AGNs}): GS3, GS5 and GS8. We used the code \textit{decompIR} by \cite{mullaney11} to subtract AGN contributions for all the galaxies. \textit{decompIR} consistently identified an AGN contribution at 8\,$\mu$m for the three power-law AGNs and at a lower level for the galaxies GS1 and GS6 which stand very close to the limit of \cite{donley12} and for which the code \textit{decompIR} found a small contribution. For all the other galaxies, \textit{decompIR} did not find any noticeable AGN contribution.
The AGN component (shown with the purple dashed line in Fig.~\ref{FIG:SEDs_GS}) was subtracted from the data in a first iteration. We then applied the CIGALE code to fit the residual emission and determine L$_{\rm IR}$ and M$_{\rm dust}$ free from any AGN contamination. We note however that both values are little affected by this AGN emission that mostly contributes to the mid-IR spectral range. 

\subsection{Star formation rates and position on the star formation main sequence}
\label{SEC:SFR}
\begin{table*}[htp]
\caption{Integrated properties of the ALMA sources.}
\begin{center}
\begin{tabular}{lcrrrrrrrrr}
\hline
ID & S & $L_{\rm IR}$  & SFR$_{\rm tot}$ & SFR$_{\rm SED}$  & $log_{10}(M_{\star})$ & $R_{\rm SB}$ &$log_{10}(L_X)$& $M_{\rm dust}^{(a)}$ & $M_{\rm gas}^{(a)}$ & $\tau_{\rm dep}$ \\
   & M & ($\times$10$^{11}$L$_{\odot}$) & (M$_{\odot}$yr$^{-1}$)  & (M$_{\odot}$yr$^{-1}$)   & (M$_{\odot}$) &  & (erg.s$^{-1}$) & ($\times$10$^8$M$_{\odot}$)&($\times$10$^{10}$M$_{\odot}$)& (Myr)   \\
(1)&(2)& (3)  &  (4) & (5) & (6) & (7)  & (8)   &  (9) & (10) & (11)  \\ 
\hline
  GS1 & S & 17.6$\pm$0.9 &  306$\pm$15 &  107 & 11.18 &  1.39 & 43.44 &  4.2$\pm$0.4 &  4.9$\pm$0.5 & 161$\pm$24 \\
  GS2 & S & 22.2$\pm$1.9 &  385$\pm$32 &  100 & 11.23 &  1.47 &    -- &  5.1$\pm$0.5 &  6.0$\pm$0.6 & 156$\pm$28 \\
  GS3 & S & 25.3$\pm$1.3 &  438$\pm$22 &  112 & 11.28 &  1.66 & 43.44 &  5.5$\pm$0.3 &  6.3$\pm$0.3 & 144$\pm$14 \\
  GS4 & S & 20.9$\pm$1.8 &  360$\pm$31 &  290 & 11.29 &  1.70 &    -- &  8.2$\pm$1.1 &  9.3$\pm$1.3 & 258$\pm$57 \\
  GS5 & S & 66.8$\pm$3.3 & 1154$\pm$57 &  339 & 11.54 &  2.41 & 43.54 & 14.5$\pm$0.7 & 15.1$\pm$0.8 & 131$\pm$13 \\
  GS6 & M & 66.1$\pm$3.3 & 1139$\pm$57 &  107 & 11.28 &  4.10 & 42.29 & 15.7$\pm$1.2 & 17.9$\pm$1.4 & 158$\pm$20 \\
  GS7 & M & 27.9$\pm$1.4 &  482$\pm$24 &  241 & 10.90 &  5.36 & 42.05 & 10.1$\pm$1.2 & 13.8$\pm$1.6 & 286$\pm$48 \\
  GS8 & U & 58.8$\pm$3.6 & 1016$\pm$61 &  567 & 11.49 &  1.49 & 43.26 & 21.3$\pm$4.1 & 22.4$\pm$4.3 & 220$\pm$55 \\
\hline
 UDF1 & U & 57.4$\pm$2.9 &  987$\pm$49 &   536 & 11.03 &  3.45 & 43.92 & 12.6$\pm$0.8 & 16.0$\pm$1.0 & 162$\pm$ 18 \\
 UDF2 & M & 26.0$\pm$1.3 &  448$\pm$22 &   120 & 11.07 &  1.71 &    -- &  8.3$\pm$1.1 & 10.3$\pm$1.4 & 231$\pm$ 42 \\
 UDF3 & M & 53.4$\pm$2.7 &  928$\pm$46 &    41 & 10.13 & 32.61 & 42.66 & 11.3$\pm$0.6 & 25.4$\pm$1.3 & 274$\pm$ 27 \\
 UDF4 & M & 10.6$\pm$0.6 &  183$\pm$11 &    48 & 10.60 &  2.17 &    -- &  2.9$\pm$0.3 &  4.7$\pm$0.5 & 257$\pm$ 43 \\
 UDF5 & M &  7.7$\pm$0.4 &  132$\pm$ 7 &   127 & 10.39 &  3.67 &    -- &  4.8$\pm$0.8 &  8.8$\pm$1.5 & 669$\pm$147 \\
 UDF6 & S &  9.1$\pm$0.5 &  157$\pm$ 8 &    -- & 10.71 &  2.88 &    -- &  4.0$\pm$0.6 &  6.0$\pm$0.9 & 383$\pm$ 76 \\
 UDF7 & M & 11.2$\pm$0.6 &  194$\pm$10 &    95 & 10.49 &  2.93 & 42.68 &  2.5$\pm$0.2 &  4.4$\pm$0.3 & 224$\pm$ 26 \\
 UDF8 & S &  9.9$\pm$0.5 &  173$\pm$ 9 &   154 & 11.12 &  1.49 & 43.70 &  3.2$\pm$0.5 &  3.9$\pm$0.6 & 227$\pm$ 44 \\
UDF11 & M & 22.5$\pm$1.1 &  396$\pm$19 &   267 & 10.80 &  3.99 & 42.38 &  4.8$\pm$0.2 &  6.9$\pm$0.3 & 173$\pm$ 17 \\
UDF13 & S &  7.4$\pm$0.5 &  128$\pm$ 9 &   166 & 10.81 &  0.96 & 42.45 &  1.9$\pm$0.3 &  2.7$\pm$0.4 & 214$\pm$ 46 \\
UDF16 & S &  3.2$\pm$0.2 &   54$\pm$ 3 &    -- & 10.80 &  1.01 &    -- &  1.7$\pm$0.5 &  2.4$\pm$0.7 & 439$\pm$145 \\
\hline\end{tabular}
\end{center}
\begin{small}
\textit{\textbf{Notes:}} 
Col.(1) Simplified ID. 
Col.(2) Visual morphological classification of the rest-frame optical images of the galaxies (from \textit{HST}-WFC3 \textit{H}-band): single/isolated galaxy (S), merger (M) and undefined (U). 
Col.(3) Total IR (8 -- 1000\,$\mu$m) luminosity measured from the fit of the data from \textit{Spitzer}, \textit{Herschel} and ALMA.
Col.(4) Total SFR=SFR$_{\rm IR}$+SFR$_{\rm UV}$ in M$_{\odot}$ yr$^{-1}$ where both SFR are defined in Eq.~\ref{EQ:sfrir} and Eq.~\ref{EQ:sfruv}.
Col.(5) SFR derived from the fit of the UV-optical-near IR SED in M$_{\odot}$ yr$^{-1}$ assuming a contant SFR history.
Col.(6) Logarithm of the stellar mass (Salpeter IMF).
Col.(7) Starburstiness, R$_{\rm SB}$=SFR/SFR$_{\rm MS}$, where SFR$_{\rm MS}$ is the MS SFR at the redshift of the galaxy.
Col.(8) Logarithm of the total 0.5--8 $keV$ X-ray luminosities in erg s$^{-1}$ from \cite{luo17}.
Col.(9) Dust mass derived from the fit of the far-IR SED (see Sect.~\ref{SEC:sed}).
Col.(10) Gas mass derived from the dust mass in Col.(10) following the recipe for the dust-to-gas ratio described in Sect.~\ref{SEC:sed}.
Col.(11) Depletion time, $\tau_{\rm dep}$ (=M$_{\rm gas}$/SFR), in Myr. 
\end{small}
\label{TAB:sample}
\end{table*}

The total SFR of the galaxies is defined as the sum of the IR (SFR$_{\rm IR}$) and uncorrected UV (SFR$_{\rm UV}$) SFR, SFR$_{\rm tot}$=SFR$_{\rm IR}$$+$SFR$_{\rm UV}$. SFR$_{\rm IR}$ and SFR$_{\rm UV}$ were computed following the conversions of \cite{kennicutt98} and \cite{daddi04} given in Eq.~\ref{EQ:sfrir} and Eq.\ref{EQ:sfruv}; where L$_{\rm UV}$ is the rest-frame 1500\,\AA\,UV luminosity computed from the best-fitting template obtained with EAzY (uncorrected for attenuation) and L$_{\rm IR}$ is the total dust luminosity given by the best-fitting \cite{draine07} model (see Section~\ref{SEC:sed}).

\begin{equation}
{\rm SFR}_{\rm IR}~[{\rm M_{\odot}~yr^{-1}}]=1.72 \times 10^{-10} \times L_{\rm IR}~[{\rm L_{\odot}}]
\label{EQ:sfrir}
\end{equation}
\begin{equation}
{\rm SFR}_{\rm UV}~[{\rm M_{\odot}~yr^{-1}}]=2.17 \times 10^{-10} \times L_{\rm UV}~[{\rm L_{\odot}}]
\label{EQ:sfruv}
\end{equation}

We also computed SFR$_{\rm SED}$ by fitting the rest-frame UV-optical-NIR data assuming a constant star formation history and a \cite{calzetti00} reddening law (Col.(5) in Table~\ref{TAB:sample}). This SFR$_{\rm SED}$ will be compared to SFR$_{\rm tot}$ in order to determine the presence of residual dust attenuation unaccounted for by the UV-optical SED fitting. To derive SFR$_{\rm SED}$, we limited ourselves to a constant SFR history in order to avoid the degeneracy between dust attenuation and stellar population ages. Rest-frame magnitudes were computed from the best-fit SED model integrated through the theoretical filters by running EAZY on the multi-wavelength catalog. 
The resulting SFR$_{\rm tot}$ and SFR$_{\rm SED}$ are listed in Table~\ref{TAB:sample} together with the total IR luminosities obtained from the SED fitting described in Section~\ref{SEC:sed}. 

 \begin{figure}
 \centering
      \includegraphics[width=9cm]{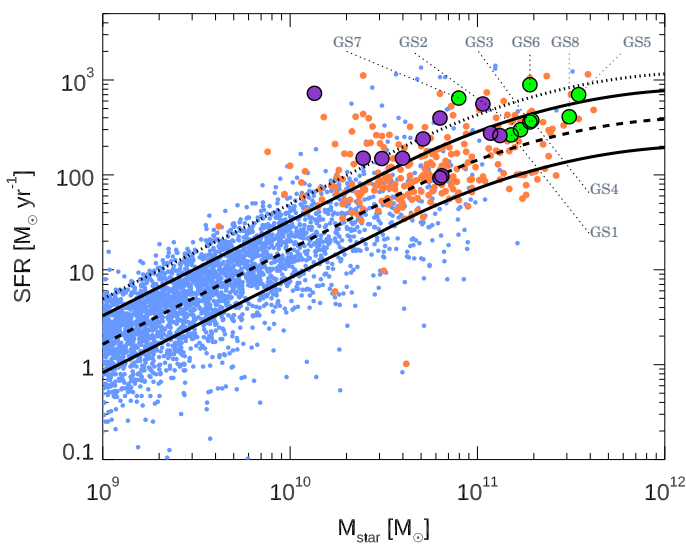}     
      \caption{SFR -- M$_{\star}$ main sequence at 1.5$<$$z$$<$2.5 as measured in GOODS-\textit{South}. Blue dots: SFR$_{\rm SED}$ derived from UV-optical-NIR SED fitting. Orange dots: SFR$_{\rm tot}$ = SFR$_{\rm IR}$ $+$ SFR$_{\rm UV}$. In order to keep the relative position of each galaxy with respect to the main sequence at its redshift, the SFR was multiplied by SFR$_{\rm MS}^{z}$/SFR$_{\rm MS}^{z=2}$ using Eq.9 from \citet{schreiber15}. The 8 GS and 11 UDF ALMA sources discussed in this paper are shown with green and purple filled symbols respectively. The $z$=2 star-formation main sequence and its rms=0.3 dex scatter are shown with a dashed line and solid lines respectively. The dotted line shows the limit above which galaxies are classified as starbursts (R$_{\rm SB}$=SFR/SFR$_{\rm MS}$$>$3).}
     \label{FIG:MS}
  \end{figure}

The positions of the ALMA galaxies in the SFR--M$_{\star}$ plane are shown in Fig.~\ref{FIG:MS} where the eight galaxies from our GOODS-\textit{South} observations are marked with green filled circles. The dashed and two solid black lines show the median and its 68\,\% standard deviation determined by \cite{schreiber15}. The full catalog of GOODS-\textit{South} galaxies at 1.5$<$$z$$<$2.6 is presented with orange and blue dots for the galaxies with and without an \textit{Herschel} detection respectively. For the galaxies with no \textit{Herschel} detection, we used SFR$_{\rm SED}$ (blue dots) while for galaxies with an \textit{Herschel} detection, we used SFR$_{\rm tot}$ (orange dots). 

The position of the MS, i.e. the median of the SFR--M$_{\star}$ values, varies rapidly with redshift. It increases by a factor 1.9 between $z$=1.5 and $z$=2.6 when using the parametrization of the redshift evolution of the MS from  \cite{schreiber15}. In order to place the galaxies at their right distance relative to the MS in Fig.~\ref{FIG:MS}, we first computed this distance at the exact redshift of each individual source and then kept this distance but this time relative to the MS at $z$=2 (dashed line in Fig.\ref{FIG:MS}) when we plotted the galaxies in Fig.~\ref{FIG:MS}. As a result, the SFR of a galaxy located at a redshift $z$=1.6 is shown on Fig.\ref{FIG:MS} with a higher SFR value equal to its actual SFR multiplied by a factor SFR$_{\rm MS}^{z=2}$/SFR$_{\rm MS}^{z=1.6}$. This normalization was only used to produce Fig.\ref{FIG:MS} with realistic galaxy positions in the SFR--M$_{\star}$ plane.

\begin{figure*}
\centering
\includegraphics*[width=0.45\textwidth]{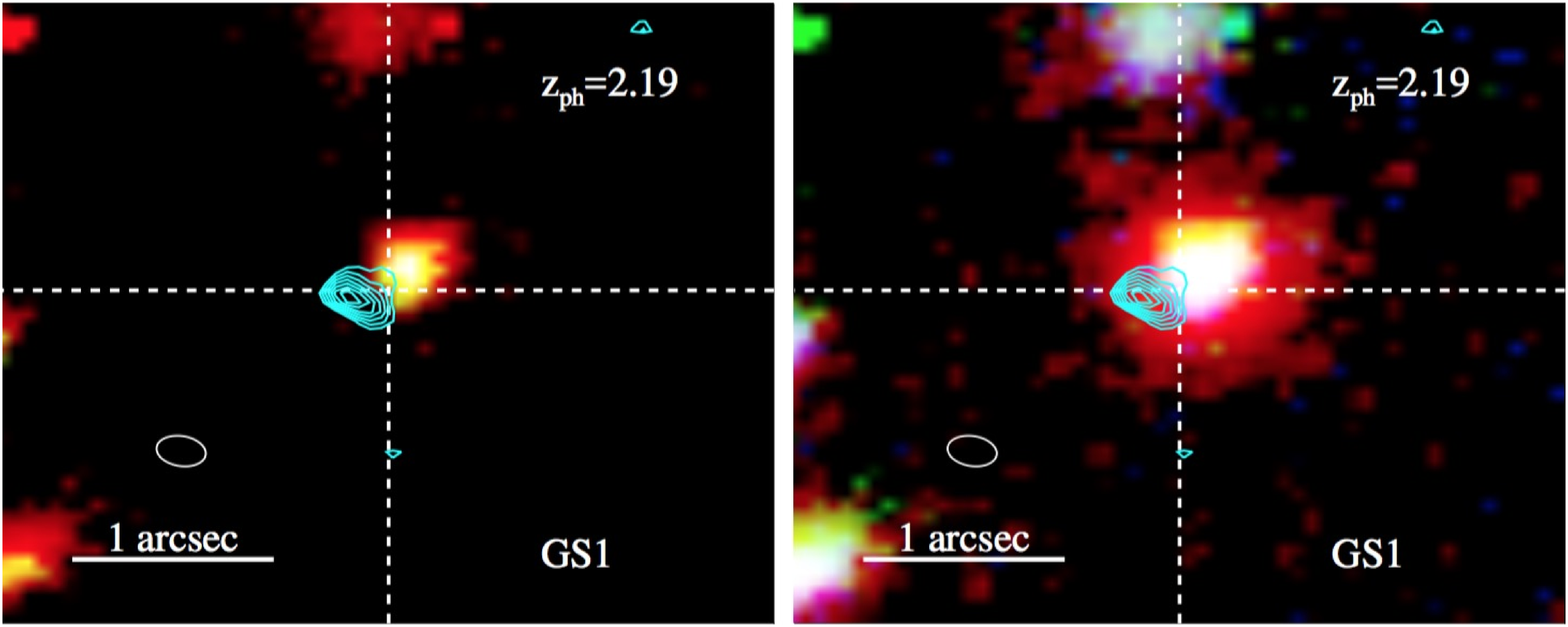}
\includegraphics*[width=0.45\textwidth]{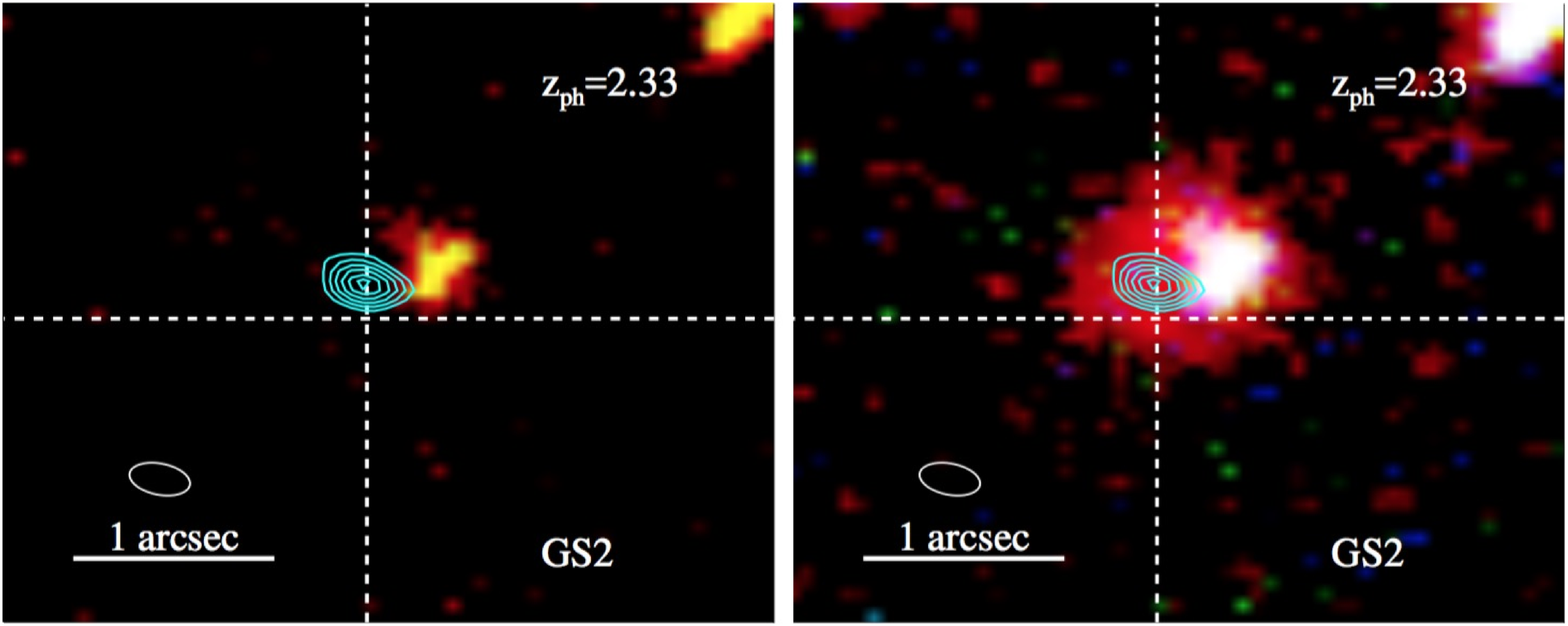}
\includegraphics*[width=0.45\textwidth]{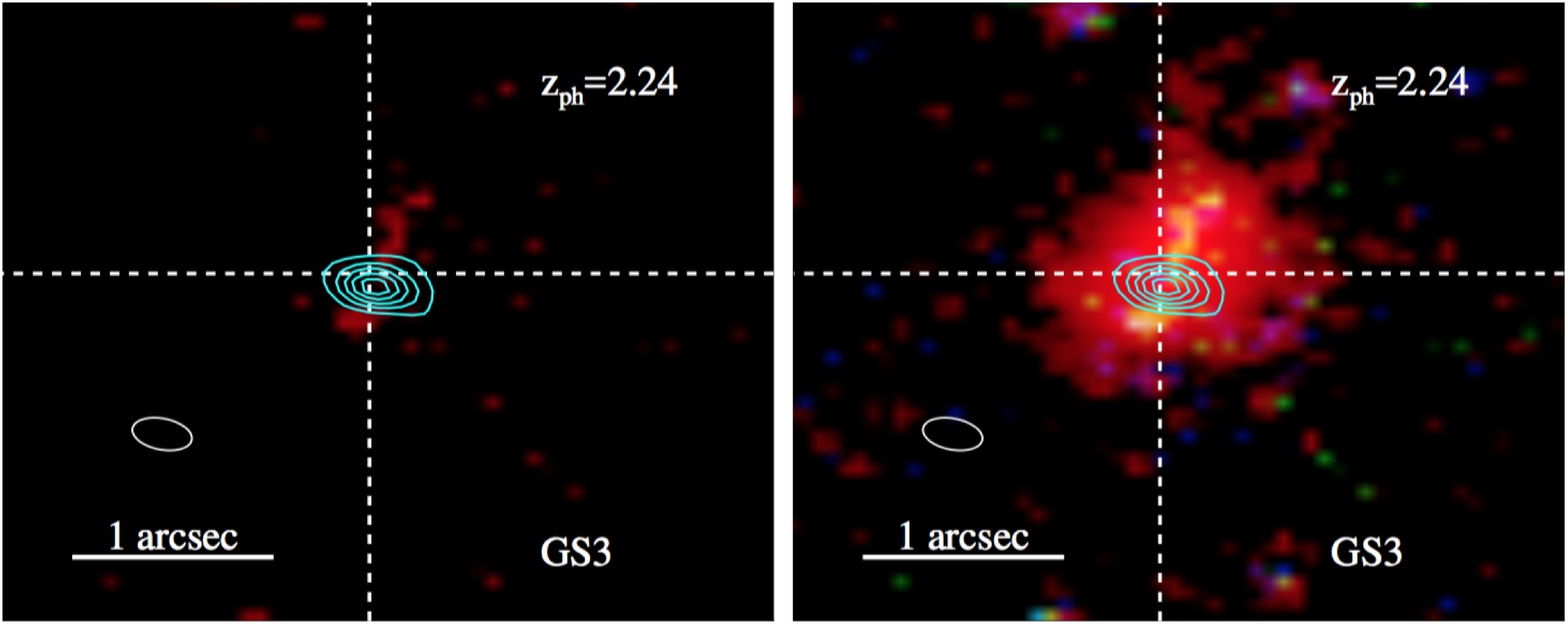}
\includegraphics*[width=0.45\textwidth]{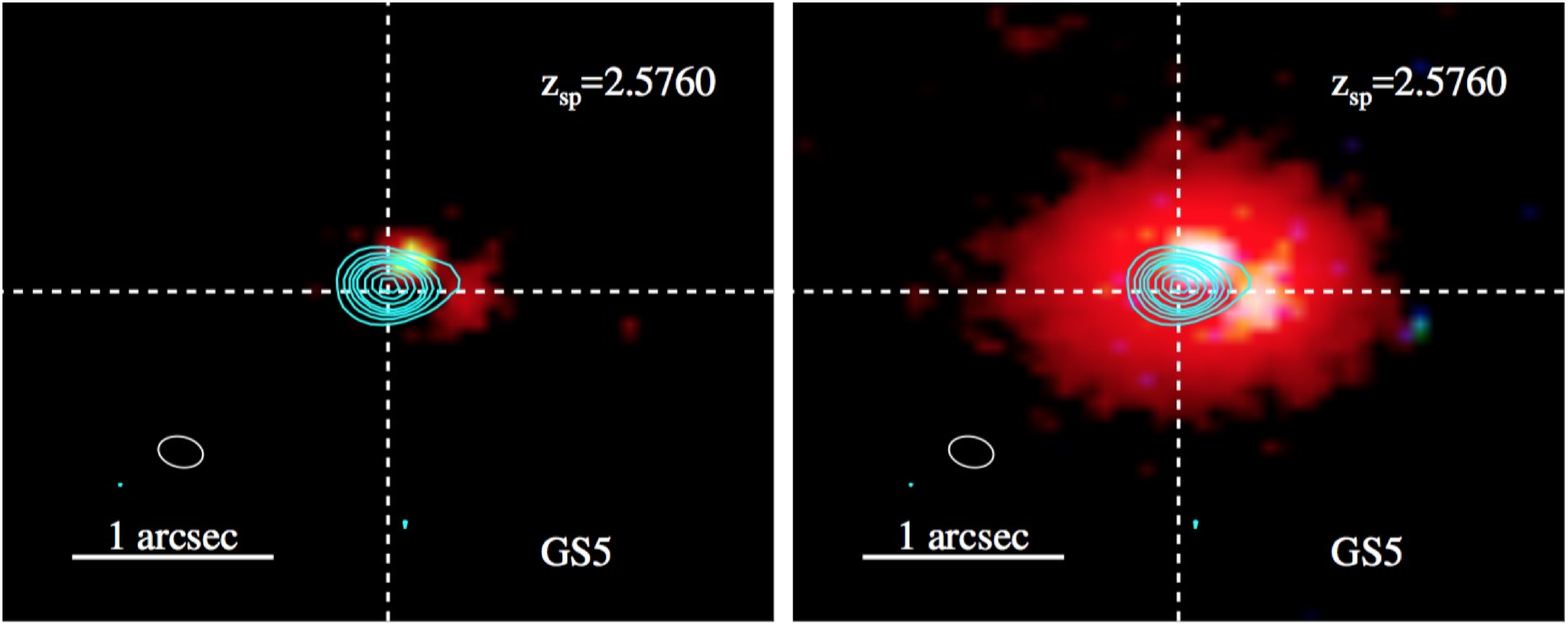}
\includegraphics*[width=0.45\textwidth]{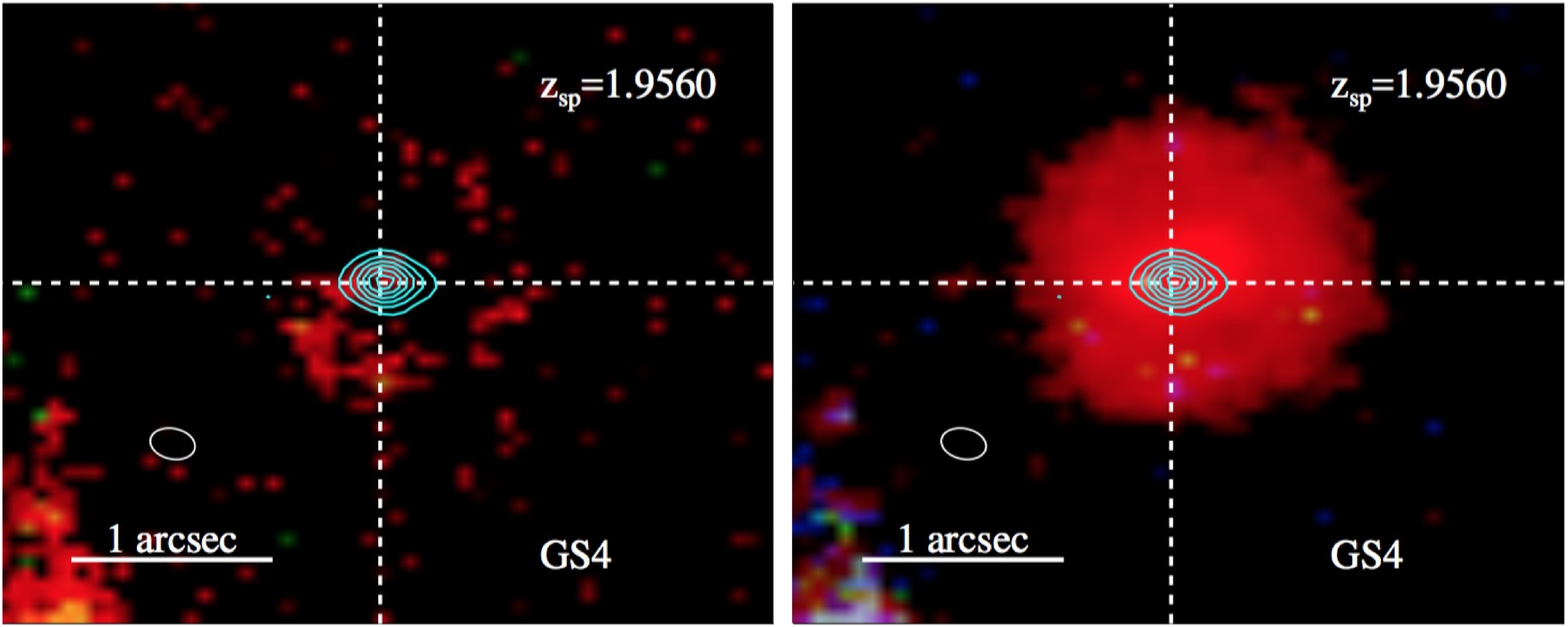}
\includegraphics*[width=0.45\textwidth]{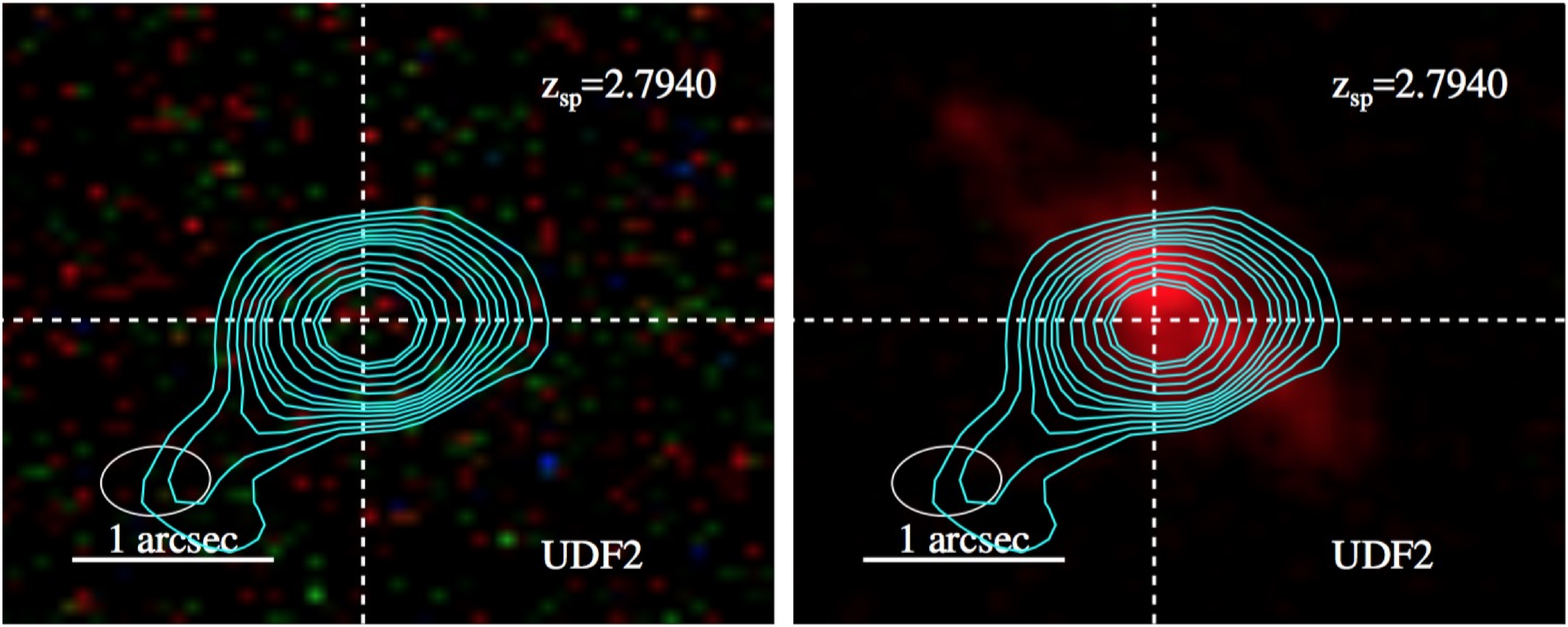}
\includegraphics*[width=0.45\textwidth]{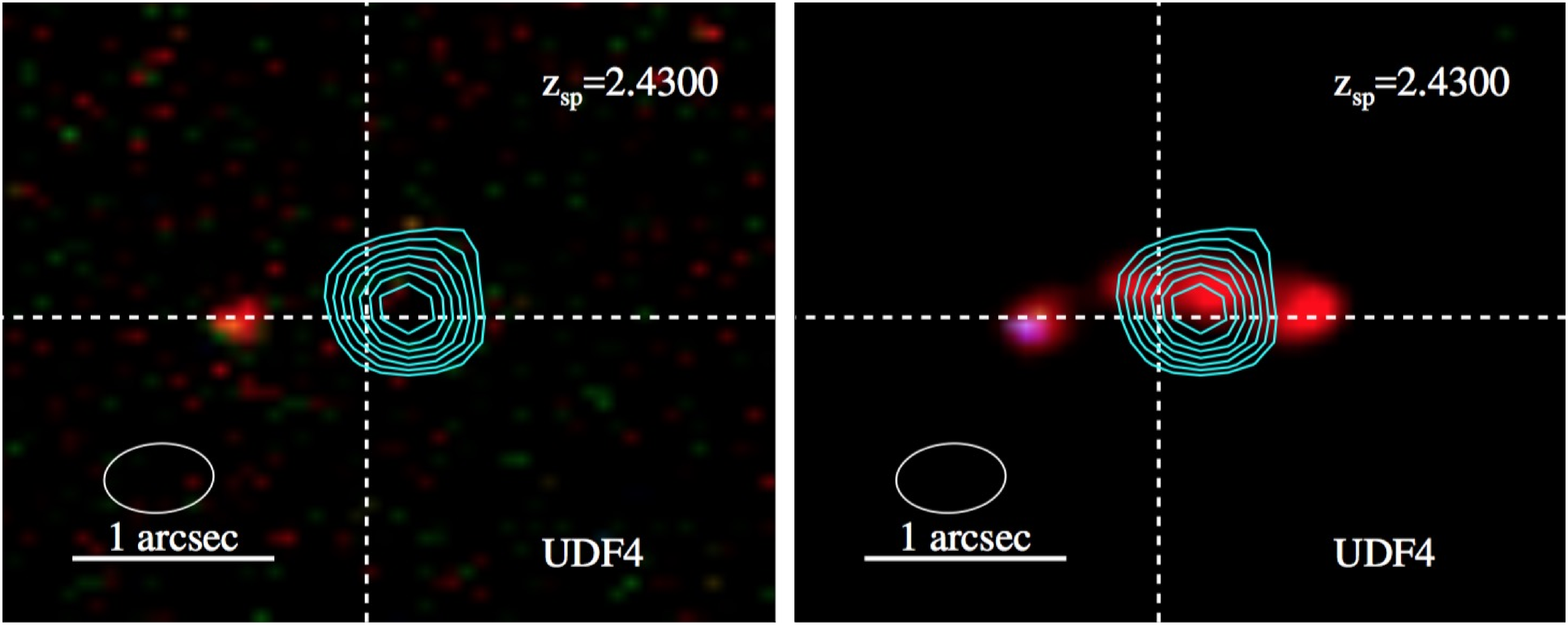}
\includegraphics*[width=0.45\textwidth]{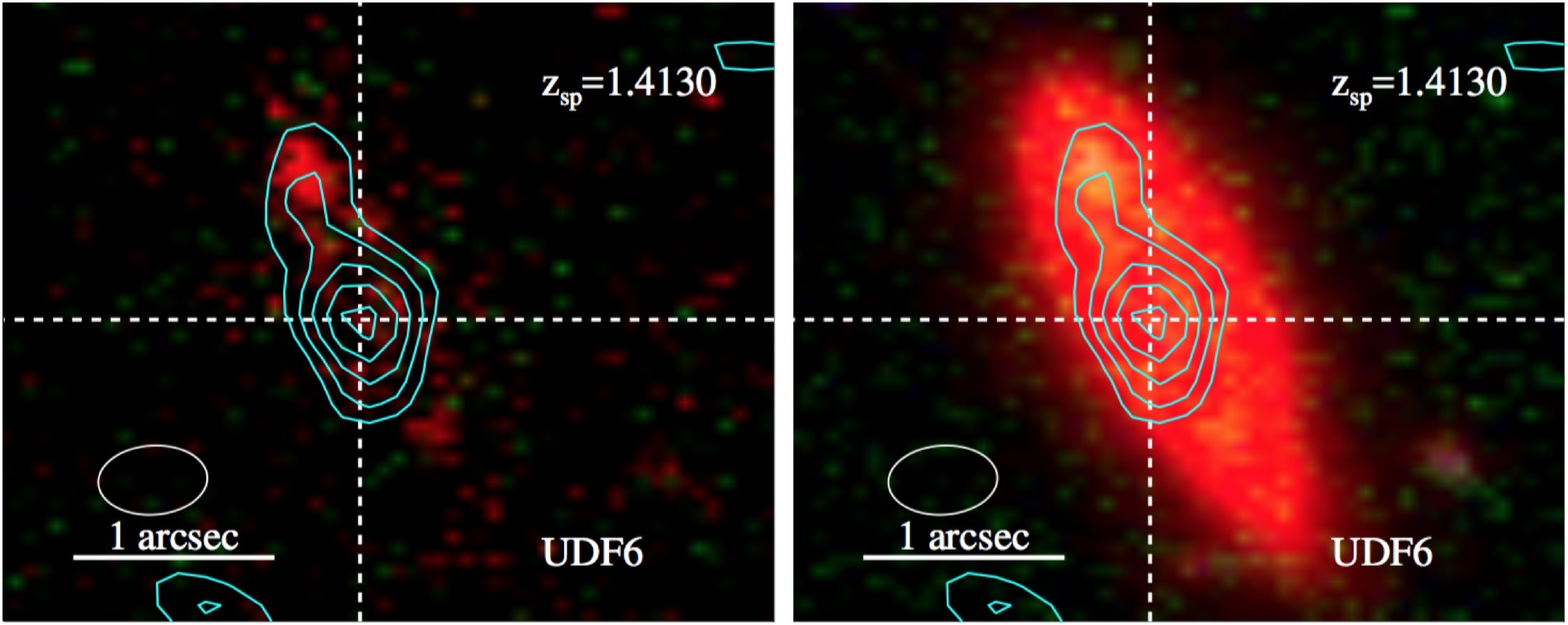}
\includegraphics*[width=0.45\textwidth]{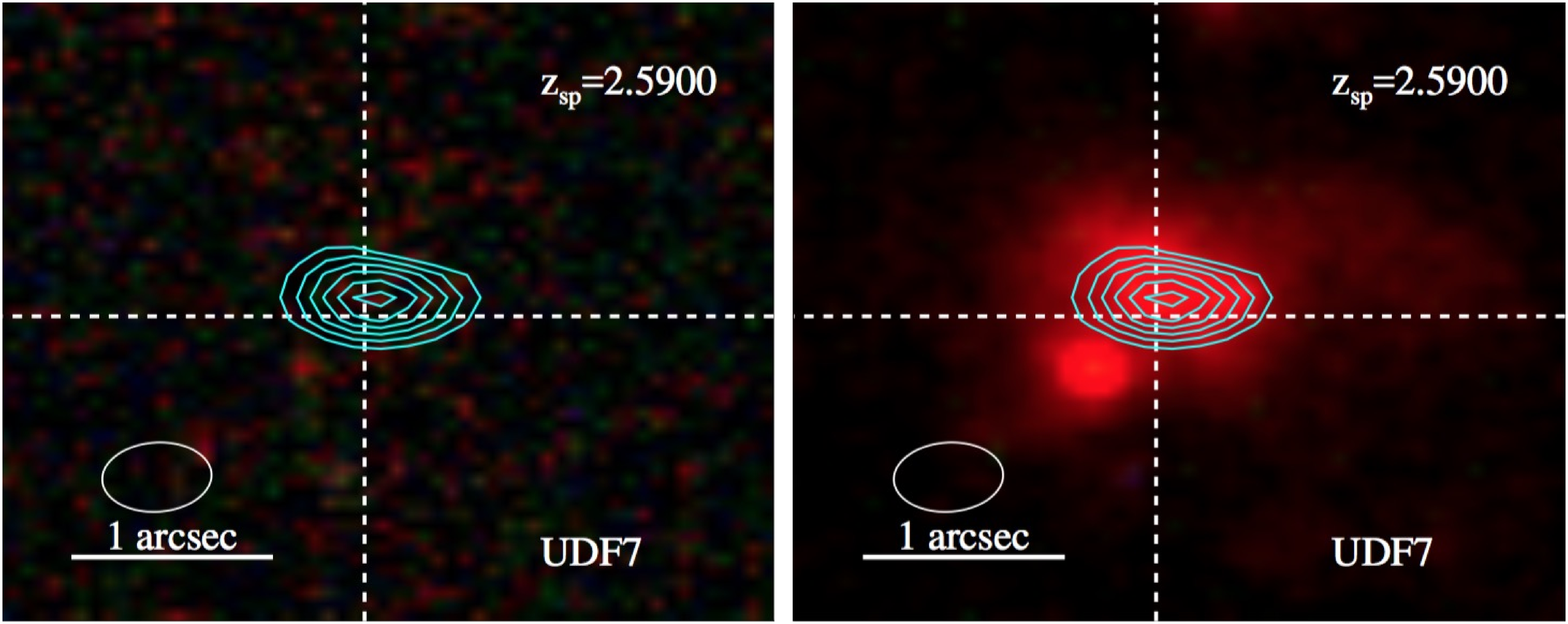}
\includegraphics*[width=0.45\textwidth]{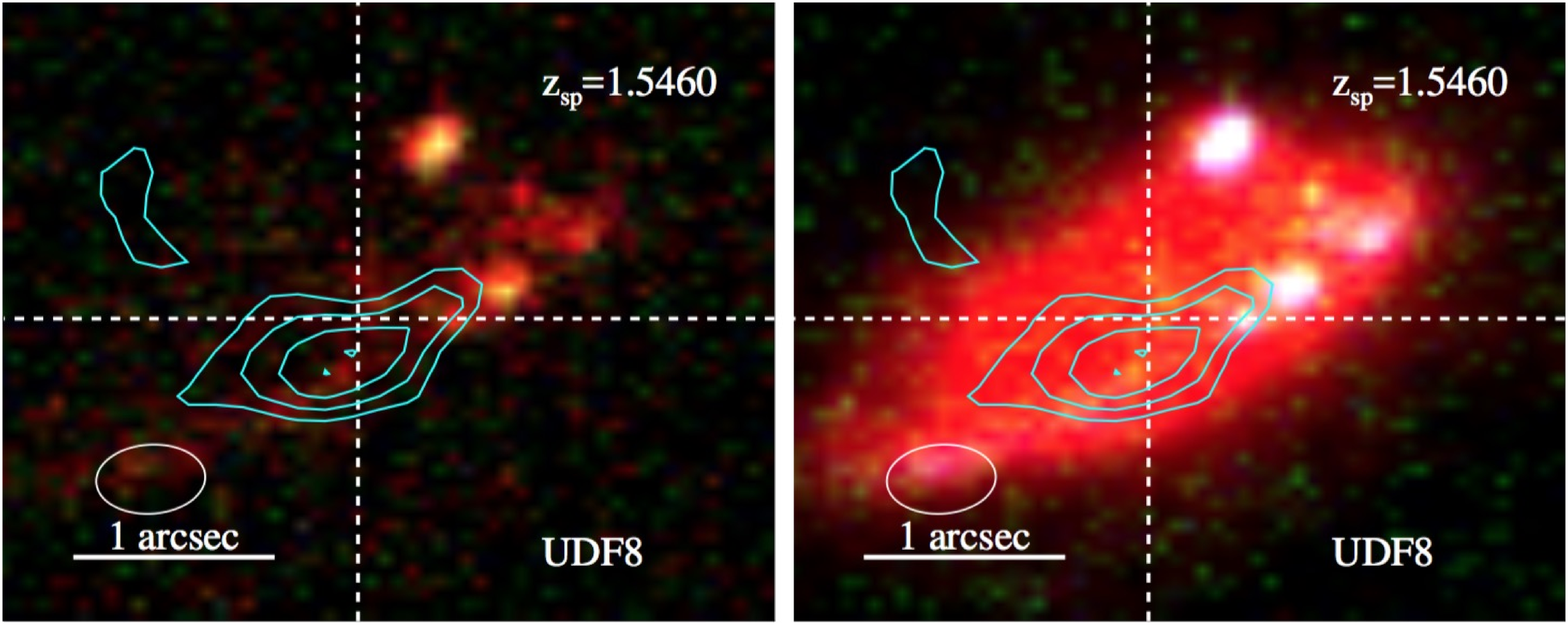}
\includegraphics*[width=0.45\textwidth]{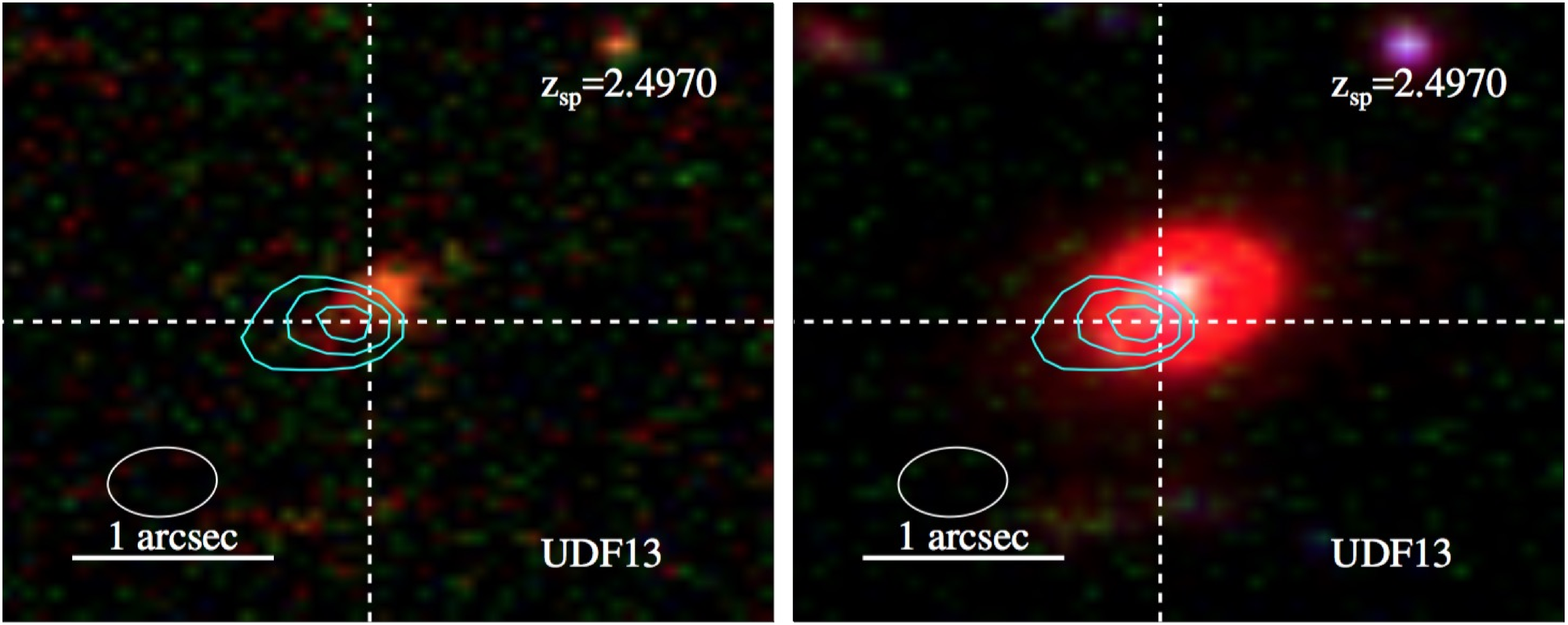}
\includegraphics*[width=0.45\textwidth]{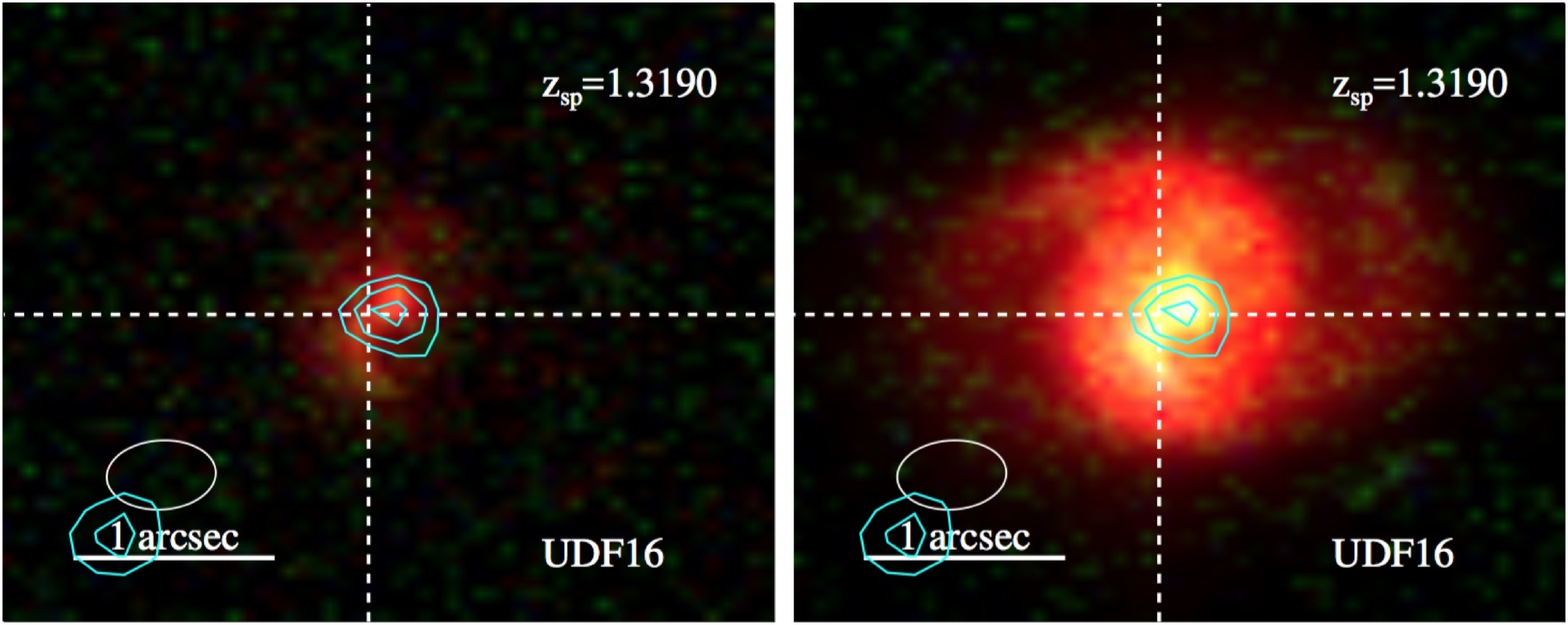}
\caption{ALMA contours on \textit{HST} images for main sequence galaxies. For each galaxy, the \textbf{\textit{Left}} column shows the rest-frame UV from ACS F606W \& F814W (equivalent to 2020 \& 2710\,\AA\, rest-frame at $z$$\sim$2), whereas the \textbf{\textit{Right}} column includes the \textit{HST}--WFC3 F160W band (1.6\,$\mu$m) sampling the rest-frame 5300\,\AA\, band at $z$$\sim$2 (here RGB = V,I,H). The ALMA contours correspond to 1.3mm for the UDF sources and 870\,$\mu$m for the GS source. Contours: for the UDF galaxies increase with steps of 0.5$\sigma$ starting from 2.2$\sigma$, for the GS galaxies: 3$\sigma$ steps up to 14$\sigma$, then 21, 29, 36$\sigma$. The dashed crosses are centered on the WFC3 F160W band center. The ALMA PSF is shown in the bottom left of each figure.}
\label{FIG:IM_MS}
\end{figure*}

\begin{figure}
\centering
\includegraphics*[width=0.45\textwidth]{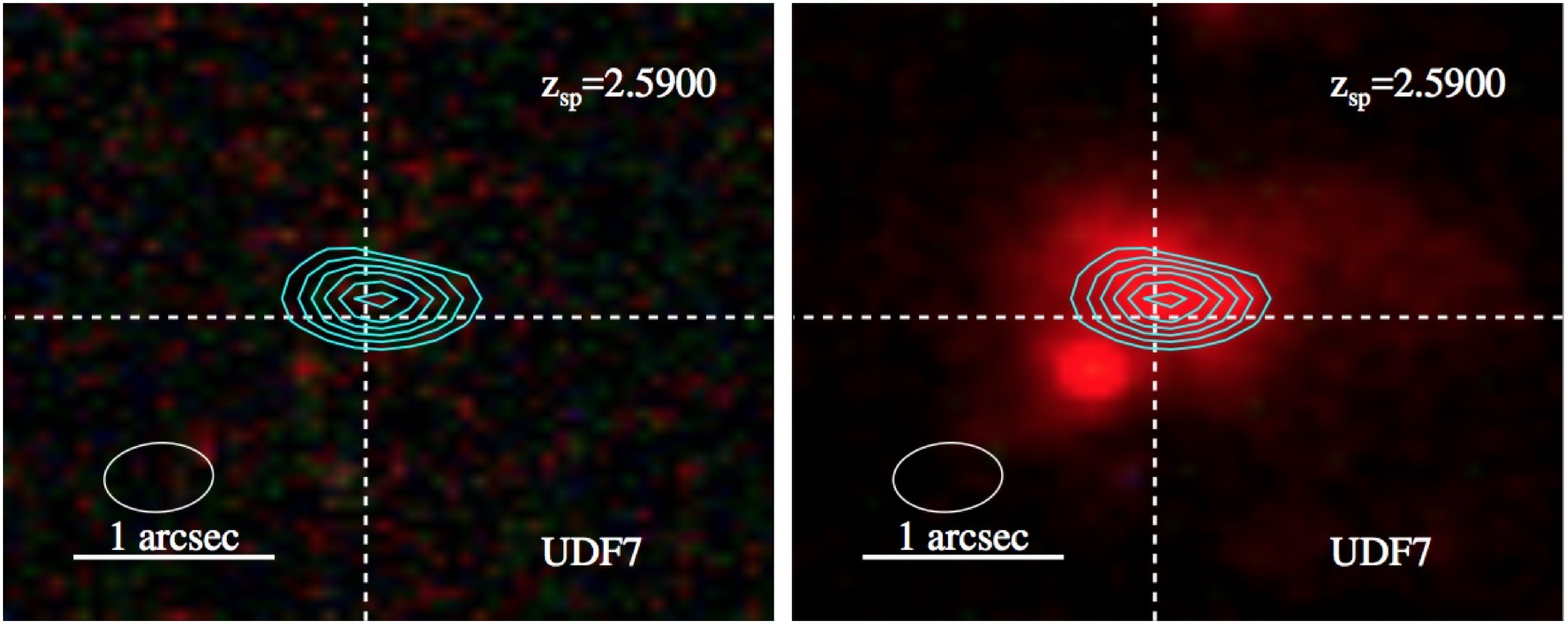}
\includegraphics*[width=0.45\textwidth]{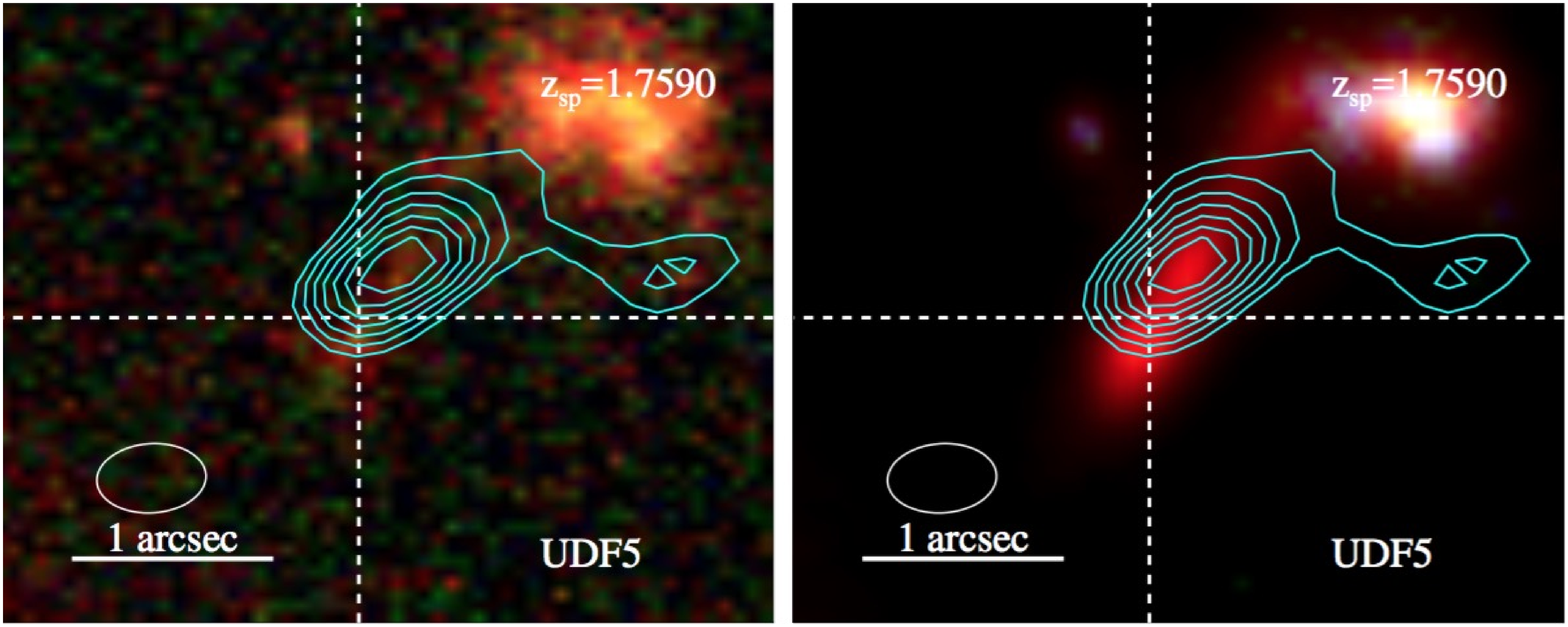}
\includegraphics*[width=0.45\textwidth]{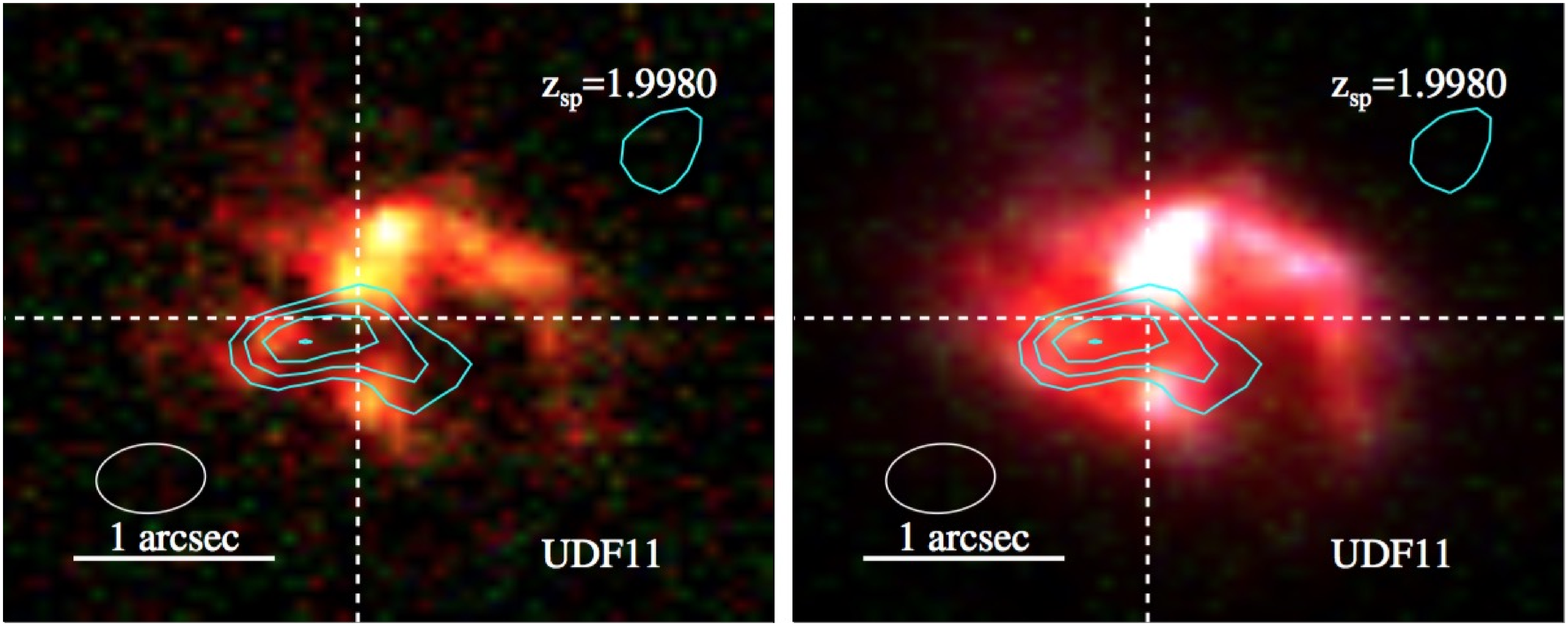}
\includegraphics*[width=0.45\textwidth]{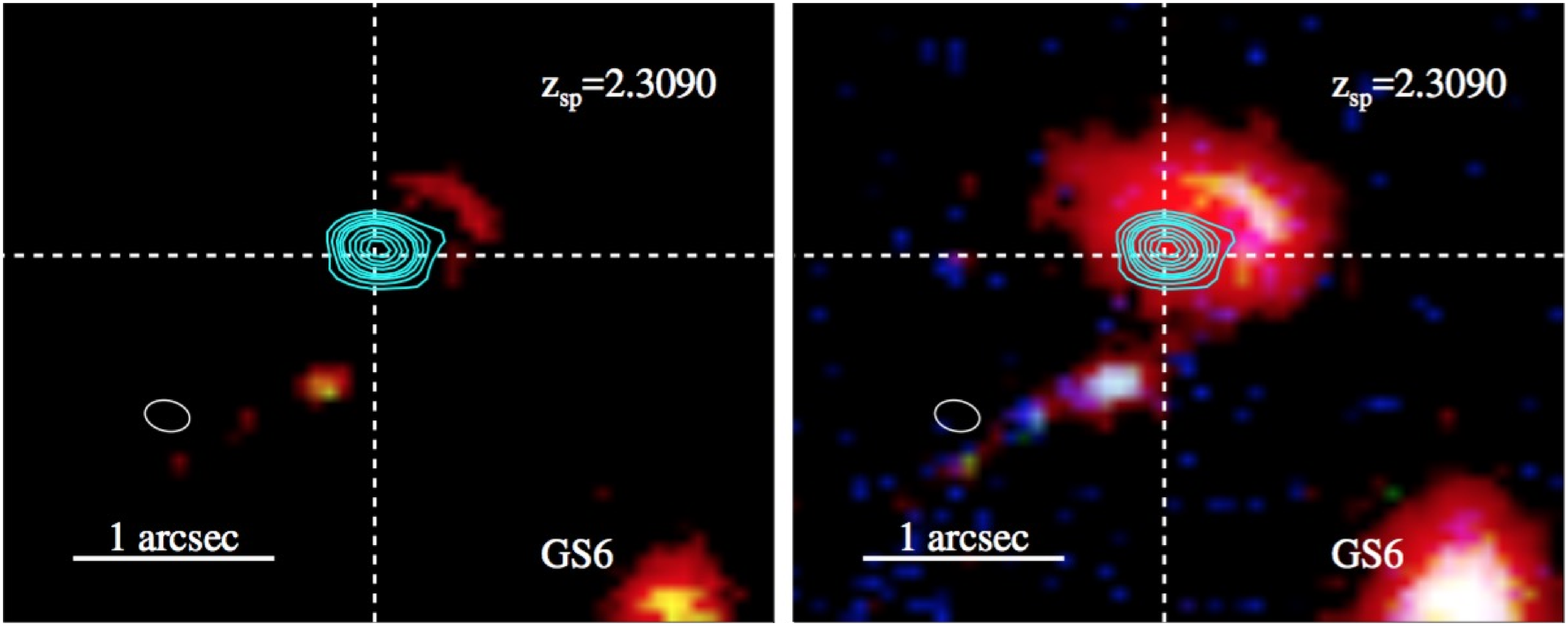}
\includegraphics*[width=0.45\textwidth]{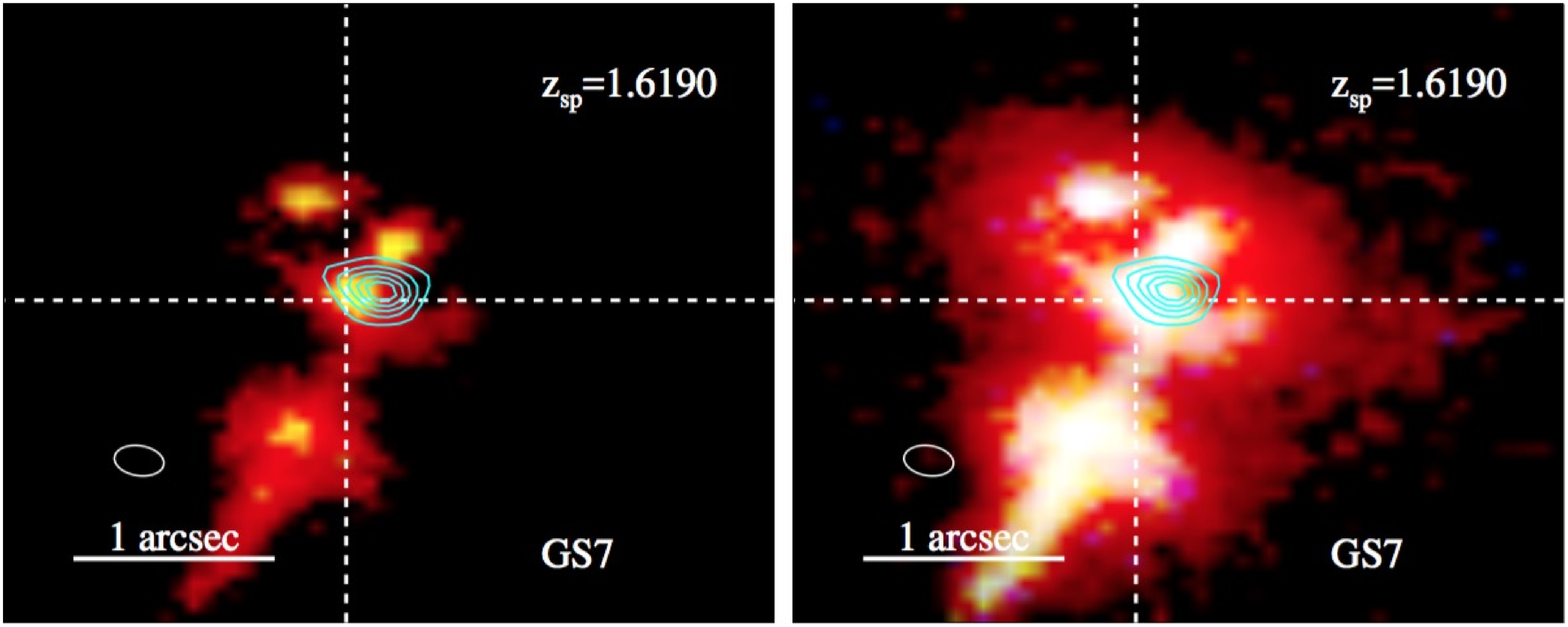}
\includegraphics*[width=0.45\textwidth]{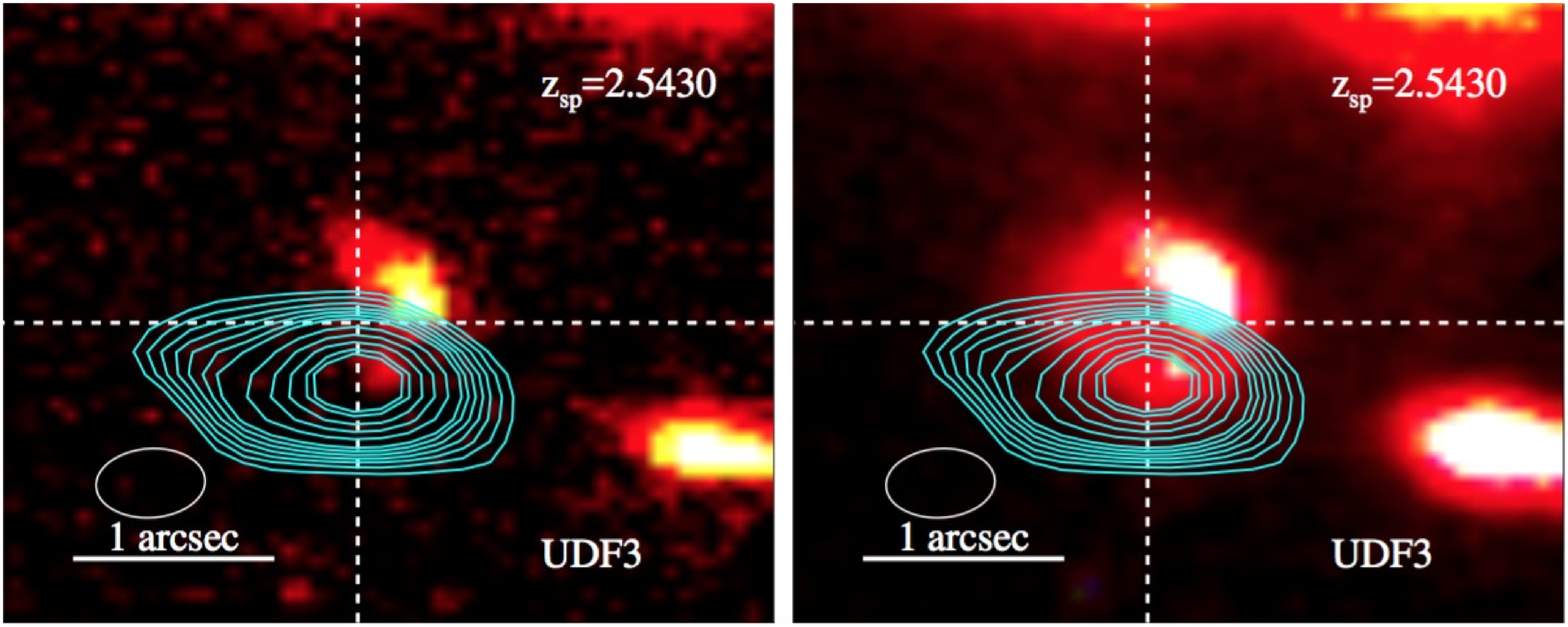}
\includegraphics*[width=0.45\textwidth]{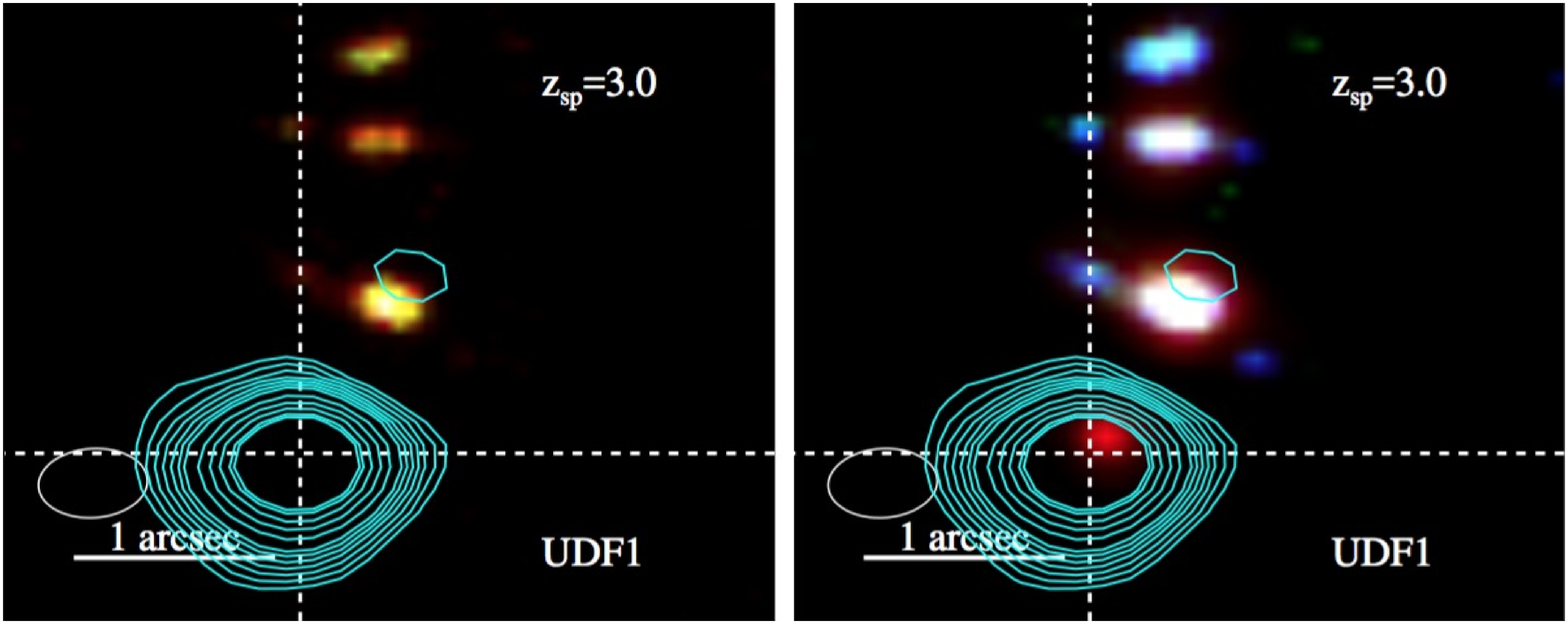}
\caption{Same as Fig.~\ref{FIG:IM_MS} for the "starburst" galaxies with R$_{\rm SB}$$>$3 sorted by increasing R$_{\rm SB}$ from top to bottom. Contours as in Fig.~\ref{FIG:IM_MS}.}
\label{FIG:IM_SB}
\end{figure}

In the following, we will use a single parameter to quantify this distance to the MS called the "starburstiness" as in \cite{elbaz11}, i.e., R$_{\rm SB}$=SFR/SFR$_{\rm MS}$. Out of the present list of ALMA targets, only 31\,\% (or 15\,\%) may be considered as "starbursts" (SB) defined as galaxies with a starburstiness R$_{\rm SB}$$>$3 (or $>$4; see Table~\ref{TAB:sample}). The remaining 69\,\% (or 85\,\%) consist of galaxies located in the upper part of the MS or slightly above the 68\,\% $rms$ of 0.3 dex of the MS. In the following, we will call MS and SB the star-forming galaxies with R$_{\rm SB}$$\leq$3 and $>$3 respectively, but we will study galaxy properties as a function of starburstiness more generally. The postage stamp images of the MS and SB galaxies (\textit{HST} images with ALMA contours) are shown in Fig.~\ref{FIG:IM_MS} and Fig.~\ref{FIG:IM_SB} respectively (except for GS8 discussed in Section~\ref{SEC:dark} for which the \textit{HST} and ALMA images correspond to two different galaxies).
\section{Results}
\label{SEC:results}
\subsection{Serendipitous detection of an "HST--dark" galaxy at $z$$\sim$3}
\label{SEC:dark}
\begin{figure}
\centering
\includegraphics*[width=0.48\textwidth]{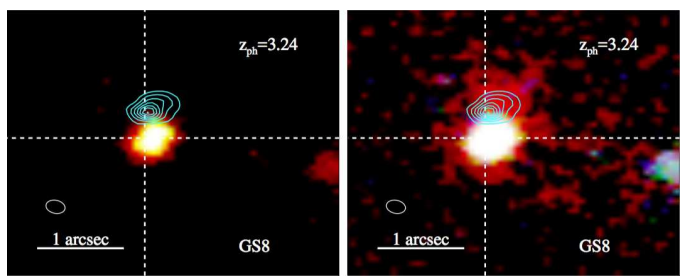}
\caption{ALMA contours overlaid on the \textit{HST} image of the galaxy GS8 (Same as Fig.~\ref{FIG:IM_MS}).}
\label{FIG:GS8}
\end{figure}

In one out of the 8 GS galaxies, GS8, we found an offset between the ALMA and $H$-band centroids that we attribute to a projection effect, the ALMA source being associated to a background  source. The red part of Fig.~\ref{FIG:GS8} showing the $H$-band contribution to the $VIH$ image shows a clear extension to the North. The offset of 0.35\arcsec between the UV and ALMA centroids is similar to those observed for some of the other galaxies studied here. However, there are three reasons why we believe that the ALMA emission is associated to another galaxy in the case of GS8. 

First, there is this offset of 0.35\arcsec between the $H$-band and the ALMA centroids, and not only with the UV. Second, the photometric redshift of the foreground galaxy associated with the UV image is $z_{\rm phot}$=1.101 whereas the far-infrared SED combining the \textit{Herschel} and ALMA photometric points peaks at 350\,$\mu$m. If the far-IR emission were associated with this galaxy, it would peak at a rest-frame $\lambda$= 167\,$\mu$m as opposed to the typical galaxies at this redshift which peak around $\lambda$= 100\,$\mu$m. Third, this galaxy (CANDELS ID = 5893) has an estimated stellar mass of M$_{\star}$=4.6$\times$10$^9$ M$_{\odot}$. At a $z_{\rm phot}$=1.101, this galaxy would have an extreme starburstiness of R$_{\rm SB}$$=$60 and if the whole far-IR emission was to be attributed to this galaxy, it would lead to a dust mass of M$_{\rm dust}$=7.9$\times$10$^9$ M$_{\odot}$, i.e., 1.7$\times$M$_{\star}$, and a gas mass of M$_{\rm gas}$=2.5$\times$10$^{12}$ M$_{\odot}$. Considering these unrealistic dust temperatures and masses, together with the spatial distribution of the ALMA and $H$-band light, we believe that the \textit{Herschel} and ALMA emission arise primarily from a background galaxy. Since the foreground galaxy has the CANDELS ID 5893, we decided to call the background galaxy 5893b. 

  \begin{figure}[!h]
  \centering
   \includegraphics[width=8.7cm]{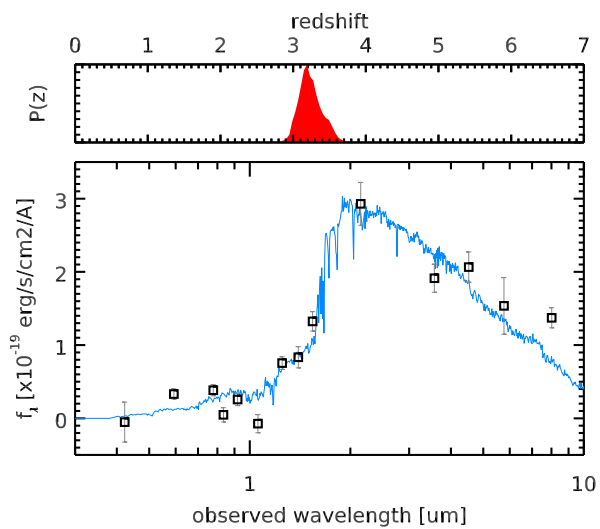}
  \caption{Spectral energy distribution of the offset source GS8 (5893b) measured from aperture photometry at the location of the ALMA source. The photometric redshift probability distribution peaks at $z$=3.24.} 
  \label{FIG:5893}
  \end{figure}
In order to determine the photometric redshift of this galaxy, we modelled the light profile of all the other surrounding galaxies within an $8\arcsec$ radius as single S\'ersic profiles using \texttt{imfit} \citep{erwin15} on the HawK-I $Ks$-band image (PSF FWHM=0.4 arcsec, \citealt{fontana14}). We then used the results of this modelling to measure the photometry on all the {\it HST} images from F435W to F160W, as well as in the {\it Spitzer} IRAC images from $3.6$ to $8\,\mu{\rm m}$. We convolved the S\'ersic profiles with the point-spread function of the corresponding image, and only varied the total flux of each galaxy to minimise the $\chi^2$ of the residuals. The resulting photometric measurements for ID5893b are shown in Fig.~\ref{FIG:5893}. After this process, ID5893b was only clearly detected in bands $J$, $H$ from {\it HST}--WFC3, $Ks$ from HawK-I, and all the {\it Spitzer} IRAC bands where it clearly dominates over ID5893.

The photometry obtained for GS8 (ID5893b) was then used to determine a photometric redshift with the program EAZY and a stellar mass with the program FAST. We found $z$$_{\rm phot}$=3.24$\pm$0.20 and M$_{\star}$=3$\times$ 10$^{11}$ M$_{\odot}$. This photometric redshift is consistent with the observed peak of the far-IR SED of ID5893b located at $\lambda_{\rm peak}$$\sim$ 350\,$\mu$m. 

The resulting starburstiness of GS8, R$_{\rm SB}$$=$1.49, corresponds to a typical MS galaxy at $z$$\sim$3.24. The fit of the far-IR SED using the Draine \& Li model gives a dust mass of M$_{\rm dust}$=2.3$\times$10$^9$ M$_{\odot}$, which is lower than the value estimated for the $z$=1.101 redshift because the dust mass is very sensitive to the dust temperature, which is much higher here since the peak emission is now located close to 85\,$\mu$m. The gas fraction, $f_{\rm gas}$=M$_{\rm gas}$/(M$_{\rm gas}$+M$_{\star}$), is also reasonable (as opposed to the low redshift option) since it now reaches a value of 40\,\%. 

To conclude, ALMA allowed us to identify a distant counterpart to a previously detected \textit{Herschel} source that was not present in the CANDELS \textit{HST} catalog \citep{guo13}. If one considers the \textit{H}-band extension that we analyzed above, then GS8 is not strictly speaking an \textit{HST}-dark galaxy, but without ALMA it would have remained \textit{HST}-dark. We can only extrapolate the implications of this finding on the GS sample of 8 galaxies because the UDF galaxies were selected in a blank field that would require an analysis of the existence of \textit{HST}-dark sources over the whole field. Extrapolating from our small sample, one may expect 10-15\,\% of the ALMA detections to be associated with an optically dark galaxy. This statement will be studied on a firmer statistical ground in a forthcoming paper discussing a 6.7 $\times$10 arcmin$^2$ extragalactic survey in GOODS-\textit{South} with ALMA at 1.1mm (PI D.Elbaz, Franco et al. in prep.).
\subsection{Compact star-formation in $z$$\sim$2 galaxies}
\label{SEC:compactness}
The ALMA images probe the dust continuum emission at typical rest-frame wavelengths of $\lambda^{\rm rest}_{\rm GS}$=260\,$\mu$m (870\,$\mu$m observed from $\bar{z}_{\rm GS}$=2.3) and $\lambda^{\rm rest}_{\rm UDF}$=380\,$\mu$m (1.3mm observed from $\bar{z}_{\rm UDF}$=2.43) for the GS and UDF galaxies respectively. We will consider that these two wavelengths are close enough to probe the same physical origin. We will assume that the median 325\,$\mu$m wavelength for the whole sample probes the location of the dust heated by the newly formed young stars and that it can therefore be used to trace the geometry of the star-formation regions.

\begin{figure}
 \centering
      \includegraphics[width=8.7cm]{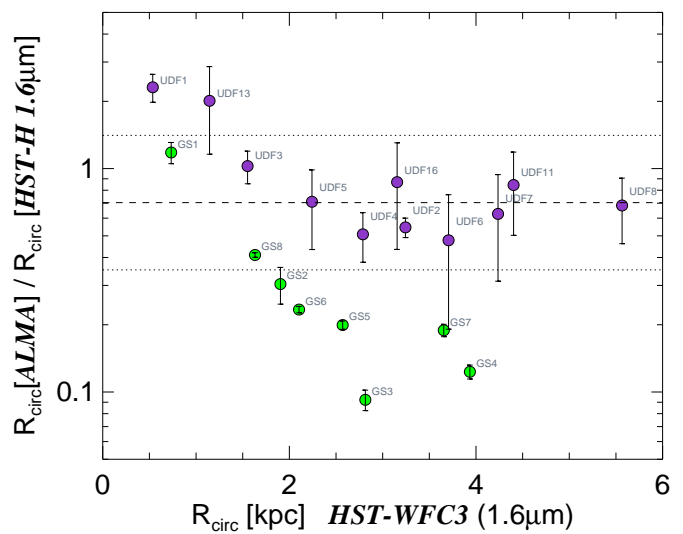}     
      \caption{ALMA circularized effective radii as a function of HST--WFC3 $H$-band effective radii from \cite{vanderwel12}.}
     \label{FIG:size}
  \end{figure}

The first remarkable result that comes out of the resolved dusty star-formation maps obtained with the high angular resolution mode of ALMA is their compactness (see Fig.~\ref{FIG:IM_MS} and Fig.~\ref{FIG:IM_SB}). 
The optical sizes measured by \cite{vanderwel12} using a S\'ersic profile fitting of the \textit{HST}-\textit{H} band images are compared to the ALMA sizes, computed using 2D Gaussian profiles, in Fig.~\ref{FIG:size}. Both are circularized as in Eq.~\ref{EQ:conv}. As discussed in Section~\ref{SEC:size}, the Gaussian and S\'ersic fits to the ALMA data provide similar sizes. 

Over the whole sample of 19 $z$$\sim$2 star-forming galaxies resolved with ALMA, we find that the ALMA sizes are systematically smaller than the rest-frame $V$-band sizes. Similar results have been systematically found by several different authors using galaxy samples selected with different strategies (e.g., \citealt{simpson15}, \citealt{hodge16}, \citealt{barro16}, \citealt{rujopakarn16}, \citealt{fujimoto17}). 
Using a compilation of ALMA observations with typical angular resolutions of $\sim$0.6 arcsec (as compared to 0.2 arcsec here), \cite{fujimoto17} measured a factor of $R_{H}^{circ}/R_{ALMA}^{circ}$$\sim$1.4 (see their Fig.12) that we have represented with a dashed line in Fig.~\ref{FIG:size}. We can see that most of our galaxies fall within a factor two around this ratio (dotted lines in Fig.~\ref{FIG:size}) except a sub-population of compact sources that we will discuss in more detail in the following.

We note however an important caveat related to our sample selection. The condition that we imposed on the GS galaxies resulted in selecting exclusively massive star-forming galaxies with a median M$_\star$$\sim$1.4$\times$10$^{11}$ M$_{\odot}$. The fact that the GS galaxies exhibit systematically smaller sizes than the UDF galaxies may be a consequence of their larger stellar masses if massive galaxies turned out to exhibit particularly compact star-formation distributions. As we will show in the next sections, massive galaxies do turn out to exhibit particularly compact star-formation as also found by \cite{barro16} and as one would expect if they were candidate progenitors of the population of compact ellipticals at $z$$\sim$2 \citep{vanderwel14}.

As discussed in Section~\ref{SEC:size} (see also \citealt{rujopakarn16}), the S/N ratio on the UDF ALMA sources in not high-enough to provide a robust S\'ersic profile fitting and derive a S\'ersic index. But for the higher-quality of the GS galaxies, we find that the ALMA profiles can be fitted by a S\'ersic profile with a median S\'ersic index is $n$=1.27$\pm$0.48, hence close to an exponential disk. The dusty star formation regions therefore seem to be disk-like, confirming what was previously found by \cite{hodge16}.

\subsection{An ALMA view on kpc clumps of star formation}
\label{SEC:clumps}
The discovery of giant star-forming regions in the high-redshift population of so-called "chain galaxies" and "clump-cluster galaxies" revealed by the first generation of deep \textit{HST} images (\citealt{cowie95}, \citealt{elmegreen05}) started a still-ongoing debate on their role in the stellar mass growth and morphological transformation of galaxies throughout cosmic time. Expected to form as a result of dynamical instabilities in high-redshift gas-rich galaxies, those kpc-size $\sim$10$^8$ M$_{\odot}$ clumps of star formation could lead to the formation of the central bulge of galaxies if they lived long enough to survive their migration from the peripheries to the centers of galaxies \citep{elmegreen08}. 

How much of the integrated stellar mass growth of a galaxy comes from these kpc clumps remains uncertain. 
Since $z$$\sim$2 galaxies with strong SFR 
systematically radiate most of their energy from star formation in the far-infrared/submm, it is only by resolving these galaxies at these long wavelengths that one will be able to determine the role of kpc-clumps. If kpc-large clumps of star formation were responsible for a large fraction of the resolved far-IR emission of galaxies, this would imply that the physical mechanism responsible for their formation plays an important role in shaping present-day galaxies.

In a recent paper, \cite{hodge16} studied the possible existence of kpc-clumps of star formation with an ALMA follow-up of a sample of 16 $z$$\sim$2.5 SMGs with S$_{870\,\mu m}$=3.4--9 mJy and $L_{\rm IR}$$\sim$4$\times$10$^{12}$ L$_{\odot}$. 
They searched for point-like sources that could be associated with kpc-clumps using a synthesized beam of 0.17\arcsec$\times$0.15\arcsec FWHM, corresponding to a physical size of 1.3 kpc at the median redshift of $z$$\sim$2.5 (the analysis was also performed at a resolution of 0.12\arcsec corresponding to 1 kpc). While marginal evidence was found for residual emission that could be associated to the kpc clumps, the authors generated some simulated ALMA images of mock galaxies with smooth profiles without any clumps and found that the analysis of the resulting mock ALMA images showed similar signatures of kpc clumps with low significance. Hence they concluded that "while there may be a hint of clump-like dust emission in the current 870\,$\mu$m data on kiloparsec scales, higher signal-to-noise observations at higher spatial resolution are required to confirm whether these clumpy structures are indeed real". 

\cite{cibinel17} used ALMA to spatially resolve the CO(5--4) transition -- which probes dense star formation -- in the $z$=1.57 clumpy galaxy UDF6462. In this galaxy, the UV clumps make individually between 10 and 40\,\% of the total UV SFR. Using the observed L'$_{\rm CO(5-4)}$--\LIR\, correlation \citep{daddi15}, they find that none of the six clumps produces more than 10\,\% of the dusty SFR (upper limit of $\sim$5\,\My for a total SFR =56\,\My). The limit goes down to less than 18\,\% for the combined contribution of the clumps after stacking all 6 clumps. If this conclusion may be generalised, it would imply that the giant clumps observed in the UV are not major contributors to the bulk of the stellar mass growth of $z$$\sim$2 galaxies.

We designed our ALMA exposures for the 8 "GS" sources (see Section~\ref{SEC:observations}) to detect individual clumps of star formation at 870\,$\mu$m assuming that a clump could be responsible for 20\,\% of the total SFR of a galaxy. In all 8 galaxies, we find that the ALMA continuum emission is concentrated in a nuclear region, with no evidence for external clump contributions, similarly to what was found by \cite{cibinel17} and \cite{hodge16}. 

Three galaxies among the closest sources of the UDF sample --  UDF6, UDF8 and UDF16 at $z_{\rm spec}$=1.413, 1.546, 1.319 respectively -- present the shape of grand design spirals with a total extent in the rest-frame 6400\,\AA\,of $\sim$20--30 kpc (observed WFC3 $H$-band; see bottom right galaxies in Fig.~\ref{FIG:IM_MS}). 
UDF16 is a face-on spiral that shows no evidence for kpc size UV clumps outside its central UV nucleus whereas UDF6 and UDF8 present clear kpc-size clumps in the rest-frame UV light distribution.

UDF8 presents three clumps in the N-W side and UDF6 one clump in the N-E side. All four UV clumps have a total extent of 0.25\arcsec, i.e., 2 kpc at $z$$\sim$1.5, hence a radius of about 1 kpc. We note that both galaxies are close to the median SFR of the MS  with a starburstiness of R$_{\rm SB}$=1.5 and 2.9 respectively and experience a similar SFR$\sim$150 M$_\odot$ yr$^{-1}$. UDF8 presents the largest number of UV clumps, it is the closest galaxy to the median SFR of the MS. 

At the typical redshift of $z$$\sim$1.5 of these galaxies, the ALMA images of the UDF at 1.3mm probe the rest-frame 520\,$\mu$m emission in the rest-frame. This emission is found to peak on the center of the WFC3-$H$ band images in both galaxies (crosses in Fig.~\ref{FIG:IM_MS}) whereas the UV clumps present a systematic offset. This offset is much larger than the astrometric uncertainty (note the perfect agreement between ALMA and \textit{HST}) and than the ALMA PSF FWHM (ellipse on the lower-left corner). It implies that there is a clear dichotomy in these two galaxies between the young massive stars inhabiting highly attenuated dust clouds and those responsible for the bulk of the UV light. The UV images of UDF6 and UDF8 (left images in Fig.~\ref{FIG:IM_MS}) show that no UV light is detected at the location of the peak ALMA emission, which implies that the UV-slope is not a good proxy for the amount of dust extinction here. 

In both galaxies, the ALMA contours do present an extension suggesting some marginal contribution of the kpc clumps to the far-infrared emission.
The rest-frame 520\,$\mu$m contours of UDF6 present a second, less pronounced, peak centered on one of the kpc clump and the same is seen, slightly less pronounced, in UDF8 with a contribution from the southern kpc clump. However due to the low S/N of the 1.3mm images of these sources, this evidence is provided only by the first and second ALMA contours at the 80 and 100\,$\mu$Jy levels, i.e., only at the 2--$\sigma$ confidence level. As shown by the simulations of \cite{hodge16} discussed above, deeper ALMA integrations would be needed to confirm the detection of the giant clumps in the far-infrared. 

Assuming that the dusty star-formation is indeed not spread in a series of clumps, then we will consider that it is well characterized by its IR luminosity surface density, $\Sigma_{\rm IR}$ (Eq.~\ref{EQ:Sir}). 
\subsection{Star formation compactness and $IR8$ color index}
\label{SEC:IR8}
 \begin{figure}[!]
 \centering
      \includegraphics[width=8.7cm]{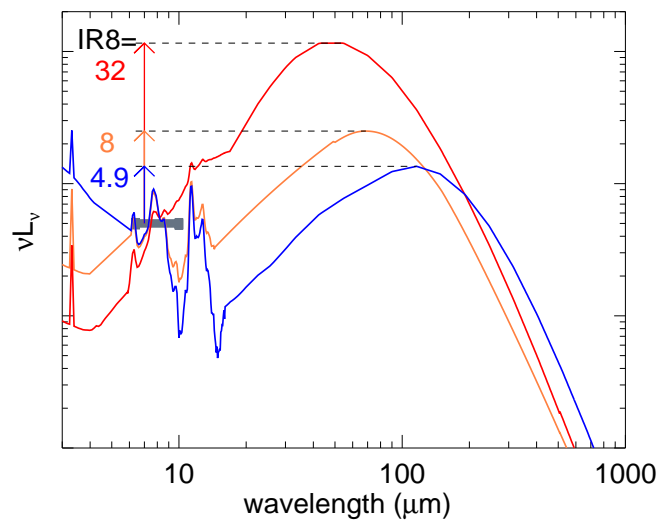}    
      \caption{Three SEDs from the \cite{chary01} library normalized to the same L$_{8\mu \rm m}$ (using the IRAC passband centered on 8\,$\mu$m). We present the SEDs 11, 58 and 95 for which $IR8$=L$_{\rm IR}$/L$_{8\mu \rm m}$= 4.9, 8 and 32 respectively. The horizontal black solid segment shows the width of the IRAC2 channel centered on 8\,$\mu$m.}
     \label{FIG:IR8}
  \end{figure}

The IR luminosity density, $\Sigma_{\rm IR}$ -- a proxy for the dusty SFR density -- has been found to correlate with an integrated property of local $z$$\sim$0 galaxies, the $IR8$ color index (see Fig.13 of \citealt{elbaz11}). Compact star-forming galaxies were found to exhibit large values of the $IR8$=L$_{\rm IR}$/L$_{8\mu \rm m}$ color index whereas galaxies with extended star-formation presented a "normal, universal" $IR8$ ratio of $IR8$$\sim$4.9. Compact and extended star-forming galaxies were distinguished as galaxies with $\Sigma_{\rm IR}$ above and below a critical density of $\Sigma_{\rm IR}^{\rm crit}$ = 3$\times$10$^{10}$ L$_{\odot}$ kpc$^{-2}$ respectively. In this Section, we present a revised version of the local $IR8$ -- $\Sigma_{\rm IR}$ relation and discuss its application to the distant Universe.

The $IR8$ color index measures the ratio of the continuum emission -- as measured by the total mid to far-IR luminosity L$_{\rm IR}$, dominantly due to the emission of big dust grains with a peak emission around 100\,$\mu$m -- over the emission due to the combination of continuum emission from very small grains (VSGs) of dust and broadband features commonly attributed to polycyclic aromatic hydrocarbons (PAHs; \citealt{leger84}, \citealt{allamandola85}) -- as measured by the 8\,$\mu$m \textit{Spitzer}-IRAC broadband filter. 

In local galaxies, strong $IR8$ values (red curve in Fig.~\ref{FIG:IR8}) are systematically associated with compact star formation taking place in merger-driven starbursts.
Instead, normal spiral galaxies exhibit strong PAH equivalent widths and a weaker contribution of warm and cold dust continuum in the mid-IR and far-IR (blue curve in Fig.~\ref{FIG:IR8}). The three template IR SEDs presented in Fig.~\ref{FIG:IR8} come from the \cite{chary01} library and present $IR8$ values of 4.9, 8 and 32 respectively. Therefore in local galaxies, $IR8$ not only correlates with star-formation compactness, as probed by $\Sigma_{\rm IR}$, but also with starburstiness, R$_{\rm SB}$ (see Fig.17 in \citealt{elbaz11}). 

In the following, we discuss the physical origin of the $IR8$ --  $\Sigma_{\rm IR}$ relation that may apply to both local and distant galaxies. The interest of this relation is twofold. On one hand, it provides an empirical method to connect integrated and resolved galaxy properties that may be used to derive one from the other. On the other hand, if the relation holds in distant galaxies, $IR8$ and $\Sigma_{\rm IR}$ may both be used to separate compact and extended star formation, and may therefore serve to unveil the role of mergers.

 \begin{figure}
 \centering
     \includegraphics[width=8.7cm]{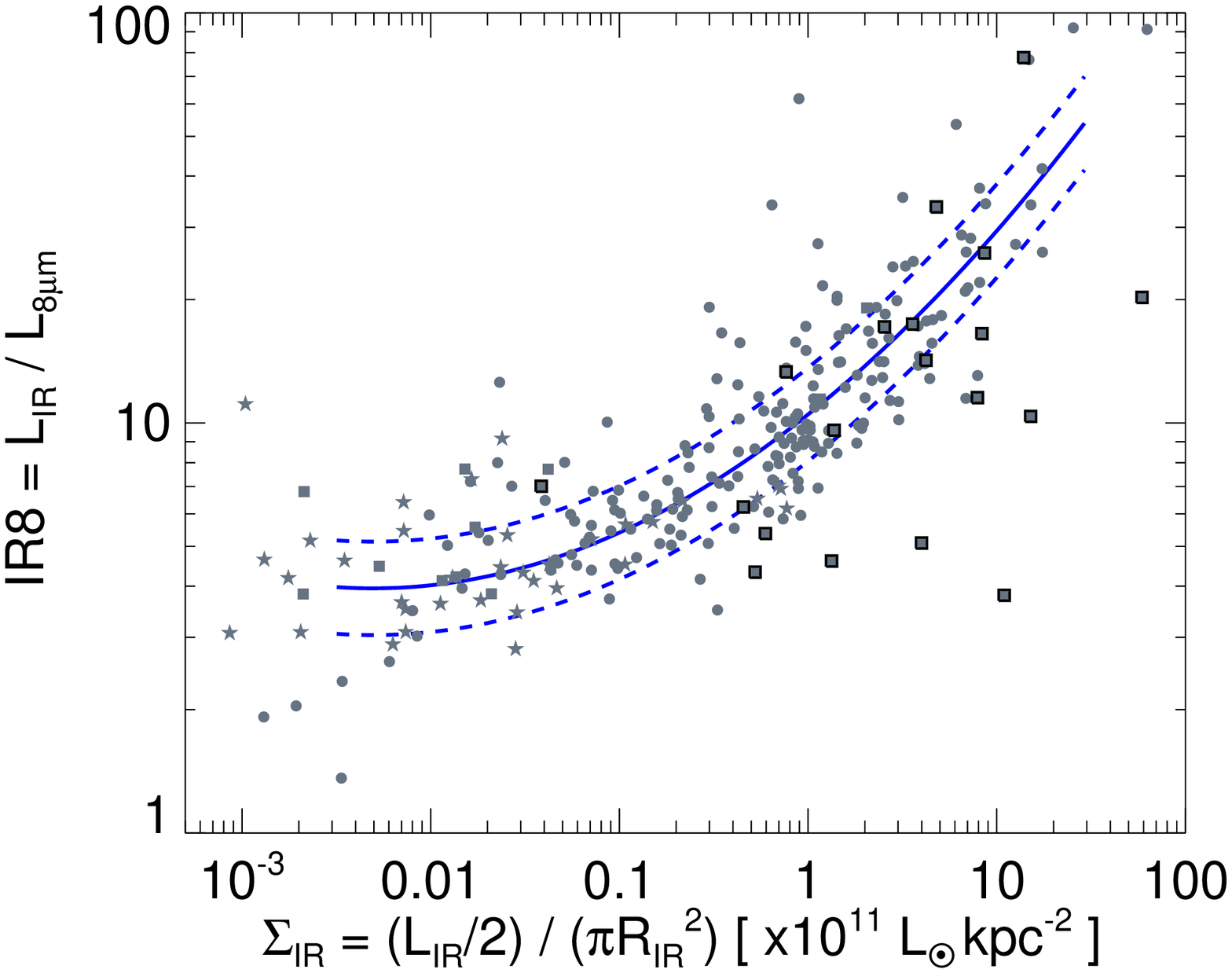}  
     \includegraphics[width=8.7cm]{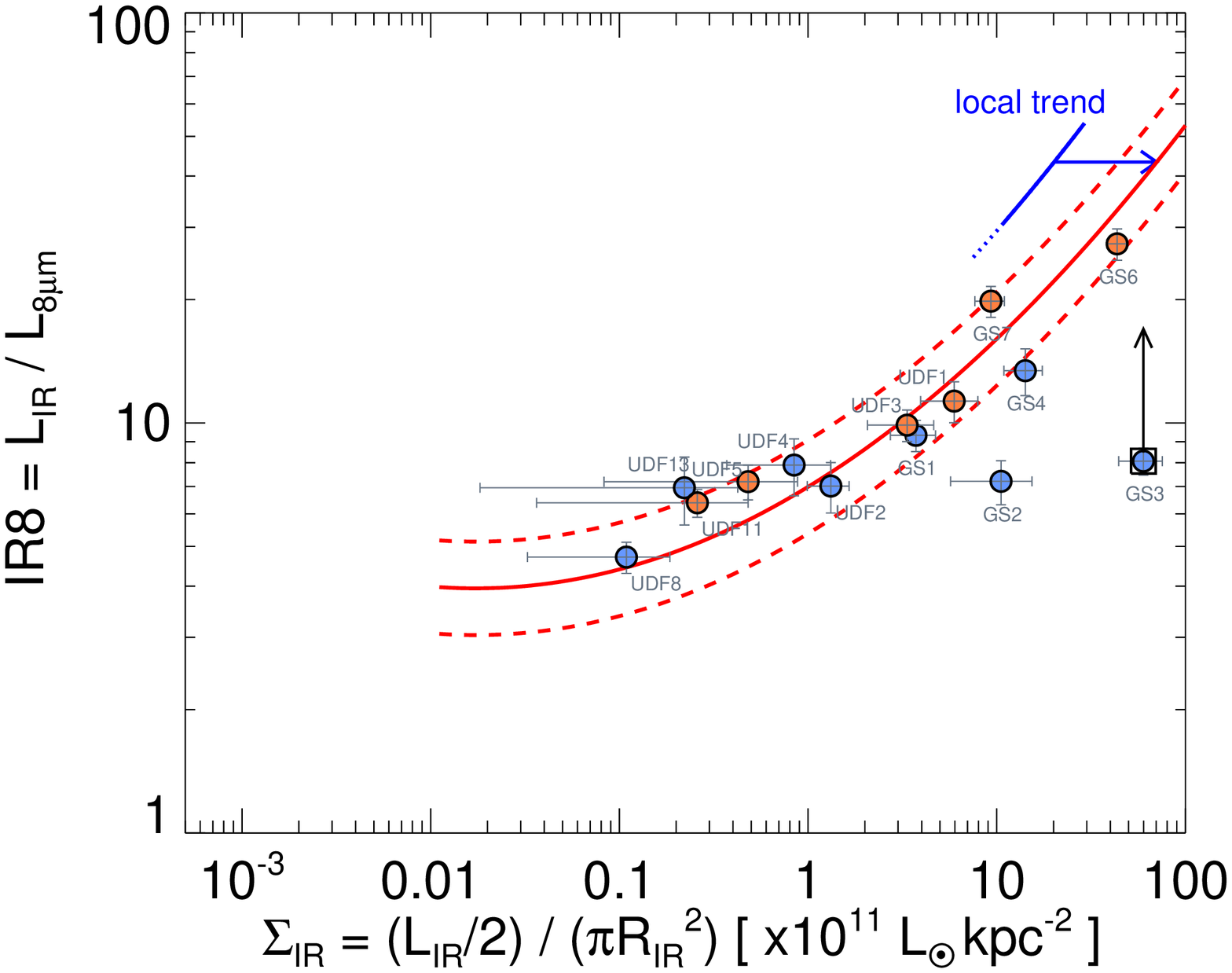}
      \caption{\textbf{\textit{Top:}} $IR8$ -- $\Sigma_{\rm IR}$ relation for local galaxies. Grey filled dots: 293 galaxies with \textit{Herschel} PACS 70\,$\mu$m sizes (R$_{\rm IR}$=R$_{70\,\mu\rm m}$) from \cite{diazsantos17}. Filled squares and stars: 11 and 39 galaxies with radio and mid-IR sizes from \cite{elbaz11}. Galaxies with mid-IR spectral signatures of AGNs (identified by \citealt{diazsantos17}) are marked with bold squares. The solid blue line is a polynomial fit to the sliding median described by Eq.~\ref{EQ:ir8_sir}. The two dashed lines 30\,\% above and below ($\pm$0.11 $dex$) encompass 68\,\% of the galaxies. \textbf{\textit{Bottom:}} position of the $z$$\sim$2 ALMA sources on the $IR8$ -- $\Sigma_{\rm IR}$ diagram. The solid and dashed red lines show the local relation shifted to the right by a factor 3.5 to account for the combined effects of the rise of the MS SFR with redshift and decrease of galaxy sizes with redshift. Blue and orange filled dots separate galaxies on or close to the MS (R$_{\rm SB}$$\leq$3) from starbursts (R$_{\rm SB}$$>$3). Only galaxies within 1.5$\leq$$z$$\leq$2.5 are shown here because of a large and uncertain extrapolation of $L_8$ from the observed 24\,$\mu$m flux density for galaxies outside this redshift range. The black square on GS3 indicates that it is a power-law for which the contribution of the AGN has been corrected in a conservative way.}
     \label{FIG:ir8_sir}
  \end{figure}
\subsubsection{Revised version of the local $IR8$ -- $\Sigma_{\rm IR}$ relation and discussion of its physical origin}
\label{SEC:localir8}

To produce their local $IR8$ -- $\Sigma_{\rm IR}$ relation, \cite{elbaz11} used IR sizes for local galaxies coming from a combination of radio data (converted to FIR sizes using a correlation observed between FIR and radio sizes) and mid-IR sizes from \textit{Spitzer}-IRS (from \citealt{diazsantos10}). 

Here we use sizes directly measured in the far-IR coming from an \textit{Herschel}-PACS 70\,$\mu$m follow-up of local LIRGs from the Great Observatories All-sky LIRG Survey (GOALS; \citealt{armus09}). Sizes were measured for a total 293 galaxies including 42 normal galaxies (10 $\leq$ log(L$_{\rm IR}$/L$_{\odot}$) $<$ 11), 175 LIRGs (11 $\leq$ log(L$_{\rm IR}$/L$_{\odot}$) $<$ 12) and 22 ULIRGs (12 $\leq$ log(L$_{\rm IR}$/L$_{\odot}$)). The sizes measured for these galaxies (as described in \citealt{diazsantos17}) were used to produce the revised version of the local $IR8$ -- $\Sigma_{\rm IR}$ relation presented in Fig.~\ref{FIG:ir8_sir}-top. We also included 11 galaxies with radio sizes (filled squares) and 39 galaxies with mid-IR sizes from AKARI (filled stars) that were originally used in \cite{elbaz11}. 

This revised $IR8$ -- $\Sigma_{\rm IR}$ relation for local galaxies is amazingly tight with a 68\,\% median absolute deviation (MAD) of only 0.11 $dex$. The polynomial fit to the sliding median (Eq.~\ref{EQ:ir8_sir}) and its MAD are represented by the solid and dashed lines in Fig.~\ref{FIG:ir8_sir}.

\begin{equation}
\begin{array}{c}
\log_{10}\left(IR8\right) =  \left(1.02\pm0.11\right)+0.37 \times \log_{10}\left(\Sigma_{\rm IR} / 10^{11}\right) \\
+  0.08 \times \left[\log_{10}\left(\Sigma_{\rm IR} / 10^{11}\right)\right]^2
\end{array}
\label{EQ:ir8_sir}
\end{equation}

We find that the trend is relatively flat up to $\Sigma_{\rm IR}^{\rm crit}$ $\sim$ 3$\times$10$^{10}$ L$_{\odot}$ kpc$^{-2}$ and then rises steeply. Although continuous, this improved relation confirms the existence of two regimes of star formation. Galaxies with extended star formation ($\Sigma_{\rm IR}$ $<$ 3$\times$10$^{10}$ L$_{\odot}$ kpc$^{-2}$) have $IR8$=4--5, whereas compact star-forming galaxies reach $IR8$ values well above 10. 

We wish to emphasize that most of the outliers below the relation are galaxies identified as power-law AGNs by \cite{diazsantos17} from their mid-IR spectra (thick black squares in Fig.~\ref{FIG:ir8_sir}-top). This trend may either be explained by \textit{(i)} a contribution to the 8\,$\mu$m emission by warm dust heated by AGN, and/or \textit{(ii)} a physical connection between star-formation compactness and AGN activity.

The physical origin of this tight correlation can now be better understood thanks to the analysis of \cite{diazsantos17} who present a similarly tight relation between L$_{\rm IR}$/L$_{\rm[CII]158\,\mu m}$ and $\Sigma_{\rm IR}$. The [CII]158\,$\mu$m emission comes from the same region as the PAHs that contribute to the 8\,$\mu$m luminosity, namely the PDR or photo-dissociation region. In both cases, there is a nearly flat $IR8$ and L$_{\rm IR}$/L$_{\rm[CII]158\,\mu m}$ ratio with increasing $\Sigma_{\rm IR}$ up to the same critical density, $\Sigma_{\rm IR}^{\rm crit}$ $\sim$ 3--5$\times$10$^{10}$ L$_{\odot}$ kpc$^{-2}$ (see their Fig.2-middle right, which shows L$_{\rm[CII]158\,\mu m}$/L$_{\rm IR}$ analog to 1/$IR8$ here). Then $IR8$ and the L$_{\rm IR}$/L$_{\rm[CII]158\,\mu m}$ ratio both increase with increasing $\Sigma_{\rm IR}$. This behavior may be understood with the concept of "dust-bounded" star-formation nebulae described in \cite{abel09}. The idea proposed in this paper and discussed in \cite{diazsantos17} involves the role of dust that absorbs part of the ionizing radiation and therefore prevents it from reaching the PDR region. This leads to a rise of the dust temperature and consequently also of the total IR luminosity emitted, proportional to $T_{\rm dust}^{4+\beta}$ (where $\beta$ is the emissivity of the dust), and a drop of the emission in [CII] and PAHs from the PDR. In the local Universe, merger-driven starbursts are systematically associated with such young and compact star-formation regions that may be considered as dust-bounded, with the consequence that $IR8$ and L$_{\rm IR}$/L$_{\rm[CII]158\,\mu m}$ rise during the merger-driven starburst. 
\cite{diazsantos17} find that the ratio of the intensity of the interstellar radiation field, \textit{G}, over the average PDR hydrogen density, $n_{\rm H}$, $G/n_{\rm H}$, remains constant below $\Sigma_{\rm IR}^{\rm crit}$ and increases rapidly with $\Sigma_{\rm IR}$ above $\Sigma_{\rm IR}^{\rm crit}$.

\subsubsection{The $IR8$ -- $\Sigma_{\rm IR}$ relation for $z$$\sim$2 ALMA galaxies}
\label{SEC:distantir8}
The positions of the $z$$\sim$2 ALMA galaxies in the $IR8$ -- $\Sigma_{\rm IR}$ plan are shown in Fig.~\ref{FIG:ir8_sir}-bottom. $IR8$ values were only computed for galaxies with 1.5$\leq$$z$$\leq$2.5 in order to include the dominant 7.7\,$\mu$m PAH feature in the observed \textit{Spitzer}-MIPS 24\,$\mu$m passband. The power-law AGN GS3 (see Fig.~\ref{FIG:AGNs}, GS5 and GS8 are out of the redshift range where $IR8$ can be computed) is marked with a bold square and an upward pointing arrow. This arrow illustrates the fact that the AGN contribution to the rest-frame 8\,$\mu$m luminosity is uncertain. Our estimate of $L_8$ was done assuming a conservative SED for the star-formation component. This assumption does not affect the determination of $L_{\rm IR}$ but does affect $L_8$. We found that using an SED like Arp 220, for example, would nearly equally well fit the IR SED if the AGN produced the bulk of $L_8$. In this case, $IR8$ would be increased by a large factor, potentially bringing the galaxy in the relation followed by the rest of the sample.

First, we note that the $z$$\sim$2 samples follows a correlation like the local galaxies. Second, one can see that galaxies \textit{on} and \textit{above} the MS are not separated as it is the case for local galaxies where "starbursts" present a compact geometry with $\Sigma_{\rm IR}^{\rm crit}$$>$$\Sigma_{\rm IR}^{\rm crit}$ while MS galaxies experience more extended star-formation. Galaxies three times above the MS (i.e., with R$_{\rm SB}$$>$3) are marked with orange symbols in Fig.~\ref{FIG:ir8_sir}-bottom, while MS galaxies below this R$_{\rm SB}$ value are represented with blue symbols. Both types span the full dynamic range in $\Sigma_{\rm IR}$. However, as we will see in the following, it is possible that some of our MS are actually experiencing a merger that enhances their star-formation by a moderate factor. 

In the framework of the dust-bounded star-formation regions interpretation that we discussed above, galaxies with a larger $IR8$ are observed in a younger stage, that precedes the destruction of the dust within the HII region. However, the typical $\Sigma_{\rm IR}$ of a galaxy at the median redshift of $z$$\sim$2.3 of our sample was larger than that of a local galaxy. We estimate this factor to be around $\sim$3.5 as a result of the fact that the SFR of MS galaxies was 20 times greater \citep{schreiber15} and the typical size of a star-forming galaxy was (1+$z$)$^{0.75}$ times smaller \citep{vanderwel12}. We find that the red solid and dashed lines that represent the local $IR8$ -- $\Sigma_{\rm IR}$ relation shifted by this factor 3.5 do provide a good fit to the ALMA data. This suggests that $IR8$ may serve as a good proxy for $\Sigma_{\rm IR}$ even in the distant Universe. 

We note that the fact that high-redshift galaxies are more metal-poor than present-day galaxies, they may naturally exhibit weaker PAH emission, hence stronger $IR8$ values, as discussed in \cite{shivaei17}. However, star-formation in high-$z$ galaxies is taking place both in more compact regions and in less metal-rich environments, and disentangling both effects might be quite complex (see \citealt{schreiber17}). 

\subsection{Star formation compactness and AGN activity}
\label{SEC:agn}
Out of our sample of 19 ALMA galaxies, 11 are detected in X-rays with the 7 Msec exposure of the \textit{Chandra} Deep Field-\textit{South} image \citep{luo17}. Following the commonly used AGN definition (also used in \cite{luo17}), we classify an X-ray source as an AGN when its total X-ray luminosity integrated over the whole 0.5 -- 7 keV range satisfies $L_X$$\geq$10$^{42.5}$ erg s$^{-1}$. With this AGN definition, we detect a total of 8 AGNs. The three power-law AGNs (see Fig.~\ref{FIG:AGNs}) are all identified as X-ray AGNs with $L_X$$\geq$10$^{43}$ erg s$^{-1}$. 

We find that the proportion of galaxies hosting an AGNs rises with increasing $\Sigma_{\rm IR}$: 75\,\% of the galaxies with $\Sigma_{\rm IR}$$\geq$3$\times$10$^{11}$  L$_{\odot}$ kpc$^{-2}$ harbor an AGN (6 out of 8 galaxies). Hence there is a clear increase of the fraction of AGNs among the galaxies with compact star-formation within the limited statistics of the present sample. This suggests that the physical mechanism responsible for the rise in star-formation compactness also efficiently feeds the central black hole. Such a relationship was also found by \cite{chang17a,chang17b}.

We also note that only 25\,\% of the AGNs are associated with galaxies the we visually classified as morphologically disturbed, i.e., mergers. The rest of the AGN population is morphologically classified as either isolated or unknown. Interestingly, the AGN fraction exhibits a tighter link with the FIR luminosity density, hence with the star-formation compactness, than with the visual identification of a merger signature. However, we cannot rule out the possibility that these galaxies are in the late-stage of a merger, which at these redshifts would not present clearly identifiable morphological perturbations. We discuss in Section~\ref{SEC:compactMS} the possibility that some of the MS galaxies may be in such a late-stage merger phase.

\section{Starbursts \textit{in} and \textit{out} of the star-formation main sequence}
\label{SEC:SBinout}
\subsection{Gas fraction and depletion time of galaxies \textit{in} and \textit{out} of the star-formation main sequence}
\label{SEC:gasfrac}
Our sample of galaxies presents some systematic biases that need to be carefully taken into account before discussing its ALMA properties. In order to dominantly include normal main sequence galaxies, our \textit{Herschel} selected sample of 8 GS galaxies is heavily biased towards massive galaxies. As a result, our discussion on $z$$\sim$2 main sequence galaxies is limited to a stellar mass of $M_{\star}$$>$10$^{11}$ M$_{\odot}$. Only three galaxies of the HUDF sample fall in the same mass range. 

\begin{figure}
 \centering
\includegraphics[width=8.7cm]{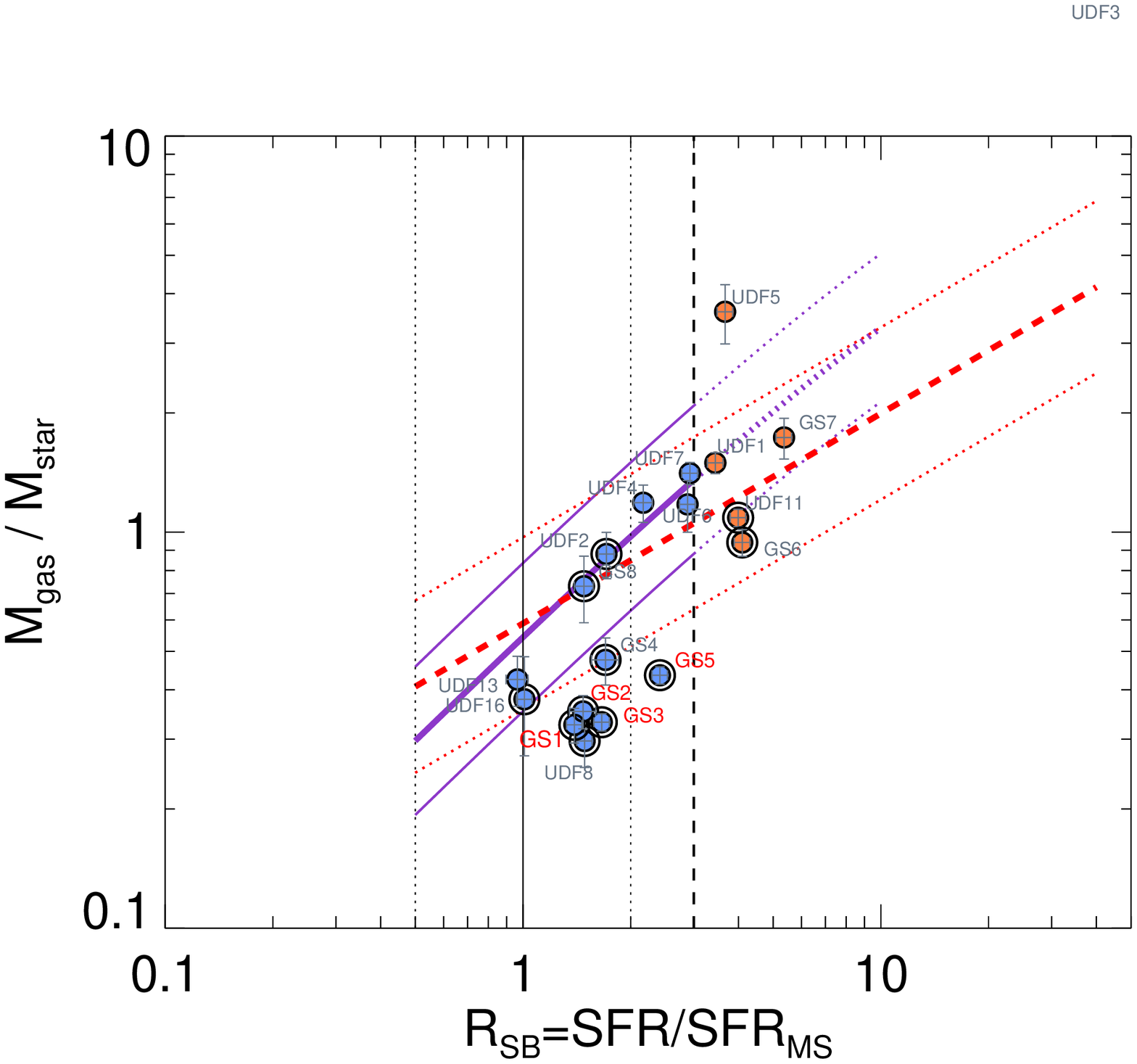}     
\includegraphics[width=8.7cm]{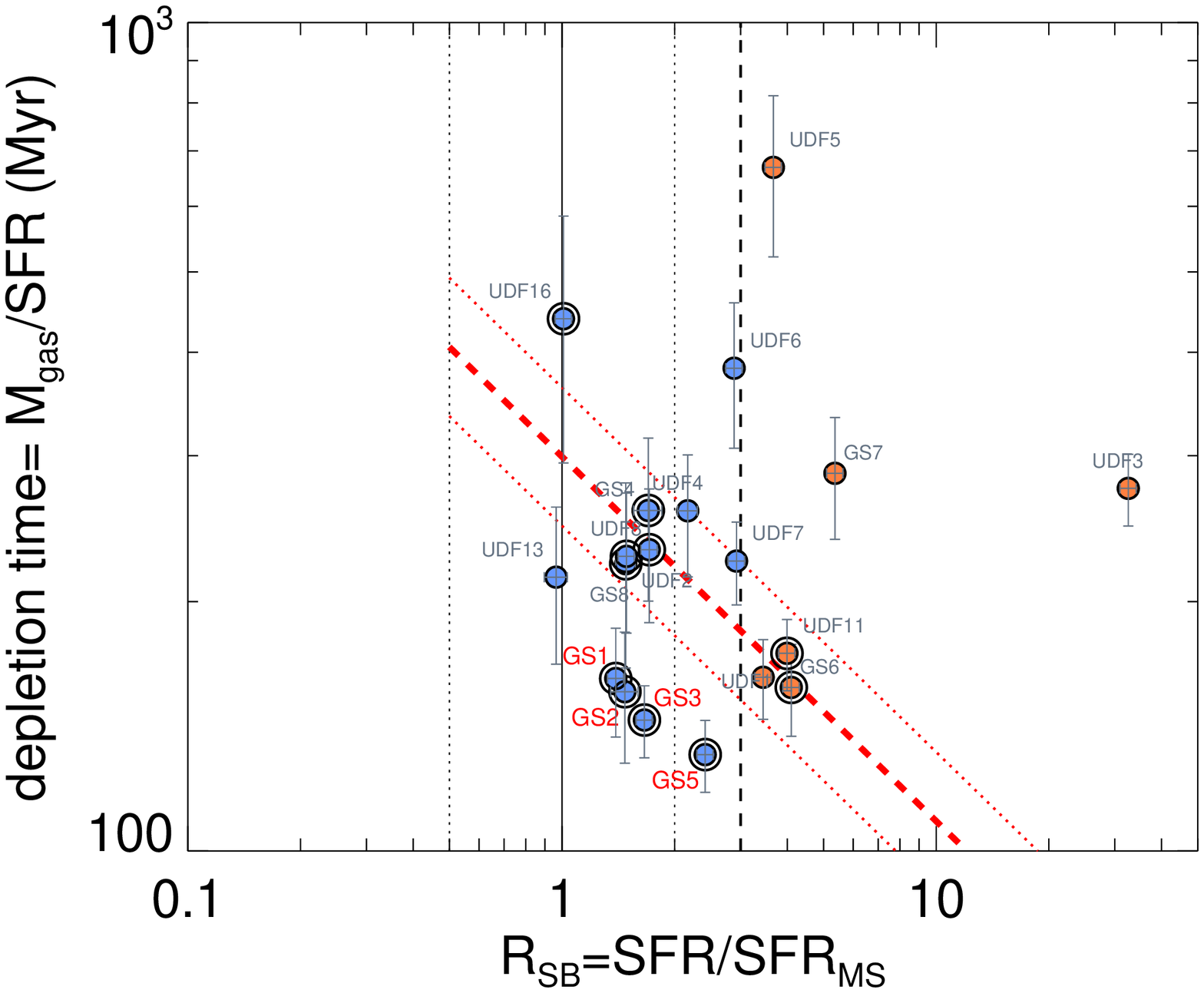}
\caption{Gas fraction (\textbf{\textit{top}}) and depletion time (\textit{\textbf{bottom}}) as a function of starburstiness, $R_{\rm SB}$ = SFR/SFR$_{\rm MS}$. The star-formation main sequence (MS) and its 68\,\% scatter are shown with a solid and dotted lines respectively. MS galaxies (R$_{\rm SB}$$\leq$3, left of vertical dashed line) are marked with cyan filled dots; starbursts above the MS (i.e., R$_{\rm SB}$$>$3) with orange filled dots. Galaxies with $M_{\star}$$>$10$^{11}$ M$_{\sun}$ are surrounded with an open circle. The dashed and dotted red lines show the relations and their scatter obtained by \cite{tacconi18} for the median $z$=2.3 and $M_{\star}^{\rm Salpeter}$=1.4$\times$10$^{11}$M$_{\odot}$ of this galaxy sample (see Eqs.~\ref{EQ:sig},\ref{EQ:tdep}). The solid purple line in the top figure shows the relation obtained by \cite{magdis12a} scaled to the median stellar mass of the sample. The four galaxies with red labels (GS1, GS2, GS3, GS5) present short depletion times typical of starbursts (they are discussed in Section~\ref{SEC:compactMS}).
}
\label{FIG:RSBrole}
\end{figure}

The characteristic mass of the main sequence galaxies discussed here is $M_{\star}$=1.4$\times$10$^{11}$M$_{\odot}$ as compared to $M_{\star}$=5$\times$10$^{10}$M$_{\odot}$ for starbursts. However, despite this mass segregation, we do find that both our main sequence and starburst galaxies follow the global trends observed using a much wider sample of nearly 1300 galaxies (combining individual detections and stacks, \citealt{tacconi18}; see Fig.~\ref{FIG:RSBrole}) as well as the original relation of \cite{magdis12a} (purple line in Fig.~\ref{FIG:RSBrole}-top, observed only up to $R_{\rm SB}$=3, extrapolated above this value). We observe a similar rise of the gas fraction and star formation efficiency (hence a drop in depletion time) as the global trends for galaxies with the median properties of our sample ($M_{\star}$=1.4$\times$10$^{11}$M$_{\odot}$, $z$=2.3) summarized in Eq.~\ref{EQ:sig} and Eq.~\ref{EQ:tdep} (derived from the Table 3 of \citealt{tacconi18}).
\begin{equation}
\sigma_{\rm gas}=\frac{M_{\rm gas}}{M_{\star}}=[0.59_{-0.24}^{+0.37}]\times R_{\rm SB}^{0.53} 
\label{EQ:sig}
\end{equation}
\begin{equation}
\tau_{\rm dep}=\frac{M_{\rm gas}}{\rm SFR}=[300^{+63}_{-52}]\times R_{\rm SB}^{-0.44}~~~{\rm [Myr]} 
\label{EQ:tdep}
\end{equation}

We note that these trends only weakly depend on stellar mass. The typical depletion time for a MS galaxy at this mass and redshift is 300 Myr (for a Salpeter IMF and 660 Myr for a Chabrier IMF as used in \citealt{tacconi18}). The agreement with the global trends apply both for our main sequence galaxies (cyan filled dots) and starbursts (orange filled dots). We can clearly see that even though our sample has a strong mass selection, it does follow very well the trend found by a sample spanning a wider dynamic range of stellar masses. Galaxies more massive than $M_{\star}$$=$10$^{11}$ M$_{\odot}$ (surrounded with an empty circle in Fig.~\ref{FIG:RSBrole}) do not depart from these global trends. We note however the presence of four MS galaxies with depletion times that are typical of starbursts ($\sim$150 Myr, labeled in red in Fig.~\ref{FIG:RSBrole}). These galaxies will be subject to a dedicated section (Section~\ref{SEC:compactMS}).

When considering those galaxies marked in red, a quick look at the depletion time of MS and starburst galaxies could be interpreted as evidence that starbursts do not form stars more efficiently than MS galaxies. But this is due to a combination of our limited statistical sample and mass range. Instead, as stated above, Fig.~\ref{FIG:RSBrole} shows that our sample is fully consistent with the fact that starbursts both exhibit higher gas fractions and shorter depletion times.

\subsection{Starbursts above the main sequence}
\label{SEC:SB}
 \begin{figure}
 \centering
      \includegraphics[width=8cm]{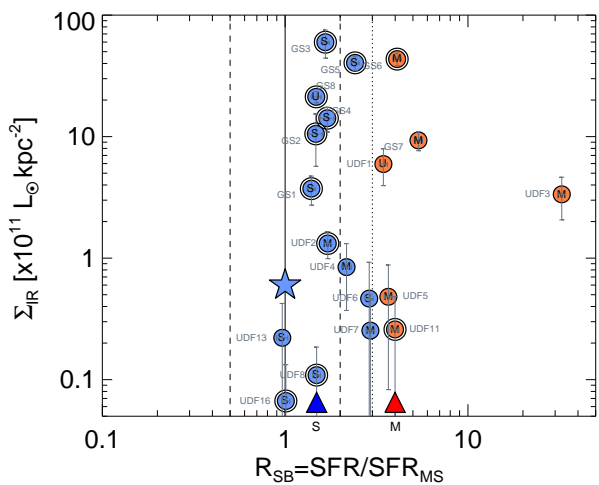}
\caption{IR luminosity surface density, $\Sigma_{\rm IR}$, as a function of starburstiness, $R_{\rm SB}$. Blue and orange filled dots separate galaxies on and 3$\times$ above the MS. The letters indicate the morphology. The median position of mergers (M) and "single/isolated" (S) galaxies are indicated at the bottom with a red and blue triangle respectively. The blue star shows the position of typical $z$$\sim$2 MS galaxies. The most massive galaxies ($M_{\star}$$>$10$^{11}$ M$_{\sun}$) are surrounded with an open circle.}
     \label{FIG:RSBSir}
  \end{figure}
 \begin{figure}
 \centering
   \includegraphics[width=8cm]{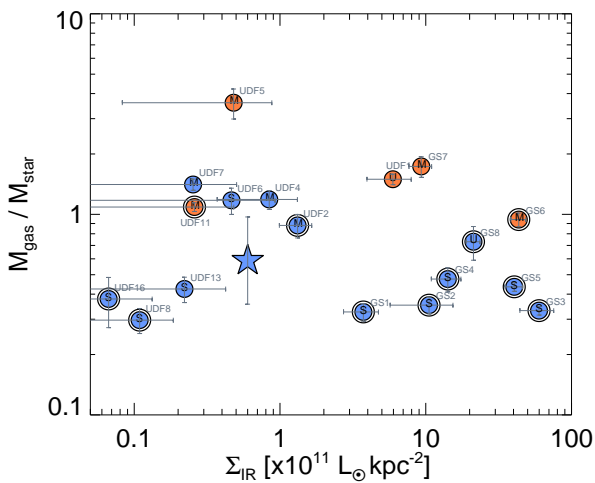}
\caption{Gas fraction (in\,\%) as a function of IR luminosity surface density, $\Sigma_{\rm IR}$. Caption as in Fig.~\ref{FIG:RSBSir}.}
     \label{FIG:gasSir}
  \end{figure}

We visually classified galaxies from the \textit{HST} \textit{H}-band images, hence in the rest-frame visible, in two broad categories: single/isolated galaxy (S), with no clear sign of perturbation, and galaxies with a perturbed morphology that we listed as mergers (M) in Table~\ref{TAB:sample}. Two cases are listed as undefined: GS8, discussed in detail in Section~\ref{SEC:dark}, and UDF1, which appears point-like even in the \textit{HST} image. 

We find that all the starbursts of our sample (here defined as galaxies with R$_{\rm SB}$$>$3) exhibit the morphologies of mergers (see Fig.~\ref{FIG:IM_SB}). Instead, the median starburstiness of the single/isolated galaxies (S-type) is R$_{\rm SB}$$\sim$1.5, hence nearly equal to the median of the MS. As noted in Sect.~\ref{SEC:gasfrac}, star-formation in starbursts is not only associated with a shorter depletion time but also with an enhanced gas fraction (see Fig.~\ref{FIG:RSBrole}). This increase of the gas fraction during a merger event could be due to the impact of the merger on the circum-galactic gas surrounding the galaxies before the merger. This gas may be driven towards the center of the galaxies. Hydrodynamic simulations accounting for the presence of circum-galactic matter, and its possible infall induced during a merger, do not exist at present to our knowledge. Such simulations should be performed to test this hypothesis.

We searched for a signature in the IR luminosity surface density, but we did not find any trend neither with starburstiness (Fig.~\ref{FIG:RSBSir}) nor gas fraction (Fig.~\ref{FIG:gasSir}). We describe below how we determined the characteristic IR luminosity surface density of a MS at this mass and redshift. The SFR of a MS galaxy at our median redshift of $z$=2.3 and stellar mass of $M_{\star}$=1.4$\times$10$^{11}$ M$_{\odot}$ is SFR$^{\rm MS}$= 224 M$_{\odot}$ yr$^{-1}$ (from \citealt{schreiber15}), which corresponds to $L_{\rm IR}^{\rm MS}$=1.3$\times$10$^{12}$ L$_{\odot}$ using the \cite{kennicutt98} relation for a Salpeter IMF. The typical $H$-band size of such galaxy is given by the relation shown in the Fig.5 of \cite{tadaki17}: log$_{10}$[R$_e^{\rm MS}(H)$]=0.14$\times$log$_{10}$[$M_{\star}^{\rm Salpeter}$/1.7]-1.11, i.e., 2.6 kpc. We used the tyical R$_e^{\rm MS}(H)$/R$_e^{\rm MS}$(870\,$\mu$m)=1.4 ratio from \cite{fujimoto17} to derive a characteristic ALMA size of R$_e^{\rm MS}$(870\,$\mu$m)= 1.85 kpc.
The resulting characteristic IR luminosity surface density is $\Sigma_{\rm IR}$ = 6$\times$10$^{10}$ L$_{\odot}$ kpc$^{-2}$.

\subsection{Spatial offset between UV and far-IR light distributions}
\label{SEC:offset}
 \begin{figure}
 \centering
      \includegraphics[width=8.7cm]{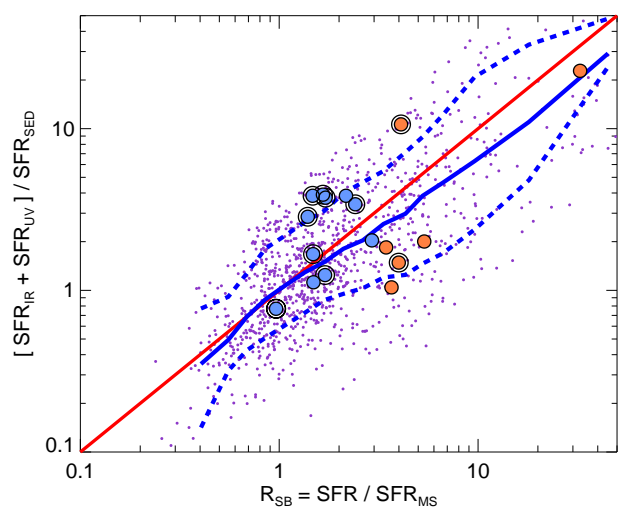}
      \caption{Excess of total SFR, SFR$_{\rm tot}$=SFR$_{\rm IR}$$+$SFR$_{\rm UV}$, with respect to the SFR$_{\rm SED}$ determined by fitting the rest-frame UV-optical-NIR, as a function of starburstiness, R$_{\rm SB}$=SFR$_{\rm tot}$/SFR$_{\rm MS}$. The purple dots show the positions of the galaxies from all four CANDELS fields with an \textit{Herschel} detection with 1.5$\leq$$z$$\leq$2.5. Sliding median of the purple points in thick solid blue line and 68\,\% median absolute deviation in dashed. Symbols for the ALMA galaxies as in Fig.~\ref{FIG:RSBrole}. The red line marks the direct proportionality. The solid and dashed blue lines indicate the sliding median and its 68\,\% absolute deviation.
      }
     \label{FIG:sfr_rsb}
  \end{figure}
We can see in Fig.~\ref{FIG:IM_SB} that there is a systematic offset between the UV and ALMA light distributions in starbursts three times above the main sequence. Such spatial separation was already noticed in the literature (see \citealt{rujopakarn16}, \citealt{barro16}, \citealt{hodge16}). This suggests that the use of the UV emission to determine the total SFR of starburst galaxies may lead to strong underestimates, even when accounting for a dust attenuation correction based on the UV slope. We show that this is indeed the case in Fig.~\ref{FIG:sfr_rsb}.

The rest-frame far-UV emission of $z$$\sim$2 galaxies is commonly used to derive total SFR after applying a correction for dust attenuation either from the determination of the UV-slope, $\beta$, or from the modelling of the full UV-optical SED with e.g., the \cite{calzetti00} attenuation law (see e.g., \citealt{meurer99}, \citealt{calzetti00}, \citealt{daddi04}, \citealt{overzier11}, \citealt{buat12}, \citealt{reddy12},  \citealt{heinis13}, \citealt{pannella15}). The average consistency for most galaxies between both ways to derive total SFR -- \textit{(1)} SFR$_{\rm tot}$=SFR$_{\rm IR}$$+$SFR$_{\rm UV}$ and \textit{(2)} SFR$_{\rm tot}$=SFR$_{\rm SED}$ or SFR$_{\rm UV}^{\rm corrected}$ -- suggests that the young massive stars responsible for the UV and FIR emission are located in the same region. Alternatively, this consistency would be difficult to explain if a spatial segregation was found for most galaxies.

To address this question, we compare the ratio of both estimates of the total SFR, [SFR$_{\rm IR}$
$+$SFR$_{\rm UV}$]$/$SFR$_{\rm SED}$, to the distance to the MS, R$_{\rm SB}$=SFR$/$SFR$_{\rm MS}$. The purple dots in Fig.~\ref{FIG:sfr_rsb} show the positions of all the galaxies with 1.5$\leq$$z$$\leq$2.5 detected in at least one \textit{Herschel} band (to ensure a robust determination of L$_{\rm IR}$) in all four CANDELS fields (GOODS-S, GOODS-N, UDS and CANDELS-COSMOS) and in the COSMOS 2 degree field. 
The sliding median (thick solid blue line, and 68\,\% dispersion in dashed) shows that SFR$_{\rm SED}$ and [SFR$_{\rm IR}$$+$SFR$_{\rm UV}$] provide the same estimate of SFR$_{\rm tot}$ within 0.3 \textit{dex} for MS galaxies, i.e., where R$_{\rm SB}$$\sim$1. We note that the density of the purple points at the [1,1] position is not representative of the actual number of MS galaxies because at the \textit{Herschel} sensitivity limit only the most massive MS galaxies are detected. Hence this result is consistent with previous studies which found that on average both SFR estimators are consistent for typical MS galaxies.

However, the sliding median of the relation between [SFR$_{\rm IR}$$+$SFR$_{\rm UV}$]/SFR$_{\rm SED}$ and R$_{\rm SB}$ (solid blue line) nearly follows the line of direct proportionality (solid red line). The resolved ALMA images offer a nice explanation for this increasing "wrongness" of SFR$_{\rm SED}$ with increasing "starburstiness" by showing a clear offset between the UV and IR light distributions in starbursts (Fig.~\ref{FIG:IM_SB}). A galaxy with R$_{\rm SB}$$\sim$4 forms stars with an intensity that is typically 4 times greater than the one derived from SED fitting. This implies that galaxies experiencing a starburst phase may wrongly be interpreted as normal MS star-forming galaxies in the absence of direct far-infrared measurements. Equivalently, R$_{\rm SB}$ is a good proxy for the wrongness of SED-derived SFR. This is particularly true in the cases of GS6 and UDF3 for which SFR$_{\rm SED}$ is wrong by a factor 11 and 23 respectively (the two highest ALMA points in Fig.~\ref{FIG:sfr_rsb}).

Among the ALMA galaxies that belong to the MS (blue symbols in Fig.~\ref{FIG:sfr_rsb}), a group of galaxies clusters around [SFR$_{\rm IR}$$+$SFR$_{\rm UV}$]/SFR$_{\rm SED}$$\sim$1 as expected for typical MS galaxies. Their three color V,I,H images (from the third to last row of Fig.~\ref{FIG:IM_MS} except UDF2 and UDF4) show that these galaxies present the shapes of large disks with no obvious sign of disturbance and no strong offset between the UV and IR light. Here the UV and far-infrared light come from the same region consistently with the fact that [SFR$_{\rm IR}$$+$SFR$_{\rm UV}$] and SFR$_{\rm SED}$ are in good agreement, although with a slight offset for UDF13.

However, our sample also includes a group of MS galaxies with a strong excess of heavily obscured star formation with [SFR$_{\rm IR}$$+$SFR$_{\rm UV}$]/SFR$_{\rm SED}$$\sim$4. This group includes the four galaxies GS1, GS2, GS3 and GS5 marked with red labels in Fig.~\ref{FIG:RSBrole} (the first four images in Fig.~\ref{FIG:IM_MS}) that will be discussed in Section~\ref{SEC:compactMS} as well as UDF2 and UDF4. These MS galaxies present an excess dusty star-formation rate as compared to the SED-fitting one similar to that of starbursts, even though they belong to the MS.

\subsection{Compact starbursts \textit{hidden} within the main sequence}
\label{SEC:compactMS}
\begin{figure}
 \centering
      \includegraphics[width=8.7cm]{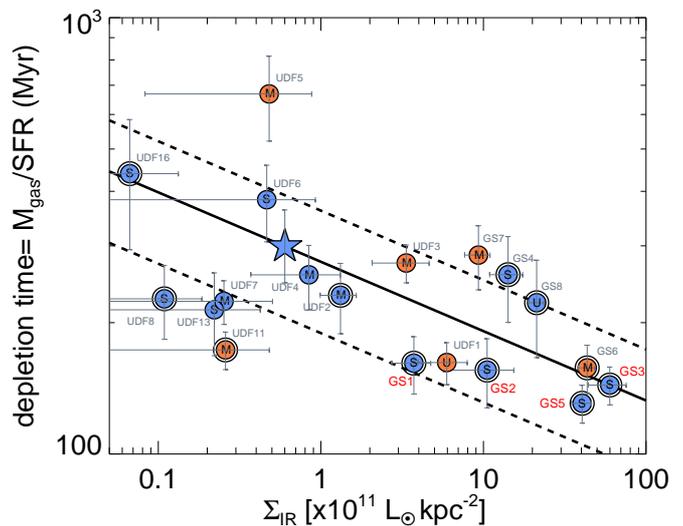}     
\caption{Depletion time as a function IR luminosity surface density, $\Sigma_{\rm IR}$. The blue star shows the typical depletion time of $z$$\sim$2.3 MS galaxies. Colors and symbols as in Fig.~\ref{FIG:RSBSir}. The four MS galaxies with the shortest depletion times are identified with red labels (GS1, GS2, GS3, GS5) as in Fig.~\ref{FIG:RSBrole}. The solid and dashed lines are the fit to the sliding median and its 68\,\% scatter.}
     \label{FIG:hiddenSB}
  \end{figure}

The depletion time, $\tau_{\rm dep}$, of our galaxy sample decreases with IR luminosity surface density following Eq.~\ref{EQ:tdep_sir} (Fig.~\ref{FIG:hiddenSB}).
\begin{equation}
\tau_{\rm dep}=[276_{-87}^{+85}]\times \Sigma_{\rm IR}^{-0.16} 
\label{EQ:tdep_sir}
\end{equation}

This implies that the galaxies hosting the most compact star-formation convert their gas reservoirs in stars more efficiently.
The six galaxies forming stars with the shortest depletion times ($\tau_{\rm dep}$$\sim$146 Myr) present a $\Sigma_{\rm IR}$=1.5$\times$10$^{12}$ L$_{\odot}$ kpc$^{-2}$, or equivalently 255 M$_{\odot}$ yr$^{-1}$ kpc$^{-2}$. Interestingly, four out of these six galaxies are MS galaxies, those that we marked with red labels in Fig.~\ref{FIG:RSBrole} (GS1, GS2, GS3 and GS5). In comparison, the  typical depletion time for a MS galaxy of similar mass and redshift is more than twice longer (blue star in Fig.~\ref{FIG:hiddenSB}) and its IR luminosity density is 25 times lower, $\Sigma_{\rm IR}^{\rm MS}$ = 6$\times$10$^{10}$ L$_{\odot}$ kpc$^{-2}$.

These four main sequence galaxies exhibit \textit{(i)} low gas fractions, \textit{(ii)} short depletion times and \textit{(iii)} extreme IR luminosity surface densities. They present all the characteristics of starbursts despite their location within the standard deviation of the MS, hence can be seen as compact starburst \textit{hidden} within the main sequence. 

We checked whether these galaxies could be mistakenly identified as part of the MS while being in reality starbursts. Only one of them, GS5, has a spectroscopic redshift whereas the other three only have photometric redshifts. In order to determine how the uncertainty on the redshifts of these galaxies may impact their starburstiness, we looked at the 16 and 84 percentile values of these redshifts from their probability distribution functions. All three have very accurately determined photometric redshifts with $\delta$$z$$/$(1+$z$) values of [$-$1.9,$+$0.9], [$-$19,$+$22] and [$-$1.3,$+$1.1]$\times$10$^{-3}$ for GS1, GS2 and GS3 respectively. If we combine these redshift uncertainties with the error bars on the photometric measurements, we find that the resulting stellar masses vary by $\pm$11, 1 and 2\,\% respectively. This would not modify their positions relative to the MS. A more important potential caveat comes from the possible contribution of an AGN component to the photometry. We discussed the contribution of AGNs to the multi-wavelength SEDs in Section~\ref{SEC:sed} but limiting ourselves to their impact on the IR luminosity and dust mass. However, AGNs may as well contaminate the rest-frame near-IR emission of the galaxies which stellar masses may then be overestimated if the AGN contribution was thought to be due to stellar emission. There is however no reliable way to cleanly extrapolate the SED of the dusty AGN component to the near-IR and optical range to our knowledge. In the absence of robust model or template AGN SEDs that would allow such extrapolation, we have decided to quantify this effect by re-computing the stellar masses of GS1, GS3 and GS5 -- where we found evidence for the presence of dust heated by an AGN -- after excluding the near-IR photometry coming from IRAC, that would be the most contaminated by hot dust from an AGN. The new stellar masses are the same for GS3, 2\,\% higher for GS1 and 65\,\% lower for GS5. GS5, which had a starburstiness of $R_{\rm SB}$=2.4 when using the full SED to determine its stellar mass, would have a revised value of $R_{\rm SB}$=4 that would make it a starburst under our definition of $R_{\rm SB}$$>$3 for starbursts. In conclusion, we find that the bulk of these compact starbursts hidden in the MS would remain MS galaxies even when including potential uncertainties on their redshift and AGN contamination with the exception of GS5 that may well be a starburst if an AGN does indeed generate the bulk of its near-IR emission.

A third common point between these galaxies is particularly striking: they all exhibit a strong dichotomy between their IR and UV light distributions (see Fig.~\ref{FIG:IM_MS}). This offset explains why SFR$_{\rm SED}$ is a bad proxy for SFR$_{\rm IR}$$+$SFR$_{\rm UV}$ for these galaxies. Hence these MS galaxies present equivalent offsets and SFR$_{\rm SED}$ "wrongness" than the "starburst" galaxies lying above the MS, but they belong to the MS. In fact, they do exhibit two important differences with the galaxies lying above the MS.

First, they all exhibit a similar morphology in the $H$-band. They look like face-on disks or alternatively like spheroids. Considering that the probability to see these galaxies only face-on is low, we believe that at least some of them are spheroids or will become so. This is at least the case of GS1 with an \textit{H}-band S\'ersic index of $n_H^{\rm Sersic}$=3.75 and possibly also GS3 with $n_H^{\rm Sersic}$=2.03 and $n_{ALMA}^{\rm Sersic}$=4.7$\pm$1.7. Second, they all have small gas fractions more than twice lower than the galaxies lying above the MS. 

The physical origin of these compact starbursts \textit{hidden} in the main sequence is not completely clear at this stage. 
They experience such rapid star formation that they will exhaust their gas reservoirs and stop forming stars in only $\sim$150 Myr. Hence if they are not replenished by diffuse intergalactic matter, these massive galaxies will become passive.
It has been debated whether $z$$\sim$2 compact early-type galaxies (ETG) were the natural result of the global shrinking of galaxies going to higher redshifts or whether their existence could be seen as the signature of a "wet compaction" mechanism. The median 5000\,\AA\,effective radius (observed $H$-band) of $z$$\sim$2 massive ETGs is indeed of only $R_{e}$$\sim$1 kpc whereas the radii of star-forming galaxies in the same mass range is close to $R_{e}$$\sim$4 kpc (see \citealt{vanderwel14} and references therein).

In a study of six massive, compact, dusty star-forming galaxies at $z$$\sim$2.5 (including three galaxies in common with the present paper), \cite{barro16} proposed to explain compact dusty star-forming galaxies as galaxies experiencing a wet compaction event. Our analysis conforts the idea that a mechanism is at play in a sub-population of $z$$\sim$2 MS galaxies that leads to a strong concentration of their gas and dust reservoirs, associated with a drop of their depletion time.

The question of which physical mechanism produced this gas concentration remains open. \cite{dekel14} proposed to explain the wet compaction as a result of a violent disk instability (VDI) in a gas-rich system. However, more recent studies indicate that the effect of VDI alone may not be strong enough to generate such strong concentrations of gas. A violent mechanism, such as a major or a minor merger, appears to be required in most cases to strongly reduce the angular momentum (A.Dekel private communication and paper in preparation). This was already coming out of the simulations presented in \cite{zolotov15} and \cite{tacchella16} although  \cite{tacchella16} "find that the high-SFR galaxies in the upper envelope of the MS are compact, with high gas fractions and short depletion times ('blue nuggets')" whereas our compact MS galaxies exhibit lower gas fractions than typical MS galaxies.

The idea that mergers may affect MS galaxies without inducing strong starbursts is reinforced by the high-resolution hydrodynamic simulations of \cite{fensch17} showing that the SFR of two nearly equal mass galaxies with 60\,\% gas fractions (mimicking the large gas fractions of high-redshift galaxies) does not vary much during the merger event. Only a mild rise of the SFR by a factor 2.5 is observed at the late stage of the merger. The starbursts hidden in the MS could be explained by this phase in which the gas fraction has dropped and the SFR has risen without bringing the galaxy above the MS. The morphology of these compact star-forming galaxies is consistent with that of a spherical galaxy having lost its disk during the encounter. This would mean that mergers could be playing an important role in both triggering star-formation and transforming disks into spheroids \textit{within} the MS.

However, our sample of $z$$\sim$2 MS galaxies is limited to massive galaxies with M$_{\star}$$>$10$^{11}$ M$_{\odot}$. Because of this strong bias in favor of massive galaxies, we cannot generalize the existence of starbursts \textit{hidden} in the main sequence to the full population of $z$$\sim$2 galaxies. It is possible that these galaxies only exist in the very high mass end of the main sequence or are less important in relative fractions at lower masses. They may even represent a large fraction of the most massive galaxies at $z$$\sim$2. Only deeper ALMA data will allow us to answer this question by probing the sizes of individual galaxies at lower stellar masses, knowing that the typical size of 5$\times$10$^{10}$ M$_{\odot}$ galaxies were found to be $\sim$5 kpc from stacking measurements \citep{lindroos16}. At this stage, it is entirely possible that this mechanism is typical of massive galaxies only and as such it may be more directly to the mass of the galaxies than to a recent major merger.

Lastly, we are left with the following question regarding starbursts above the MS. If major mergers of gas-rich systems do not produce starbursts, then why do we see starbursts at all at $z$$\sim$2 ? The answer to this question may be found in the gas fraction measured in starbursts. As we showed in Fig.~\ref{FIG:RSBrole} and as discussed in e.g. \cite{tacconi18}, starbursts are not only forming stars efficiently, they exhibit an equivalently strong excess in their gas fraction. In that sense, they are different from the late-stage merger phase that we discussed above if their enhanced gas content is not artificially created by an underestimation of their metallicity (underestimating the metallicity of a dusty starbursts would imply underestimating their dust-to-gas ratio hence overestimating their gas mass). A possible explanation for these gas enriched starbursts could be that they are fed by the infall of circum-galactic material induced during the merger in some specific conditions. How else can one understand that two galaxies with a given gas fraction end up forming a merged system with an higher gas fraction ?
Idealized hydrodynamic simulations would be needed in order to determine if indeed there exists specific conditions that may induce an efficient infall of circum-galactic matter during a major merger. Simulations that would allow a quantitative comparison of the same merger conditions with and without circum-galactic matter do not exist at present to our knowledge. Such simulations should be performed to study the effects of circum-galactic gas and its possible infall during a merger. Simulations in full cosmological context naturally include intergalactic infall (e.g., \citealt{martin17}) but lack comparison cases of galaxies in the exact same conditions with and without a circum-galactic reservoir of gas.

\section{Conclusions}
\label{SEC:conclusions}
In this paper, we have presented the properties of a sample of 19 $z$$\sim$2 star-forming galaxies located in the GOODS-\textit{South} field 
with high angular resolution imaging and photometry from the UV to the mm range, combining data from the ground, \textit{HST}, \textit{Spitzer}, \textit{Herschel} and ALMA.
The ALMA data combine information from deep integrations at 870\,$\mu$m on 8 $z$$\sim$2 sources selected among the brightest \textit{Herschel} sources with data on 11 galaxies coming from a published 1.3 mm survey of the Hubble Ultra-Deep Field, HUDF (\citealt{dunlop17}, \citealt{rujopakarn16}). These galaxies were selected in the high stellar mass range of the star-forming galaxies at 1.5$<$$z$$<$2.5 in order to include normal star-forming galaxies within the standard deviation of the star formation main sequence, as well as starbursts well above the main sequence.

Our main results are listed below:
\begin{enumerate}

\item \textit{Heavily obscured $z$$\sim$2 massive galaxies:} out of 8 ALMA pointings targeting \textit{Herschel} sources with an optical counterpart at $z$$\sim$2, one of the ALMA detection presents two properties that led us to identify a background \textit{HST}-dark galaxy as the real \textit{Herschel} and ALMA counterpart. These properties are \textit{(i)} an offset between ALMA and \textit{HST} larger than the astrometric error, \textit{(ii)} a far-IR SED that peaks at $\sim$400\,$\mu$m. We identified a potential counterpart at a redshift that we estimate to be of $z$$\sim$3.24. A 67 armcin$^2$ survey of GOODS-\textit{South} at 1.1mm (P.I. D.Elbaz) appears to confirm the possibility that typically 10-15\,\% of the ALMA sources are associated with "optically dark" galaxies (Franco et al., in prep.).

\item \textit{Compact star-formation in $z$$\sim$2 massive galaxies:} while dusty star-formation is resolved in all ALMA galaxies, we find a common point among these massive $z$$\sim$2 star-forming galaxies: their star formation appears to be concentrated towards the mass center of the galaxies and to be 1.45$\pm$1.0 times more compact than at 1.6\,$\mu$m, i.e., observed 
\textit{HST}--WFC3 \textit{H}-band.

\item \textit{Minor contribution of kpc-clumps of star-formation:} kpc-clumps of star formation seen in the UV 
do not contribute a large fraction of the total star formation rate measured in these massive $z$$\sim$2 galaxies. This is consistent with the small SFR attributed to the giant UV clumps (see \citealt{elmegreen09}). In one case, we see marginal evidence for the ALMA detection of a kpc-clump.

\item \textit{The $IR8$ color index as a probe of star-formation compactness:} we present an updated version of the $IR8$-$\Sigma_{\rm IR}$ relation (introduced in \citealt{elbaz11}) for local galaxies, now including resolved \textit{Herschel} images for local galaxies, and discuss its extension to $z$$\sim$2. The $IR8$ color index (=L$_{\rm IR}$/L$_8$) and the IR luminosity surface density, $\Sigma_{\rm IR}$ (a proxy for the dusty star formation density) present a tight correlation for local galaxies with a standard deviation of only \textit{rms}=0.11 dex. The $z$$\sim$1.5-2.5 galaxies resolved with ALMA appear to follow the local relation, although galaxies both \textit{within} and \textit{above} the main sequence can exhibit high $\Sigma_{\rm IR}$ values.

\item \textit{A connection between AGN activity and star-formation compactness:} galaxies hosting an AGN appear to be outliers to the $IR8$-$\Sigma_{\rm IR}$ relation both locally and at $z$$\sim$2. While a correction of the contribution of hot dust continuum due to AGN heating may explain in part or possibly completely the position of these galaxies, by lowering their $IR8$, galaxies hosting an AGN are found to be systematically associated with the most compact star-forming galaxies. This suggests that the mechanism responsible for the most compact star-forming galaxies also switches on the AGN, or possibly that the AGN plays a role in triggering the compact star-formation through positive feedback (see e.g., \citealt{silk13}, \citealt{molnar17}, \citealt{elbaz09}).

\item \textit{On the origin of merger-driven starbursts above the star formation main sequence:} galaxies above the SFR-M$_\star$ main sequence systematically exhibit the visual morphology of perturbed galaxies as expected in the case of major mergers. We confirm that their increased efficiency of star-formation is accompanied with a rise of their gas fraction. If it is not artificially created by a larger metallicity in starbursts (which would lead to an overestimate of their gas content), this increase in gas content may be explained by the infall of circum-galactic material induced during the merger in some specific conditions that remain to be determined. 

\item \textit{Compact starbursts hidden in the high-mass end of the main sequence:} we find that the 
depletion time, $\tau_{\rm dep}$ (the time for a galaxy to consume its molecular gas reservoir), drops with increasing IR luminosity surface density, $\Sigma_{\rm IR}$. The galaxies with the shortest depletion times ($\tau_{\rm dep}$$\sim$150 Myr) and highest IR luminosity surface densities ($\Sigma_{\rm IR}$$\sim$1.5$\times$10$^{12}$ L$_{\odot}$ kpc$^{-2}$) are massive galaxies (M$_{\star}$$>$10$^{11}$ M$_{\odot}$) dominantly located in the upper part of the main sequence. Hence they are starbursts "hidden" within the main sequence. Due to our \textit{Herschel} selection, we do not know whether similar galaxies exist at lower masses, hence these compact starbursts in the MS may represent the late stage of star-formation in massive galaxies before their passivization. The low gas fraction of these galaxies also favors the possibility that they are experiencing a last stage of star-formation prior to become passive. The physical origin of the compact starbursts within the main sequence remains uncertain at this stage. Understanding it will be crucial in the future to unveil the origin of the compactness of the stars of early-type galaxies observed at $z$$\sim$2.

\item \textit{Starburstiness and star-formation compactness measure the wrongness of UV-corrected SFR:}  we present a near proportionality between the ratio of SFR$_{\rm tot}$ (=SFR$_{\rm IR}$$+$SFR$_{\rm UV}$) over SFR$_{\rm SED}$ (determined by fitting the rest-frame UV-optical-NIR) and the distance to the main sequence or starburstiness (R$_{\rm SB}$=SFR$_{\rm tot}$/SFR$_{\rm MS}$). Hence, SED-fitting underestimates the SFR with increasing starburstiness. This can be explained by the fact that the regions responsible for the bulk of rest-frame UV and far-IR emission occupy very distinct locations in starbursts as shown by our ALMA images. Interestingly, this is also the case of the "starbursts hidden in the MS", i.e. MS galaxies experiencing compact star-formation and short depletion times. This suggests that SFR derived from SED-fitting will most probably miss this population which may be a problem when searching for causes of star-formation variations such as environment effects.

\end{enumerate}

\begin{acknowledgements}
This paper makes use of the following ALMA data:
ADS/JAO.ALMA\#2012.1.00983.S. ALMA is a partnership of ESO (representing
its member states), NSF (USA) and NINS (Japan), together with NRC
(Canada), MOST and ASIAA (Taiwan), and KASI (Republic of Korea), in
cooperation with the Republic of Chile. The Joint ALMA Observatory is
operated by ESO, AUI/NRAO and NAOJ.
R.L. acknowledges support from Comit\'e Mixto ESO-GOBIERNO DE CHILE, GEMINI-CONICYT FUND 32130024, FONDECYT Grant 3130558, and CEA-Saclay. This research was supported by the French Agence Nationale de la Recherche (ANR) project ANR-09-BLAN-0224. DE acknowledges the contribution of the FP7 SPACE project ASTRODEEP (Ref.No: 312725), supported by the European Commission. We acknowledge financial support from the "Programme National de Cosmologie et des Galaxies" (PNCG) of CNRS/INSU, France. GEM acknowledges support from the Carlsberg Foundation, the ERC Consolidator Grant funding scheme (project ConTExt, grant number No. 648179), and a research grant (13160) from Villum Fonden. W.R. is supported the Thailand Research Fund/Office of the Higher Education Commission Grant Number MRG6080294. T.D-S. acknowledges support from ALMA-CONICYT project 31130005 and FONDECYT regular project 1151239.

\end{acknowledgements}

\bibliographystyle{aa} 
\bibliography{delbaz_masterbibtex.bib}

\end{document}